\documentclass[journal]{IEEEtran}

\usepackage[numbers]{natbib}
\usepackage{graphicx}
\usepackage{subfigure}
\usepackage{pifont}
\usepackage{url}
\usepackage{hyperref}
\usepackage{amssymb}
\usepackage{amsmath}
\usepackage[ruled,linesnumbered]{algorithm2e}
\usepackage{multirow}
\usepackage{threeparttable}
\usepackage{makecell}
\usepackage{tabu}
\usepackage{tablefootnote}

\newcommand{\PiPar}{\texttt{PiPar}}

\begin{document}

\title{\texttt{PiPar}: Pipeline parallelism for collaborative machine Learning}


\author{
    Zihan~Zhang,
    Philip~Rodgers,
    Peter~Kilpatrick,
    Ivor~Spence,
    and~Blesson~Varghese
    \IEEEcompsocitemizethanks{
        \IEEEcompsocthanksitem Z. Zhang and B. Varghese are with the School of Computer Science, University of St Andrews, UK. Corresponding author: zz66@st-andrews.ac.uk
        \IEEEcompsocthanksitem P. Rodgers is with Rakuten Mobile, Inc., Japan.
        \IEEEcompsocthanksitem P. Kilpatrick and I. Spence are with the School of Electronics, Electrical Engineering and Computer Science, Queen's University Belfast, UK.
    }
}

\markboth{}{Zhang \MakeLowercase{\textit{et al.}}: \texttt{PiPar}: Pipeline Parallelism for Collaborative Machine Learning}
%


\maketitle

\begin{abstract}
    Collaborative machine learning (CML) techniques, such as federated learning, have been proposed to train deep learning models across multiple mobile devices and a server. CML techniques are privacy-preserving as a local model that is trained on each device instead of the raw data from the device is shared with the server. However, CML training is inefficient due to low resource utilization. We identify idling resources on the server and devices due to sequential computation and communication as the principal cause of low resource utilization. A novel framework \PiPar\ that leverages pipeline parallelism for CML techniques is developed to substantially improve resource utilization. A new training pipeline is designed to parallelize the computations on different hardware resources and communication on different bandwidth resources, thereby accelerating the training process in CML. A low overhead automated parameter selection method is proposed to optimize the pipeline, maximizing the utilization of available resources. The experimental results confirm the validity of the underlying approach of \PiPar\ and highlight that when compared to federated learning: (i) the idle time of the server can be reduced by up to 64.1$\times$, and (ii) the overall training time can be accelerated by up to 34.6$\times$ under varying network conditions for a collection of six small and large popular deep neural networks and four datasets without sacrificing accuracy. It is also experimentally demonstrated that \PiPar\ achieves performance benefits when incorporating differential privacy methods and operating in environments with heterogeneous devices and changing bandwidths.
\end{abstract}


\IEEEpeerreviewmaketitle

\section{Introduction}
\label{sec:intro}
Deep learning has found application across a range of fields including computer vision~\cite{Kendall2017, maskrcnn}, natural language processing~\cite{devlin-etal-2019-bert, NEURIPS2020_1457c0d6} and speech recognition~\cite{DBLP:journals/corr/abs-1211-3711,DBLP:journals/corr/HannunCCCDEPSSCN14}. However, there are important data privacy and regulatory concerns in sending data generated on mobile devices to geographically distant cloud servers for training deep learning models. A new class of machine learning techniques has therefore been developed under the umbrella of collaborative machine learning (CML) to mitigate these concerns~\cite{Thapa2021}. CML does not require data to be sent to a server for training deep learning models; rather the server shares models with devices that are then locally trained on the device. 

CML is used in many real-world use-cases comprising a central server and multiple homogeneous mobile devices. Smartphone manufacturers, for example, analyze user data to improve the performance of a specific smartphone model~\cite{gboard, 9354925}. For instance, CML can be employed to analyze the battery usage patterns of individual users on their phones to offer personalized plans for optimizing battery life. Similarly, CML can be used to analyze the typing habits of the users and then automatically complete and correct the typing of the users.

There are three notable CML techniques reported in the literature, namely federated learning (FL)~\cite{DBLP:journals/corr/KonecnyMR15, DBLP:journals/corr/KonecnyMRR16, DBLP:journals/corr/KonecnyMYRSB16, pmlr-v54-mcmahan17a}, split learning (SL)~\cite{DBLP:journals/corr/abs-1810-06060, DBLP:journals/corr/abs-1812-00564} and split federated learning (SFL)~\cite{DBLP:journals/corr/abs-2004-12088, DBLP:journals/corr/abs-2107-04271}. 
However, these techniques under-utilize both compute and network resources, which results in training times that do not meet real-world requirements. The cause of resource under-utilization and the resulting performance inefficiency in the three CML techniques is considered next. 

In FL, each device trains a local model of a deep neural network (DNN) using the data it generates. Local models are uploaded to the server and aggregated as a global model at a pre-defined frequency. However, the workload of the devices and the server is usually imbalanced~\cite{9492062, 9260194, DBLP:journals/corr/abs-2107-04271}. This is because the server is only employed when the local models are aggregated and is idle for the rest of the time.  

In SL, a DNN is usually decomposed into two parts, such that the initial layers of the DNN are deployed on a device and the remaining layers on the server. A device trains the partial model and sends the intermediate outputs to the server where the rest of the model is trained. The training of the model on devices occurs in a round-robin fashion. Hence, only one device or the server will utilize its resources while the rest of the devices or the server are idle~\cite{DBLP:journals/corr/abs-2004-12088, Thapa2021}. 

In SFL, which is a hybrid of FL and SL, the DNN is split across devices and the server. The devices, however, unlike SL, train the local models concurrently. Nevertheless, the server must wait while the devices train the model and transfer data, and vice versa.

Therefore, the following two challenges need to be addressed for improving resource utilization in CML:

\textit{a) Sequential execution on devices and the server causes resource under-utilization:} For FL, the server aggregates the models obtained from all devices after they complete training; for SL, after the training of the initial layers is completed on the devices, the remaining layers of the DNN are trained on the server. Since device-side and server-side computations in CML techniques occur in sequence, there are long idle times on both the devices and the server. 

\textit{b) Communication between devices and the server results in resource under-utilization:} Data transfer in CML techniques is time consuming~\cite{DBLP:journals/corr/abs-2107-09786, DBLP:journals/corr/abs-1909-09145, 9252066}, during which time no training occurs on both the server and devices. This increases the overall training time.

Although low resource utilization of CML techniques makes training inefficient, there is currently limited research that is directed at addressing this problem. This paper aims to address the above challenges by developing a framework, \PiPar\ (pronounced as `piper'), that leverages \textit{pipeline parallelism} to improve the resource utilization of devices and servers in CML techniques when training DNNs, thereby increasing training efficiency. The framework distributes the computation of DNN layers on the server and devices, balances the workload on both the server and devices and reorders the computation for different inputs in the training process. \PiPar\ overlaps the device and server-side computations with communication between the devices and server, thereby improving resource utilization, which in turn accelerates CML training. 

\PiPar\ redesigns the training process of DNNs. Traditionally, training a DNN involves the forward propagation pass (or forward pass) and backward propagation pass (or backward pass). In the forward pass, one batch of input data (also known as a mini-batch) is used as input for the first DNN layer and the output of each layer is passed on to subsequent layers to compute the loss function. In the backward pass, the loss function is passed layer by layer from the last layer to the first layer to compute the gradients of the DNN model parameters. 

\PiPar\ divides the DNN into two parts and deploys them on the server and devices as in SFL. Then the forward and backward passes are reordered for multiple mini-batches to reduce idle time. Each device executes the forward pass for multiple mini-batches in sequence. The immediate result of each forward pass (activations) is transmitted to the server, which executes the forward and backward passes for the remaining layers and transfers the gradients of the activations back to the device. The device then sequentially performs the backward passes for the mini-batches. The devices operate in parallel, and the local models are aggregated at a set frequency. Since many forward passes occur sequentially on the device, the communication for each forward pass overlaps the computation of the subsequent forward passes. Also, in \PiPar, the server and device computations occur simultaneously for different mini-batches. Thus, \PiPar\ reduces the idle time of devices and servers by overlapping server and device-side computations and server-device communication. 

This paper makes the following contributions: 

(1) The development of a novel framework \PiPar\ to accelerate collaborative training of DNNs by improving resource utilization. \PiPar\ is the first work to reduce resource idling in CML by reordering training tasks across a server and the participating devices. Idle time is reduced by leveraging pipeline parallelism to overlap device and server computations and device-server communication.

(2) Development of an low overhead automated parameter selection approach for further optimizing CML workloads across devices and servers to maximize the overall training efficiency.

\PiPar\ and the automated parameter selection approach are evaluated on a lab-based testbed. The experimental results demonstrate that: a) compared to FL, \PiPar\ can accelerate the training process by up to 34.6$\times$, and the idle time of hardware resources is reduced by up to 64.1$\times$. b) the automated parameter selection approach can find the optimal or near-optimal parameters in less time than an exhaustive search. It is also experimentally demonstrated that \PiPar\ achieves performance benefits when incorporating differential privacy methods and operating in environments with heterogeneous devices and changing bandwidths.

The rest of this paper is organized as follows. Section~\ref{sec:bgrw} considers the background and work related to this research. The \PiPar\ framework and the two approaches that underpin the framework are detailed in Section~\ref{sec:pipar}. Section~\ref{sec:convergence} provides a theoretical analysis of model convergence and accuracy using \PiPar. Experiments in Section~\ref{sec:exp} demonstrate the effectiveness of the \PiPar\ framework. Section~\ref{sec:concl} concludes this article.

\section{Background and Related Work}
\label{sec:bgrw}
Section~\ref{subsec:bg} provides the background of collaborative machine learning (CML), and Section~\ref{subsec:rw} introduces the related research on improving the training efficiency in CML.

\subsection{Background}
\label{subsec:bg}

The training process of three popular CML techniques, namely federated learning (FL), split learning (SL) and split federated learning (SFL), and their limitation due to resource under-utilization are presented.

\begin{figure*}[htbp]
	\centering
	\subfigure[Federated Learning]{
	    \includegraphics[width=0.32\textwidth]{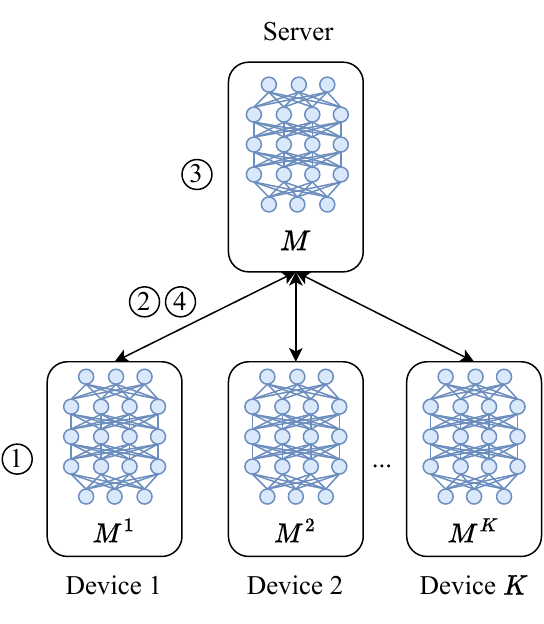}
	    \label{fig:fl}
	    }
	\centering
	\subfigure[Split Learning]{
	    \includegraphics[width=0.31\textwidth]{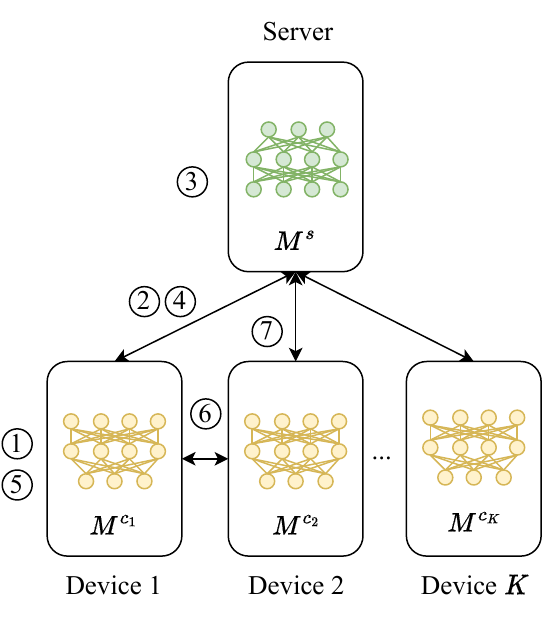}
	    \label{fig:sl}
	    }
	\centering
	\subfigure[Split Federated Learning]{
	    \includegraphics[width=0.31\textwidth]{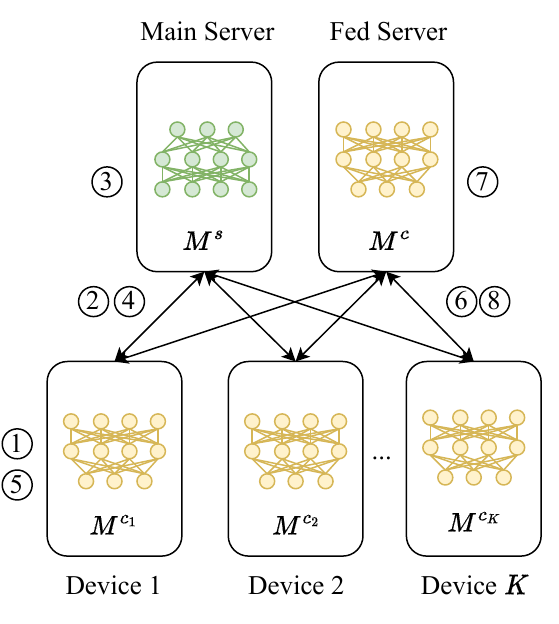}
	    \label{fig:sfl}
	    }
	\centering
	\caption{Training of CML methods, assuming $K$ devices. The training steps (circled numbers) are explained in Section~\ref{subsec:bg}}
	\label{fig:dcml}
\end{figure*}

\subsubsection{Federated learning}
FL~\cite{DBLP:journals/corr/KonecnyMR15, DBLP:journals/corr/KonecnyMRR16, DBLP:journals/corr/KonecnyMYRSB16, pmlr-v54-mcmahan17a} uses a set of devices coordinated by a central server to train deep learning models collaboratively. 

Assume $K$ devices participate in the training process as shown in Figure~\ref{fig:fl}. In Step~\ding{172}, the devices train the complete model $M^k$ locally, where $k=1,2,...,K$. In each iteration, the local model trains on a mini-batch of data by completing the forward and backward passes to compute gradients of all model parameters and then update the parameters with the gradients. A training epoch involves training over the entire dataset, which consists of multiple iterations. In Step~\ding{173}, after a predefined number of local epochs, the devices send the local models $M^k$ to the server, where $k=1,2,...,K$. In Step~\ding{174}, the server aggregates the local models to obtain a global model $M$, using the FedAvg algorithm~\cite{pmlr-v54-mcmahan17a}; $M=\sum_k\frac{|\mathcal{D}^k|}{\sum_k|\mathcal{D}^k|}M^k$, where $\mathcal{D}_k$ is the local dataset on device $k$ and $|\boldsymbol{\cdot}|$ is the function to obtain the set size. In Step~\ding{175}, the global model is downloaded to the devices. The next round of training continues until the model converges.

Typically, local model training on devices (Step~\ding{172}) takes most of the time, while the server with more significant compute performance is idle. Therefore, \PiPar\ utilizes the idle resources on the server during training.

\subsubsection{Split learning}
SL~\cite{DBLP:journals/corr/abs-1810-06060, DBLP:journals/corr/abs-1812-00564} is another privacy-preserving CML method. Since a DNN consists of consecutive layers, SL splits the entire DNN $M$ into two parts at the granularity of layers and deploys them on the server ($M^s$) and the devices ($M^{c_k}$, where $k=1,2,...,K$). 

As shown in Figure~\ref{fig:sl}, the devices train the initial layers of the DNN and the server trains the remaining layers, and the devices work in a round-robin fashion. In Step~\ding{172}, the first device executes the forward pass of $M^{c_1}$ on its local data, and in Step~\ding{173}, the intermediate results, also known as activations, are sent to the server. In Step~\ding{174}, the server uses the activations to complete the forward pass of $M^s$ to obtain the loss. The loss is then used for the backward pass on the server to compute the gradients of the parameters of $M^s$ and the gradients of the activations. In Step~\ding{175}, the gradients of the activations are sent back to the device, and in Step~\ding{176}, the gradients of the parameters of $M^{c_1}$ are computed in the device-side backward pass. Next, the parameters of the server-side model and device-side model are updated by their gradients. In Step~\ding{177}, after a device trains for a certain number of epochs, the next device gets the latest model from the previous device and starts training its model in Step~\ding{178}. 

Compared to FL, device-side computation is significantly reduced because only a few layers are trained on devices. However, since the devices work in sequence (instead of in parallel as FL), the overall training efficiency decreases as the number of devices increases.

\subsubsection{Split federated learning}
Since FL is computationally intensive on devices and SL works inefficiently on the device, SFL~\cite{DBLP:journals/corr/abs-2004-12088} was developed to alleviate both limitations. Similar to SL, SFL also splits the DNN across the devices ($M^{c_k}$, where $k=1,2,...,K$) and the server ($M^s$) and collaboratively trains the DNN. However, in SFL, the devices train in parallel and utilize a `main' server for training the server-side model and a `Fed' server for aggregation.

The training process is shown in Figure~\ref{fig:sfl}. In Step~\ding{172}, the forward pass of $M^{c_k}$, where $k=1,2,...,K$, are executed on the devices in parallel, and in Step~\ding{173}, the activations are uploaded to the main server. In Step~\ding{174}, the main server trains $M^s$, and in Step~\ding{175}, the gradients are sent back to all devices before they complete the backward pass in Step~\ding{176}. At a pre-defined frequency, the models $M^{c_k}$, where $k=1,2,...,K$, are uploaded to the Fed server in Step~\ding{177}. In Step~\ding{178}, the models are aggregated to the global model $M^c$. In Step~\ding{179}, $M^c$ is downloaded to the devices and used for the next round of training.

SFL utilizes device parallelism to improve the training efficiency of SL~\cite{Thapa2021}. However, the server still waits while the devices are training the model (Step~\ding{172}) and transmitting data~(Step~\ding{173}), and vice versa, which leads to resource under-utilization. \PiPar\ addresses this problem by parallelizing the steps performed on the server and the devices.

\subsection{Related work}
\label{subsec:rw}

Existing research to improve the training efficiency of CML focuses on the three aspects considered below.

\subsubsection{Improving resource utilization using pipeline parallelism}

Approaches employing pipeline parallelism have been proposed to improve the compute and network utilization of resources. GPipe~\cite{gpipe} and PipeDream~\cite{pipedream} use pipeline parallelism when a DNN is distributed to multiple computing nodes and parallelizes the computations on different nodes. They both reduce the idle time on computing resources. To further improve the hardware utilization, PipeMare~\cite{pipemare} implements asynchronous training, and Chimera~\cite{Chimera} uses bidirectional pipelines instead of a unidirectional one. PipeFisher~\cite{pipefisher} takes advantage of idle resources to execute second-order optimization to accelerate model convergence. However, these approaches for distributed DNN training cannot be directly applied to CML for three reasons.

Firstly, the context in which current pipeline parallelism approaches were designed to operate in is completely different from CML. They were designed for GPU clusters with substantial compute resources where the data flow is sequential (data is provided as input to one node and then the output goes to the next node and so on). However, in CML, the data generated on end-user devices is not shared with other devices or servers to preserve privacy. \PiPar\ is therefore proposed to tackle the problem of distributed training of devices in centralized topologies.

Secondly, in existing pipeline parallelism approaches, different layers of the DNN are mapped onto different nodes in a cluster and they do not share weights. However, in CML, each device trains a local model on its data, and the model weights are subsequently synchronized with other devices. \PiPar\ splits the model across the server and all the devices to alleviate the computational burden on the devices and proposes a method to synchronize the server-side and client-side models.

Thirdly, the bandwidth between different devices and servers in CML will be variable as seen in real-world mobile environments. However, the bandwidth between the nodes of a GPU cluster is relatively less prone to such variability. Since communication time between nodes of a GPU cluster is relatively small, existing pipeline parallelism approaches tend to hide communication behind computation. However, the communication of activations and gradients in CML is a substantial volume, which is not handled by existing approaches. \PiPar\ takes this into account, and hence, a parameter selection method is proposed to overlap the communication and computation.

Given the above limitations, we propose a novel framework, \PiPar\, that fully utilizes the computing resources on the server and devices and the bandwidth available between them to improve the training efficiency of CML.

\subsubsection{Reducing the impact of stragglers}

Stragglers among the devices used for training increase the overall training time of CML. A device selection method was proposed based on the resource availability of devices to minimize the impact of stragglers~\cite{DBLP:journals/corr/abs-1804-08333}. Certain neurons of the DNN on a straggler are masked to accelerate computation~\cite{9586241}. Local gradients were aggregated hierarchically to accelerate FL on heterogeneous devices~\cite{9699080}. To balance workloads across heterogeneous devices, FedAdapt~\cite{DBLP:journals/corr/abs-2107-04271} offloaded DNN layers from devices to a server. An adaptive asynchronous federated learning mechanism~\cite{aafl} was proposed to mitigate stragglers.

These methods alleviated the impact of stragglers but did not address the fundamental challenge of sequential computation and communication between the devices and server that results in low resource utilization.

\subsubsection{Reducing communication overhead}

In limited bandwidth environments, communication overhead limits the training efficiency of CML techniques. To reduce the communication traffic in FL, a relay-assisted two-tier network was developed~\cite{9439928}. Models and gradients were transmitted simultaneously and aggregated on the relay nodes. Pruning, quantization and selective updating were used to reduce the model size and thus reduce the computation and communication overhead~\cite{9366879}. The communication involved in the backward pass of SFL was improved by averaging the gradients on the server-side model and broadcasting them to the devices instead of unicasting the unique gradients to devices~\cite{DBLP:journals/corr/abs-2112-05929}. Overlap-FedAvg~\cite{10.1109/TPDS.2021.3090331} was proposed to decouple the computation and communication during training and overlap them to reduce idle resources. However, the use of computing resources located at the server was not fully leveraged.

These methods are effective in reducing the data volume transferred over the network, thus reducing the communication overhead. However, this reduces model accuracy.

\section{\texttt{PiPar}}
\label{sec:pipar}
This section develops \PiPar, a framework to improve the resource utilization of CML in the context of FL and SFL. \PiPar\ accelerates the execution of sequential DNNs for the first time by leveraging pipeline parallelism to improve the overall resource utilization in centralized CML. 

The \PiPar\ framework is underpinned by two approaches, namely \textit{pipeline construction} and \textit{automated parameter selection}. 
The first approach constructs a training pipeline to balance the overall training workload by (a) reallocating the computations for different DNN layers on the server and devices, and (b) reordering the forward and backward passes for multiple mini-batches of data by scheduling them onto idle resources. Consequently, not only is the resource utilization improved by using \PiPar, but also the overall training of the DNN. 
The second approach of \PiPar\ enhances the performance of the first approach by automatically selecting the optimal control parameters (such as the point at which the DNN must be split across the device and the server and the number of mini-batches that can be executed concurrently in the pipeline).

\subsection{Motivation}
\label{subsec:motivate}

The following three observations on low resource utilization when training DNNs in CML motivate \PiPar.

\textit{(1) The server and devices need to work simultaneously}: The devices and server work in an alternating manner in the current CML methods, which is a limitation that must be addressed to improve resource utilization. In FL, the server starts to aggregate local models only after all devices have completed training their local models. In SL/SFL, the sequential computation of DNN layers results in the sequential working of the devices and the server. The dependencies between server-side and device-side computations need to be eliminated to reduce the resulting idle time on the resources. \PiPar\ attempts to make the server and the devices work simultaneously by reallocating and reordering training tasks.

\textit{(2) Compute-intensive and I/O-intensive tasks need to be overlapped}: Compute-intensive tasks, such as model training, involve large-scale computations performed by computing units (CPU/GPU), while IO-intensive tasks refer to input and output tasks of disk or network, such as data transmission, which usually do not have a high CPU requirement. A computationally intensive and an I/O-intensive task can be executed in parallel on the same resource without mutual dependencies. However, in current CML methods, both server-side and device-side computations are paused when communication is in progress, which creates idle time on compute resources. \PiPar\ improves this by overlapping compute-intensive and I/O-intensive tasks.

\textit{(3) Workloads on the server-side and client-side need to be balanced}: Idle time on resources is also caused due to imbalanced workloads on the server and devices. \PiPar\ balances the workloads on the server and device sides by splitting the DNN carefully.

\subsection{Pipeline construction}
\label{subsec:construct}

Assume that $K$ devices and a server train a sequential DNN collaboratively by using data residing on each device. Conventionally, the dataset on each device is divided into multiple mini-batches that are fed to the DNN in sequence. Training on each mini-batch involves a forward pass that computes a loss function and a backward pass that computes the gradients of the model parameters. A training epoch ends after the entire dataset has been fed to the DNN. To solve the problem of low resource utilization faced by the current CML methods, \PiPar\ constructs a training pipeline that reduces the idle time on resources during collaborative training.

Each forward and backward pass of CML methods comprises four tasks: (i) the device-side compute-intensive tasks, such as model training; (ii) the device-to-server I/O-intensive task, such as data uploading; (iii) the server-side compute-intensive task, such as model training (only in SL and SFL) and model aggregation; (iv) the server-to-device I/O-intensive task, such as data downloading. The four tasks can only be executed in sequence in current CML methods, resulting in idle resources. To solve this problem, a pipeline is developed to balance and parallelize the above tasks. The pipeline construction approach involves three phases, namely DNN splitting, training stage reordering and multi-device parallelization.

\subsubsection*{Phase 1 - DNN splitting}
\label{subsubsec:split}

The approach aims to overlap the above-mentioned four tasks to reduce idle time on computing resources on the server and devices as well as idle network resources. Since this approach does not reduce the computation and communication time of each task, it needs to balance the time required by the four tasks to avoid straggler tasks from increasing the overall training time. For example, in FL, the device-side compute-intensive task is the most time-consuming, while the other three tasks require relatively less time. In this case, overlapping the four tasks will not significantly reduce the overall training time. Therefore, it is more appropriate to split the DNN and divide the training task across the server and the devices (similar to previous works \cite{DBLP:journals/corr/abs-2004-12088,DBLP:journals/corr/abs-2107-04271}). In addition, since the output of each DNN layer has a variable size, different split points of the DNN will result in different volumes of transmitted data. Thus, changing the split point based on the computing resources and bandwidth can also balance the I/O-intensive tasks with compute-intensive tasks. The selection of the best splitting point is presented in Section~\ref{subsec:optimise}.

Splitting DNNs does not affect model accuracy, since it does not alter computations but rather the resource on which they are executed. In FL, each device $k$, where $k=1,2,...,K$, trains a complete model $M^k$. \PiPar\ splits $M^k$ to a device-side model $M^{c_k}$ and a server-side model $M^{s_k}$ represented as: 
\begin{equation}
\label{eq:model-split}
    M^k = M^{s_k} \oplus M^{c_k}
\end{equation}
where the binary operator $\oplus$ stacks the layers of two partitions of a DNN as a complete DNN.

There are $k$ pairs of $\{M^{c_k}, M^{s_k}\}$, where $M^{c_k}$ is deployed on device $k$ while all of $M^{s_k}$ are deployed on the server. This is different from SL and SFL where only one model is deployed on the server-side. Assume the complete model $M^k$ contains $Q$ layers, $M^{c_k}$ contains the initial $P$ layers and $M^{s_k}$ contains the remaining layers, where $1 \leq P \leq Q$. 

Splitting the DNN maintains the consistency of the training process and does not change the model accuracy; this is demonstrated in Section~\ref{subsec:split-nn}. 

\subsubsection*{Phase 2 - Training stage reordering}
\label{subsubsec:reorder}

After splitting the DNNs and balancing the four tasks, idle resources in the training process need to be utilized. This is achieved by reordering the computations for different mini-batches of data. 

\begin{figure}[tp]
	\centering
	\subfigure[Conventional training pipeline when using a split DNN]{
	    \includegraphics[width=\linewidth]{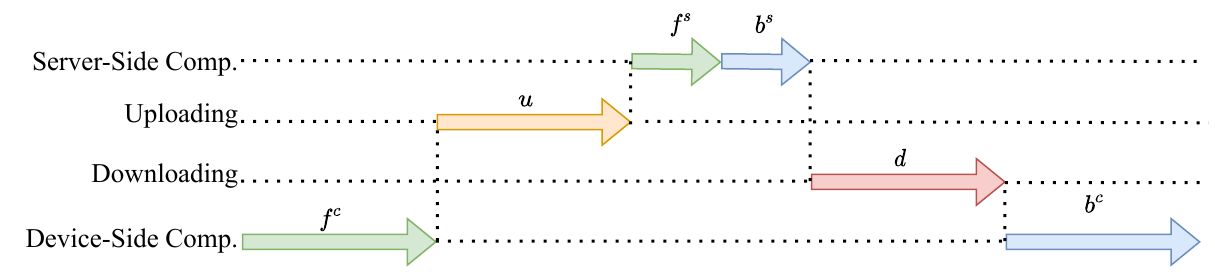}
	    \label{fig:slpp}
	    }
	\centering
	\subfigure[Training pipeline in \PiPar. $N$ mini-batches are trained in parallel in each training iteration; subscripts indicate the index of the mini-batch.]{
	    \includegraphics[width=\linewidth]{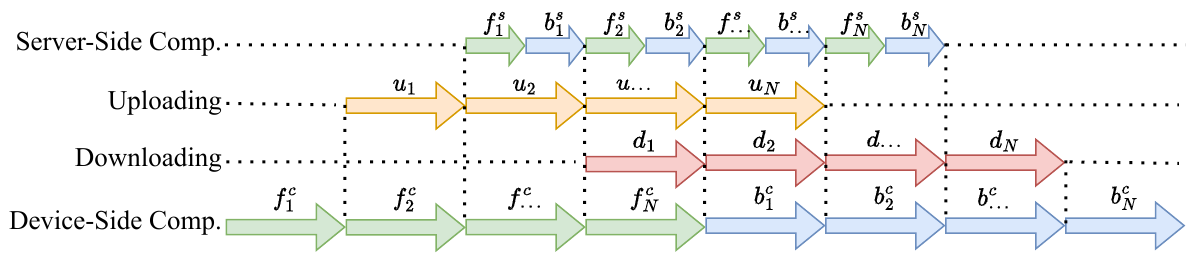}
	    \label{fig:pipefed}
	    }
	\centering
	\caption{Pipelines for one training iteration in conventional training and \PiPar\ when using a split DNN. ``Comp" is an abbreviation for computation. $f$, $b$, $u$ and $d$ represent forward pass, backward pass, upload and download, respectively. Superscripts indicate server-side ($s$) or client-side ($c$) computation or communication.}
	\label{fig:trainiteration}
\end{figure}

Figure~\ref{fig:slpp} shows the pipeline of one training iteration of a split DNN for one pair of $\{M^{s_k}, M^{c_k}\}$ (the device index $k$ is not shown). Any forward pass ($f$), backward pass ($b$), upload task ($u$) and download task ($d$) for each mini-batch is called a \emph{training stage}.

The idle time on the device exists between the forward pass $f^c$ and the backward pass $b^c$ of the device-side model. Thus, \PiPar\ inserts the forward pass of the next few mini-batches into the device-side idle time to fill up the pipeline. As shown in Figure~\ref{fig:pipefed}, in each training iteration, the forward passes for $N$ mini-batches, $f^c_1$ to $f^c_N$, are performed on the device in sequence. The activations of each mini-batch are sent to the server ($u_1$ to $u_N$) once the corresponding forward pass is completed, which utilizes idle network resources. Once the activations of any mini-batch arrive, the server performs the forward and backward passes, $(f^s_1, b^s_1)$ to $(f^s_N, b^s_N)$, and sends the gradients of the activations back to the device ($d_1$ to $d_N$). After completing the forward passes of the mini-batches and receiving the gradients, the device performs the backward passes, $b^c_1$ to $b^c_N$. Then the model parameters are updated and the training iteration ends. A training epoch ends when the entire dataset has been processed, which involves multiple training iterations. 

Figure~\ref{fig:pipefed} shows that compared to conventional training (Figure~\ref{fig:slpp}), the four tasks can be considerably overlapped and it is possible to significantly reduce the idle time of the server and the devices.

To guarantee a similar model accuracy as classic FL, the gradients must be obtained from the same number of data samples when the model is updated. This requires that the number of data samples involved in each training iteration in \PiPar\ should be the same as the original batch size in FL. Since $N$ mini-batches are used in each training iteration, the size of each mini-batch $B^\prime$ is reduced to $1/N$ of the original batch size $B$ in FL.
\begin{equation}
\label{eq:batch-size}
    B^\prime = \lfloor B/N \rfloor
\end{equation}

Reordering training stages does not impact model accuracy, which is demonstrated in Section~\ref{subsec:reorder}.

\subsubsection*{Phase 3 - Multi-device parallelization}
\label{subsubsec:multiple}

The workloads of multiple devices involved in collaborative training need to be coordinated. On the device-side, each device $k$ is responsible for training its model $M^{c_k}$, and \PiPar\ allows them to train in parallel for efficiency. On the server-side, the counterpart $K$ models ($M^{s_1}$ to $M^{s_K}$) are deployed and trained simultaneously. However, this may result in contention for compute resources. 

\begin{figure}[tp]
	\centering
	\subfigure[\PiPar\ with single device (same as Figure~\ref{fig:pipefed})]{
	    \includegraphics[width=\linewidth]{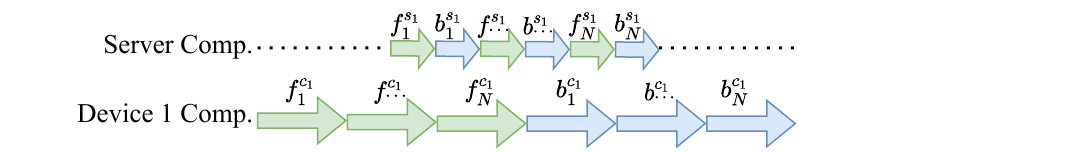}
	    \label{fig:sc}
	    }
	\centering
	\subfigure[\PiPar\ with two devices in sequence]{
	    \includegraphics[width=\linewidth]{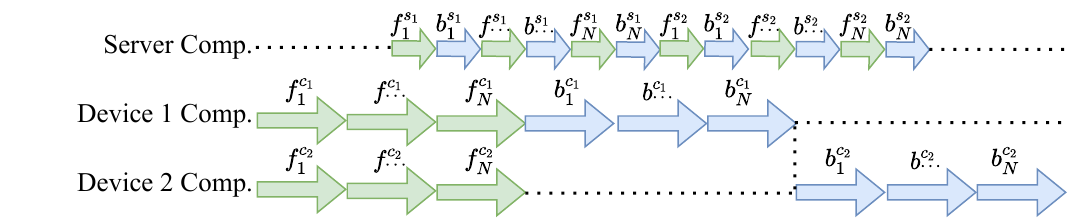}
	    \label{fig:smc}
	    }
	\centering
	\subfigure[\PiPar\ with two devices in parallel]{
	    \includegraphics[width=\linewidth]{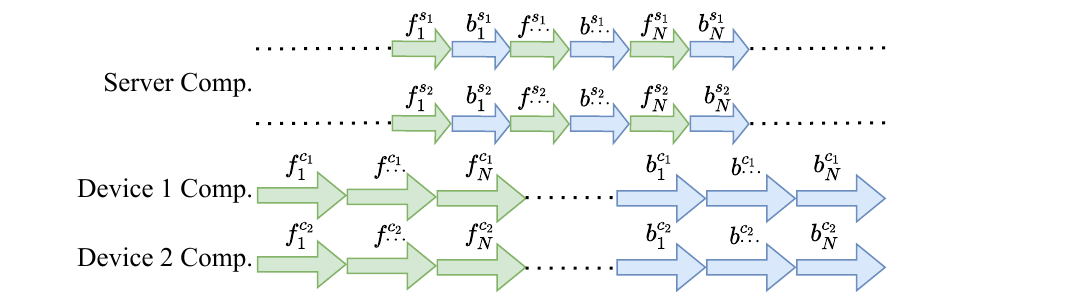}
	    \label{fig:pmc}
	    }
	\centering
	\caption{\PiPar\ using single and multiple devices. Comp, $f$, $b$, $u$ and $d$ represent computation, forward pass, backward pass, upload and download, respectively. The superscripts $s_k$ and $c_k$ represent the index of the model $M^{s_k}$ and $M^{c_k}$, $k=1,2$, respectively.
	}
	\label{fig:servertrain}
\end{figure}

Figure~\ref{fig:sc} shows the case of a single device (same as Figure~\ref{fig:pipefed} but does not show communication), whereas Figure~\ref{fig:smc} and Figure~\ref{fig:pmc} show the case of multiple devices. Figure~\ref{fig:smc} offers a solution to train the server-side models sequentially. However, the server-side models that are trained relatively late will cause a delay in the backward passes for the corresponding device-side models, for example, $b_n^{c_2}$, where $n=1,2,...,N$, in Figure~\ref{fig:smc}. 

Alternatively, data parallelism can be employed. The activations from different devices are deemed as different inputs and the server-side models are trained in parallel on these inputs. This is shown in Figure~\ref{fig:pmc}. It is worth noting that, compared to training a single model, training multiple models at the same time may result in longer training time for each model on a resource-limited server. This approach, nonetheless, mitigates stragglers on devices. 

At the end of each training epoch, the device-side models $M^{c_k}$ are uploaded to the server and will constitute the entire models $M^{k}$ when combined with the server-side models $M^{s_k}$ (Equation~\ref{eq:model-split}). The complete model $M^{k}$ of each device is aggregated to obtain a complete global model $M$, using the FedAvg algorithm~\cite{pmlr-v54-mcmahan17a}.
\begin{equation}
    M=\sum_{k=1}^K\frac{|\mathcal{D}^k|}{\sum_{k=1}^K|\mathcal{D}^k|}M^k
\end{equation}
where $\mathcal{D}^k$ is the local dataset on device $k$ and $|\boldsymbol{\cdot}|$ is the function to obtain the size of the given dataset. The server-side global model $M^{s}$ and device-side global model $M^{c}$ are split from $M$ using
\begin{equation}
    M = M^{s} \oplus M^{c}
\end{equation}

The devices download $M^c$ to update the local models for the subsequent training epochs, and the server-side models are updated by $M^s$.

It has been shown in the previous phases that the accuracy of each local model $M^k$ in \PiPar\ is not affected. The FedAvg algorithm is used in \PiPar\ to generate the global model $M$ by aggregating $M^k$, where $k=1,2,...,K$, which is the same as in classic FL. Therefore, \PiPar\ maintains a similar model accuracy to FL.

\subsubsection*{Training overview}

The entire training process of \PiPar\ is shown in Algorithm~\ref{alg:client-pipeline} and Algorithm~\ref{alg:server-pipeline}. 

All devices train simultaneously using Algorithm~\ref{alg:client-pipeline}. On device $k$, the device-side model $M^{c_k}$ is initially built given the split point (Line 1). Line 2 to Line 19 shows the complete training process until the model converges. In each training epoch (Line 3 to Line 18), the entire dataset is processed. A training epoch consists of multiple training iterations, each processing $B^\prime N^k$ data samples. In each training iteration (Line 4 to Line 13), the forward passes of $N^k$ mini-batches are executed in sequence (Line 6), and the activations are sent to the server (Line 7). Their gradients are then received from the server (Line 10), and the backward passes are executed sequentially to compute the gradients of the weights of $M^{c_k}$ (Line 11). At the end of a training iteration, the model is updated based on the gradients (Line 13). After all training iterations are completed, the signal `stop epoch', and $M^{c_k}$ is sent to the server (Line 15 to Line 16). The device then receives a global device-side model ${M^{c_k}}^\prime$ from the server (Line 17) and uses it to update the current model (Line 18). When the model converges, the client sends a `stop training' signal to the server, thus completing the training process (Line 20). Since all device-side models are synchronized, all devices will send a `stop training' signal to the server simultaneously.

\begin{algorithm}[t]
    \caption{Device-Side Training in \PiPar}
	\label{alg:client-pipeline}
	\tcc{Run on Client $k$.}
	\KwIn{local dataset $\mathcal{D}^k$;
	    batch size $B^\prime$;
	    learning rate $\eta$;
	    model split point $P^k$;
	    number of mini-batches in each iteration $N^k$ 
	    }
	\KwOut{Device-side models $M^{c_k}$}
	\BlankLine
	Build $M^{c_k}$ based on $P^k$
	
	\While{model has not converged}{
	    \tcp{Start a training epoch}
	    \For{$i=1$ \KwTo $\lfloor |\mathcal{D}^k|/B^\prime N^k \rfloor$}{
	        \tcp{Start a training iteration}
    	    \For{$n=1$ \KwTo $N^k$}{
    	        Load a mini-batch $\mathbf{x}_n$ of the size $B^\prime$ from $\mathcal{D}^k$
    	        
    	        Compute the activation $\mathbf{a}_n$ using Equation~\ref{eq:fp-client}
    	        
    	        Send $\mathbf{a}_n$ and labels $\mathbf{y}_n$ to the server
    	    }
    	    \For{$n=1$ \KwTo $N^k$}{
    	        Receive $g(\mathbf{a}_n)$ from the server\
    	        
    	        Compute the gradients of model weights $g(M^{c_k}|g(\mathbf{a}_n))$\ using Equation~\ref{eq:bp1}
    	    }
    	    Update $M^{c_k} \leftarrow M^{c_k} - \frac{\eta}{N^k} \sum_{n=1}^{N^k}g(M^{c_k}|g(\mathbf{a}_n))$\
	    }
	    Send `stop epoch' signal to the server
	    
	    Send $M^{c_k}$ to the server
	    
	    Receive ${M^{c_k}}^\prime$ from the server
	    
	    Update $M^{c_k} \leftarrow {M^{c_k}}^\prime$
	}
	Send `stop training' signal to the server
	
	\textbf{Return} $M^{c_k}$
\end{algorithm}

Algorithm~\ref{alg:server-pipeline} is executed on the server-side. The server first builds $K$ models $M^{s_k}$, where $k=1,2,...,K$ (Line 1), and starts training the models until a signal `stop training' is received from all devices (Line 2). In each training epoch (Line 3 to Line 23), the $K$ models are trained simultaneously (Line 3 to Line 19) and aggregated into a global model (Line 20 to Line 23). A training epoch of model $k$ does not end until a signal `stop epoch' (Line 5) is received from device $k$, which involves multiple training iterations. During a training iteration (Line 6 to Line 15), the server receives the activations and labels from device $k$ (Line 7) and uses them to compute the loss function (Line 9 to Line 10). After that, the gradients of activations and model weights are computed (Line 11 to Line 12). The former is then sent to device $k$ (Line 13), and the latter is used to update $M^{s_k}$ at the end of the training iteration (Line 15). After receiving the `stop epoch' signal, the server receives the device-side model $M^{c_k}$ from device $k$ (Line 17) and makes up a complete model $M^k$ (Line 18). The $K$ models $M^k$, where $k=1,2,...,K$, are aggregated into a global model $M$ (Line 20). $M$ is then split into a server-side model ${M^{s_k}}^\prime$ and a device-side model ${M^{c_k}}^\prime$ (Line 21). ${M^{c_k}}^\prime$ is sent to device $k$ (Line 22), and ${M^{s_k}}^\prime$ is used to update $M^{s_k}$ (Line 23). A training epoch ends. Training is completed when the `stop training' signal is received from all devices.

If a given device disconnects from the server, the aggregation carried out on the server will exclude the model of the device. If the device reconnects to the server, then it will download the latest global model and continue training.

Algorithm~\ref{alg:client-pipeline} and Algorithm~\ref{alg:server-pipeline} have the same computational complexity as the FedAvg algorithm~\cite{pmlr-v54-mcmahan17a}. However, \PiPar\ introduces parallelism within training.

\begin{algorithm}[t]
	\caption{Server-Side Training in \PiPar}
	\label{alg:server-pipeline}
	\tcc{Run on the server.}
	
	\KwIn{Number of devices $K$;
	    structure of the DNN with $Q$ layers;
	    learning rate $\eta$;
	    model split point $P^k$ and
	    number of mini-batches in each iteration $N^k$, where $k=1,2,...,K$
	    }
	\KwOut{Server-side models $M^{s_k}$, where $k=1,2,...,K$}  
	\BlankLine
	Build $M^{s_k}$, where $k=1,2,...,K$ based on $P^k$
	
	\While{\textnormal{`stop training'} signal not received}{
	    \tcp{Start a training epoch.}
	    \For{$k=1$ \KwTo $K$ in parallel}{
	        Initial the size of dataset on device $k$: $D^k \leftarrow 0$
	        
            \While{\textnormal{`stop epoch'} signal not received from all devices}{
                \tcp{Start a training iteration}
                \For{$n=1$ \KwTo $N^k$}{
                    Receive activations $\mathbf{a}_n$ and labels $\mathbf{y}_n$
                    
                    Update $D^k\leftarrow D^k+|\mathbf{a}_n|$
                    
                    Compute the output $\mathbf{\hat{y}}_n$ using Equation~\ref{eq:fp-server}
                    
                    Compute loss function $l(\mathbf{y}_n,\mathbf{\hat{y}}_n)$
                    
                    Compute the gradients of activation $g(\mathbf{a}_n)\leftarrow \frac{\partial l}{\partial \mathbf{\hat{y}}_n}\tilde{b}_Q(\tilde{b}_{Q-1}(...\tilde{b}_{Q+1}(\mathbf{a}_n)))$ 
                    
                    Compute the gradients of model weights $g(M^{s_k}|a_n)$ using Equation~\ref{eq:bp1}
                    
                    Send $g(\mathbf{a}_n)$ to device $k$
                }
                Update $M^{s_k} \leftarrow M^{s_k} - \frac{\eta}{N^k} \sum_{n=1}^{N^k}g(M^{s_k}|a_n)$
            }
            Receive $M^{c_k}$ from device $k$
            
            Make up complete model $M^k \leftarrow M^{s_k} \oplus M^{c_k}$
        }
        Calculate global model $M \leftarrow \sum_{k=1}^K\frac{D^k}{\sum_{k=1}^KD^k}M^k$
        
        Split $M$ to ${{M^{s_k}}^\prime, {M^{c_k}}^\prime}$, where $k=1,2,...,K$ based on $P^k$
        
        Send ${M^{c_k}}^\prime$ to device $k$, where $k=1,2,...,K$
        
        Update $M^{s_k} \leftarrow {M^{s_k}}^\prime$
	}
	\textbf{Return} $M^{s_k}$, where $k=1,2,...,K$
\end{algorithm}

\subsection{{Automated parameter selection}}
\label{subsec:optimise}

To maximize the utilization of idle resources, two parameters of \PiPar\ that impact the performance of the training pipeline are considered:

\textit{a) Split point} of a DNN is denoted as $P$. All layers with indices less than or equal to $P$ are deployed on the device and the remaining layers are deployed on the server. The number of layers determines the amount of computation on a server/device, and the volume of data output from the split layer determines the communication traffic. Therefore, finding the most suitable value for $P$ for each device will balance the time required for computation on the server and the device as well as the communication between them.

\textit{b) Parallel batch number} denoted as $N$ is the number of mini-batches used for concurrent training in each iteration. The computations of the mini-batches fill up the pipeline, so the number of mini-batches for each training iteration must be determined.

The naive choice of $\{P,N\}$ makes the results of \PiPar\ no worse than FL and SFL. When $P$ is the layer number and $N=1$, \PiPar\ is the same as FL; when $P$ is the same split point as SFL and $N=1$, \PiPar\ is the same as SFL. However, carefully selected $\{P,N\}$ values can further optimize the performance of \PiPar. The optimal values of $\{P,N\}$ can be obtained by an exhaustive search. The model will be trained with all parameter combinations, and then the optimal parameter combination with the shortest training time can be selected. This is unsuitable to be adopted in \PiPar\ in practical as it is time consuming. In addition, we can also select $\{P,N\}$ values empirically. Empirical selections will make \PiPar\ a better solution than FL and SFL, but cannot make it achieve its optimal performance as the exhaustive search. Therefore, we propose an automated parameter selection approach that identifies an optimal or near-optimal combination of parameters in a shorter time than exhaustively searching. These parameters vary with DNNs, server/device combinations, and network conditions. Therefore, the developed approach relies on estimating the training time for different parameters given the DNN and the network condition.

The approach aims to select the best pair of $\{N^k,P^k\}$ for each device $k$ to minimize the idle resources in the three phases. Firstly, we need to know how much they affect the pipeline. Several training iterations are profiled to identify the size of the output data and the training time for each layer of the DNN. Secondly, the training time for each epoch can be estimated using dynamic programming, given a pair of $\{N^k,P^k\}$. Thirdly, the candidates for $\{N^k,P^k\}$ are shortlisted. Since the training time can be estimated for every candidate, the one with the lowest training time will be selected. The three phases are explained in detail as follows.

\subsubsection*{Phase 1 - Profiling}
\label{subsubsec:profile}

In this phase, an additional training period is required. The complete model is trained on each device and server separately for a predefined number of iterations. If the entire model cannot fit in the memory of the devices, the devices train as many layers as possible and the server trains the complete model. The following information is empirically collected: 

\textit{a) Time spent in the forward/backward pass of each layer deployed on each device and server.} Assume that $\tilde{f}^{c_k}_q$, $\tilde{b}^{c_k}_q$, $\tilde{f}^{s}_q$ and $\tilde{b}^{s}_q$ denote the forward and backward pass of layer $q$ on device $k$ and server, and $t()$ denotes time. Then, $t(\tilde{f}^{c_k}_q)$, $t(\tilde{b}^{c_k}_q)$, $t(\tilde{f}^{s}_q)$ and $t(\tilde{b}^{s}_q)$ are the time taken for the forward and backward pass on the devices and server, which are measured and recorded during training.

\textit{b) Output data volume of each layer in the forward and backward pass.} $\tilde{v}^f_q$ and $\tilde{v}^b_q$ denote the output data volume for layer $q$ in the forward and backward passes. The data volumes are measured and recorded during training.

\subsubsection*{Phase 2 - Training time estimation}
\label{subsubsec:estimate}

To estimate the time spent in each training epoch of $\{M^{c_k}, M^{s_k}\}$, given the pairs of $\{N^k,P^k\}$ for device $k$, the time for each training stage must be estimated.

Assume that $f^{c_k}_n$, $b^{c_k}_n$, $f^{s_k}_n$ and $b^{s_k}_n$ is the time spent in the forward and backward passes of $M^{c_k}$ and $M^{s_k}$ for mini-batch $n$, where $n=1,2,...,N^k$. The time spent in each stage is the sum of the time spent in all relevant layers. Since the size of each mini-batch in \PiPar\ is reduced to $1/N^k$, the time required for each layer is reduced to $1/N^k$. The time of each training stage is estimated by the following:
\begin{equation}
    t(f^{c_k}_n)=\sum_{q=1}^{P^k} \frac{t(\tilde{f}^{c_k}_q)}{N^k}
\end{equation}
\begin{equation}
    t(f^{s_k}_n)=\sum_{q={P^k}+1}^{Q} \frac{t(\tilde{f}^{s_k}_q)}{N^k}
\end{equation}
\begin{equation}
    t(b^{c_k}_n)=\sum_{q=1}^{P^k}\frac{t(\tilde{b}^{c_k}_q)}{N^k}
\end{equation}
\begin{equation}
    t(b^{s_k}_n)=\sum_{q={P^k}+1}^Q \frac{t(\tilde{b}^{s_k}_q)}{N^k}
\end{equation}

Assume that $u^k_n$ and $d^k_n$ are the time required for uploading and downloading between device $k$ and the server for mini-batch $n$, where $n=1,2,...,N^k$, and $w_u^k$ and $w_d^k$ are the uplink and downlink bandwidths. Since the size of transmitted data is reduced to $1/N^k$:
\begin{equation}
    t(u^k_n)=\frac{\tilde{v}^f_{P^k}}{w_u^kN^k}
\end{equation}
\begin{equation}
    t(d^k_n)=\frac{\tilde{v}^b_{P^k}}{w_d^kN^k}
\end{equation}

The time required by all training stages is estimated using the above equations. The training time of each epoch can be estimated using dynamic programming. Within each training iteration, a given training stage has previous and next stages (exclusions for the first and last stages) as shown in Table~\ref{table:stages}. The first stage is $f^{c_k}_1$ and the last stage is $b^{c_k}_N$. We use $T(r)$ to denote the total time from the beginning of the training iteration to the end of stage $r$, and $t(r)$ to denote the time spent in stage $r$. Thus, the overall training time is $T(b^{c_k}_N)$. Since any stage can start only if all of its previous stages have been completed, we have:
\begin{equation}
\label{eq:recursion1}
    T(b^{c_k}_N)=t(b^{c_k}_N)+\max_{r\in prev(b^{c_k}_N)}T(r)
\end{equation}
\begin{equation}
\label{eq:recursion2}
    T(r)=t(r)+\max_{r^\prime\in prev(r)}T(r^\prime)
\end{equation}
\begin{equation}
\label{eq:recursion3}
    T(f^{c_k}_1)=t(f^{c_k}_1)
\end{equation}
where $prev()$ is the function to obtain all previous stages of the input stage. Since $t(b^{c_k}_N)$ is already obtained in Phase 2, Equation~\ref{eq:recursion1} to Equation~\ref{eq:recursion3} can be solved by recursion. The overall time of one training iteration can then be estimated.

\subsubsection*{Phase 3 - Parameter selection}
\label{subsubsec:param}

In this phase, the candidates of $\{N^k,P^k\}$ are shortlisted. Since the training time can be estimated for each candidate, the one with the shortest training time can be selected.

Assume that the DNN has $Q$ layers, such as dense, convolutional and pooling layers, and that the memory of devices can only accommodate the training of $Q^\prime$ layers, where $Q^\prime \leq Q$. The range of $P^k$ is $\{P^k|1 \leq P^k \leq Q^\prime,P^k\in \mathbb{Z}^+\}$, where $\mathbb{Z}^+$ is the set of all positive integers. 

Given $P^k$, the idle time of the device $k$ between the forward pass and backward pass of each mini-batch (the blank timeline between $f^c$ and $b^c$ in Figure~\ref{fig:slpp}) needs to be filled up by the forward passes of the following multiple mini-batches. As a result, the original mini-batch and the following mini-batches are executed concurrently in one training iteration. 

For example, as shown in Figure~\ref{fig:slpp}, the device idle time between $f^c$ and $b^c$ is equal to $t(u) + t(f^s) + t(b^s) + t(d)$. Thus, the forward passes or backward passes of the subsequent $\lceil \frac{t(u)+t(f^{s})+t(b^{s})+t(d)}{\min\{t(f^c), t(b^c)\}} \rceil$ mini-batches can be used to fill in the idle time, making the parallel batch number $N=1 + \lceil \frac{t(u)+t(f^{s})+t(b^{s})+t(d)}{\min\{t(f^c), t(b^c)\}} \rceil$. Since the batch size used in \PiPar\ is reduced to $1/N$, the time required for forward and backward passes of each layer, uploading and downloading is reduced to $1/N$. The parallel batch number for device $k$ is estimated as:
\begin{equation}
\begin{split}
    N^k &= 1 + \lceil \frac{t(u^k_n)+t(f^{s_n}_n)+t(b^{s_k}_n)+t(d^k_n)}{\min\{t(f^{c_k}_n), t(b^{c_k}_n)\}} \rceil \\
    &= 1 + \lceil \frac{\frac{\tilde{v}^f_{P^k}}{w_u^kN^k} + \sum_{q=1}^{P^k} \frac{t(\tilde{f}^{s}_q)}{N^k} + \sum_{q={P^k}+1}^Q \frac{t(\tilde{b}^{s}_q)}{N^k} + \frac{\tilde{v}^b_{P^k}}{w_d^kN^k}}{\min\left\{\sum_{q=1}^{P^k} \frac{t(\tilde{f}^{c_k}_q)}{N^k}, \sum_{q=1}^{P^k} \frac{t(\tilde{b}^{c_k}_q)}{N^k}\right\}} \rceil \\
    &= 1 + \lceil \frac{\frac{\tilde{v}^f_{P^k}}{w_u^k} + \sum_{q=1}^{P^k} t(\tilde{f}^{s}_q) + \sum_{q={P^k}+1}^Q t(\tilde{b}^{s}_q) + \frac{\tilde{v}^b_{P^k}}{w_d^k}}{\min\left\{\sum_{q=1}^{P^k} t(\tilde{f}^{c_k}_q), \sum_{q=1}^{P^k} t(\tilde{b}^{c_k}_q)\right\}} \rceil
\end{split}
\end{equation}

For each device $k$, the best $\{N^k,P^k\}$ can be selected from the shortlisted candidates by estimating the training time.

Since the training time of \PiPar\ with parameter pair $\{N^k,P^k\}$ is estimated based on profiling data from training complete models with the original batch size, this approach does not guarantee the selection of optimal parameters. However, our experiments in Section~\ref{subsec:op-exp} show that the parameters selected by this approach are similar to optimal values.

\begin{table}[tp]
    \centering
    \caption{Stages of a training iteration indicating the previous and next stages}
    \begin{tabular}{ccc}
        \Xhline{2\arrayrulewidth}
        \textbf{Stage} & \textbf{Previous Stages} & \textbf{Next Stages}  \\
        \Xhline{2\arrayrulewidth}
        $f^{c_k}_1$ & $n/a$ & $f^{c_k}_2$, $u^k_1$ \\
        \hline
        $f^{c_k}_n,1<n<N$ & $f^{c_k}_{n-1}$ & $f^{c_k}_{n+1}$, $u^k_n$ \\
        \hline
        $f^{c_k}_N$ & $f^{c_k}_{N-1}$ & $b^{c_k}_1$, $u^k_N$ \\
        \hline
        $u^k_1$ & $f^{c_k}_1$ & $u^k_2$, $f^{s_k}_1$ \\
        \hline
        $u^k_n,1<n<N$ & $u^k_{n-1}$, $f^{c_k}_n$ & $u^k_{n+1}$, $f^{s_k}_n$ \\
        \hline
        $u^k_N$ & $u^k_{N-1}$, $f^{c_k}_N$ & $f^{s_k}_N$ \\
        \hline
        $f^{s_k}_1$ & $u^k_1$ & $b^{s_k}_1$ \\
        \hline
        $b^{s_k}_1$ & $f^{s_k}_1$ & $f^{s_k}_2$, $d^k_1$ \\
        \hline
        $f^{s_k}_n,1<n<N$ & $u^k_n$, $b^{s_k}_{n-1}$ & $b^{s_k}_{n+1}$ \\
        \hline
        $b^{s_k}_n,1<n<N$ & $f^{s_k}_n$ & $f^{s_k}_{n+1}$, $d^k_n$ \\
        \hline
        $f^{s_k}_N$ & $u^k_N$, $b^{s_k}_{N-1}$ & $b^{s_k}_N$ \\
        \hline
        $b^{s_k}_N$ & $f^{s_k}_N$ & $d^k_N$ \\
        \hline
        $d^k_1$ & $b^{s_k}_1$ & $d^k_2$, $b^{c_k}_1$ \\
        \hline
        $d^k_n,1<n<N$ & $d^k_{n-1}$, $b^{s_k}_n$ & $d^k_{n+1}$, $b^{c_k}_n$ \\
        \hline
        $d^k_N$ & $d^{k}_{N-1}$, $b^{s_k}_N$ & $b^{c_k}_N$ \\
        \hline
        $b^{c_k}_1$ & $f^{c_k}_N$, $d^k_1$ & $b^{c_k}_2$ \\
        \hline
        $b^{c_k}_n, 1<n<N$ & $b^{c_k}_{n-1}$, $d^k_n$ & $b^{c_k}_{n+1}$ \\
        \hline
        $b^{c_k}_N$ & $b^{c_k}_{N-1}$, $d^k_N$ & $n/a$ \\
        \Xhline{2\arrayrulewidth}
    \end{tabular}
    \label{table:stages}
\end{table}

\section{Convergence analysis}
\label{sec:convergence}
This section analyzes the impact of splitting neural network and reordering training stages on model convergence and final accuracy.

\subsection{Splitting DNNs and model accuracy}
\label{subsec:split-nn}

We will demonstrate that splitting a DNN does not impact model accuracy. Assuming that $\mathbf{x}^0$ is a mini-batch of data and $\mathbf{y}$ is the corresponding label set, $\tilde{f}_q$ denotes the forward pass function of layer $q$ and $\mathbf{x}^q$ denotes the output of layer $q$, where $q=1,2,...,Q$.
\begin{equation}
    \mathbf{x}^q = \tilde{f}_q(\mathbf{x}^{q-1})
\end{equation}

The forward pass of the complete model $M^k$ in FL is:
\begin{equation}
    \mathbf{\hat{y}}=\tilde{f}_Q(\tilde{f}_{Q-1}(...\tilde{f}_1(\mathbf{x}^0)))
\end{equation}
where $\mathbf{\hat{y}}$ is the output of the final layer. If the model is split, then the training that occurs on the device and server is also split into two phases.
\begin{equation}
\label{eq:fp-client}
    \mathbf{a}=\mathbf{x}^P=\tilde{f}_P(\tilde{f}_{P-1}(...\tilde{f}_1(\mathbf{x})))
\end{equation}
\begin{equation}
\label{eq:fp-server}
    \mathbf{\hat{y}^\prime}=\tilde{f}_Q(\tilde{f}_{Q-1}(...\tilde{f}_{P+1}(\mathbf{a})))
\end{equation}
where $\mathbf{a}$ is the activations that are transferred from device $k$ to the server and $\mathbf{\hat{y}^\prime}$ is the final output. 
\begin{equation}
\begin{split}
    \mathbf{\hat{y}^\prime} &=\tilde{f}_Q(\tilde{f}_{Q-1}(...\tilde{f}_{P+1}(\mathbf{a}))) \\
    &=\tilde{f}_Q(\tilde{f}_{Q-1}(...\tilde{f}_1(\mathbf{x}))) \\
    &=\mathbf{\hat{y}}
\end{split}
\end{equation}

Thus, the loss function when splitting the model is the same as the original loss function when the model is not split.
\begin{equation}
    l(\mathbf{y}, \mathbf{\hat{y}}) = l(\mathbf{y}, \mathbf{\hat{y}}^\prime)
\end{equation}

We use $\tilde{b}_q$ to denote the backward pass function of layer $q$, which is the derivative of $\tilde{f}_q$.
\begin{equation}
\label{eq:bp}
    \tilde{b}_q(\mathbf{x}^{q-1}) = \frac{\partial \tilde{f}_q(\mathbf{x}^{q-1})}{\partial\mathbf{x}^{q-1}}=\frac{\partial\mathbf{x}^{q}}{\partial\mathbf{x}^{q-1}}
\end{equation}

The weights in layer $q$ of the original model and the split model are denoted as $\mathbf{w}_q$ and $\mathbf{w}_q^\prime$, respectively. Assume $g$ is the gradient function, then:
\begin{equation}
\label{eq:bp1}
\begin{split}
    g(\mathbf{w}_q^\prime) = \frac{\partial l(\mathbf{y}, \mathbf{\hat{y}}^\prime)}{\partial \mathbf{\hat{y}}^\prime}\tilde{b}_Q(\tilde{b}_{Q-1}(...\tilde{b}_{q+1}(\mathbf{x}_q))) \\ 
\end{split}
\end{equation}
\begin{equation}
\label{eq:bp2}
\begin{split}
    g(\mathbf{w}_q) &= \frac{\partial l(\mathbf{y}, \mathbf{\hat{y}})}{\partial \mathbf{\hat{y}}}\tilde{b}_Q(\tilde{b}_{Q-1}(...\tilde{b}_{q+1}(\mathbf{x}_q))) \\
\end{split}
\end{equation}

Based on Equation~\ref{eq:bp1} and Equation~\ref{eq:bp2}:
\begin{equation}
    g(\mathbf{w}_q^\prime) = g(\mathbf{w}_q)
\end{equation}

Since splitting a DNN does not change the gradients, it consequently does not impact model accuracy.

\subsection{Reordering training stages and model accuracy}
\label{subsec:reorder}

We will demonstrate that the model accuracy of a DNN remains the same before and after reordering the training stages. The dataset on client $k$ is denoted as $\mathcal{D}^k$. $\mathcal{B}^k$ denotes a mini-batch in the original training process and $\mathcal{B}^k_n$, where $n=1,2,...,N$, to denote mini-batches in a training round after reordering training stages, where $\mathcal{B}^k = \bigcup_{n=1}^N \mathcal{B}^k_n$.

In the original training process, the model is updated after the backward pass of each mini-batch $\mathcal{B}^k$. Assuming $M^k$ is the original model and $\eta$ is the learning rate, then the updated model is:
\begin{equation}
    M^k_{new} = M^k - \frac{\eta}{B}\sum_{\mathbf{x}\in\mathcal{B}^k} g(M^k|\mathbf{x})
\end{equation}

In \PiPar, the model is updated after the backward pass of the last mini-batch $\mathcal{B}^k_n$ in each training round. The updated model is:
\begin{equation}
    {M^k_{new}}^\prime = M^k - \frac{\eta}{N}\sum_{n=1}^N g(M^k|\mathcal{B}^k_n)
\end{equation}

We have:
\begin{equation}
    \begin{split}
        {M^k_{new}}^\prime &= M^k - \frac{\eta}{N}\sum_{n=1}^N\left(\frac{1}{B^\prime}\sum_{\mathbf{x}\in\mathcal{B}_n^k}g(M^k|\mathbf{x})\right) \\
        &= M^k - \frac{\eta}{NB^\prime}\sum_{n=1}^N\sum_{\mathbf{x}\in\mathcal{B}_n^k}g(M^k|\mathbf{x}) \\
        &= M^k - \frac{\eta}{NB^\prime}\sum_{\mathbf{x}\in\mathcal{B}^k} g(M^k|\mathbf{x}) \\
        &\approx M^k - \frac{\eta}{B}\sum_{\mathbf{x}\in\mathcal{B}^k} g(M^k|\mathbf{x}) \\
        &= M^k_{new}
    \end{split}
\end{equation}

Therefore, the updated models with and without reordering the training stages are nearly the same (and the same if $NB^\prime=B$). Thus, reordering training stages does not impact model accuracy.

\section{Experimental Studies}
\label{sec:exp}
This section quantifies the benefits of \PiPar\ and demonstrates its superiority over existing CML techniques. We first consider the experimental environment in Section~\ref{subsec:setup}. The training efficiency and the model accuracy and convergence of \PiPar\ are compared against existing CML techniques in Section~\ref{subsec:eff-exp} and Section~\ref{subsec:acc-exp}, respectively. In Section~\ref{subsec:op-exp}, the performance of the proposed automated parameter selection approach is evaluated. Section~\ref{subsec:batch-size} analyses the impact of batch size on the performance of \PiPar. Section~\ref{subsec:robust} explores the impact on performance when using heterogeneous devices, when using differential privacy methods and when the bandwidth changes.

\subsection{Setup}
\label{subsec:setup}

The test platform consists of one server and 100 devices. An 8-core i7-11850H processor with 32GB RAM is used as the server that collaboratively trains DNNs with 100 Raspberry Pi 3B devices, each with 1GB RAM.

The network conditions considered are: 
(1) \textit{4G:} 10Mbps uplink bandwidth and 25Mbps downlink bandwidth;
(2) \textit{4G+:} 20Mbps uplink bandwidth and 40Mbps downlink bandwidth;
(3) \textit{WiFi:} 50Mbps uplink bandwidth and 50Mbps downlink bandwidth.
A regular network with a normal error rate is used in the experiments. The TCP/IP protocol used will handle packet loss. When the protocol detects packet loss, it will re-transmit the packet.

Two settings, the first using small DNNs and the second using large DNNs, are used in the experiments. The small DNNs, namely VGG-5~\cite{brusilovsky:simonyan2014very}, ResNet-18~\cite{DBLP:journals/corr/HeZRS15} and MobileNetV3-Small~\cite{DBLP:journals/corr/abs-1905-02244} (Table~\ref{table:structure}) are trained on the MNIST~\cite{deng2012mnist} and CIFAR-10~\cite{cifar10,Krizhevsky09} datasets. The large DNNs, namely VGG-16\cite{brusilovsky:simonyan2014very}, ResNet-101~\cite{DBLP:journals/corr/HeZRS15} and MobileNetV3-Large~\cite{DBLP:journals/corr/abs-1905-02244} (Table~\ref{table:large-structure}) are trained on the CIFAR-100~\cite{Krizhevsky09} and Tiny ImageNet~\cite{tinyimagenet} datasets. 
VGG, ResNet and MobileNet series models are convolutional neural networks (CNN) and are representative of high-performing models from the computer vision community for testing CML methods on devices~\cite{splitgp,fedgkt,DBLP:journals/corr/abs-2004-12088}.
Since the Raspberry Pis have limited memory, the large DNNs cannot be trained using FL as the entire model needs to fit on the device memory. The small and large DNNs can be trained using SFL and \PiPar\ since the models are split across the device and server, and the device only executes a few layers. 
We have chosen a range of small and large DNNs to demonstrate that \PiPar\ can work across a range of settings.
MNIST and CIFAR-10 have ten classes, while CIFAR-100 and Tiny ImageNet have 100. Each dataset is split into training, validation and test datasets, as shown in Table~\ref{table:dataset}. During training, the data samples are provided to the DNN as mini-batches. The size of each mini-batch (referred to as batch size), unless otherwise specified, is 100 for each device in FL and SFL. The batch size in \PiPar\ is $\lfloor 100/N^k \rfloor$, where $N^k$ is the parallel batch number for device $k$ and $k=1,2,...,100$ (refer to Equation~\ref{eq:batch-size}).

\begin{table}[tp]
    \centering
    \caption{Architecture of VGG-5, ResNet-18 and MobileNetV3-Small\tablefootnote{`CONV-A-B' represents a convolutional layer of A$\times$A kernel size and of B output channels. `FC-A' represents a fully connected layer with the output size A. `RES-A-B' denotes a residual block that consists of two convolutional layers of A$\times$A kernel size and of B output channels. The output of each residual block is the output of the last inner convolutional layer plus the input of the residual block. ``BNECK-A-B" denotes a bottleneck residual block that consists of an expansion layer, a convolutional layer with the kernel size A$\times$A and a projection layer with the output channel number B. The number of classes is denoted as X. The activation, batch normalization and pooling layers are not shown for simplicity.}}
    
    \begin{tabular}{ccc}
        \Xhline{2\arrayrulewidth}
        \textbf{VGG-5}  & \textbf{ResNet-18} & \textbf{MobileNetV3-Small} \\
        \Xhline{2\arrayrulewidth}
        CONV-3-32 & CONV-7-64 & CONV-3-16 \\
        \hline
        CONV-3-64 & RES-3-64 $\times$ 2 & BNECK-3-16 \\
        \hline
        CONV-3-64 & RES-3-128 $\times$ 2 & BNECK-3-24 $\times$ 2 \\
        \hline
        FC-128 & RES-3-256 $\times$ 2 & BNECK-5-40 $\times$ 3 \\
        \hline 
        FC-X & RES-3-512 $\times$ 2 & BNECK-5-48 $\times$ 2 \\
        \hline
         & FC-X & BNECK-5-96 $\times$ 3 \\
        \cline{2-3}
         & & CONV-1-576 \\
        \cline{3-3}
         & & CONV-1-1024 \\
        \cline{3-3}
         & & FC-X \\
        \Xhline{2\arrayrulewidth}
    \end{tabular}
    \label{table:structure}
\end{table}

\begin{table}[tp]
    \centering
    \caption{Architecture of VGG-16, ResNet-101 and MobileNetV3-Large}
    \begin{tabular}{ccc}
        \Xhline{2\arrayrulewidth}
        \textbf{VGG-16}  & \textbf{ResNet-101} & \textbf{MobileNetV3-Large} \\
        \Xhline{2\arrayrulewidth}
        CONV-3-64 $\times$ 2 & CONV-7-64 & CONV-3-16 \\
        \hline
        CONV-3-128 $\times$ 2 & RES-3-64 & BNECK-3-16 \\
        \hline
        CONV-3-256 $\times$ 3 & RES-3-256 $\times$ 3 & BNECK-3-24 $\times$ 2 \\
        \hline
        CONV-3-512  $\times$ 6 & RES-3-512 $\times$ 4 & BNECK-5-40 $\times$ 3 \\
        \hline 
        FC-4096 $\times$ 2 & RES-3-1024 $\times$ 23 & BNECK-3-80 $\times$ 4 \\
        \hline
        FC-X & RES-3-2048 $\times$ 2 & BNECK-3-112 $\times$ 2 \\
        \hline
         & FC-X & BNECK-5-160 $\times$ 3 \\
        \cline{2-3}
         & & CONV-1-960 \\
        \cline{3-3}
         & & CONV-1-1280 \\
        \cline{3-3}
         & & FC-X \\
        \Xhline{2\arrayrulewidth}
    \end{tabular}
    \label{table:large-structure}
\end{table}

\begin{table}[tp]
    \centering
    \caption{Training, validation and test data sizes used in the experiments}
    \begin{tabular}{ccccc}
        \Xhline{2\arrayrulewidth}
        \textbf{Dataset} & \makecell{\textbf{Training} \\ \textbf{Set Size}} & \makecell{\textbf{Validation} \\ \textbf{Set Size}} &  \makecell{\textbf{Test} \\ \textbf{Set Size}}\\
        \Xhline{2\arrayrulewidth}
        MNIST & 60,000 & 2,000 & 8,000 \\
        CIFAR-10 & 50,000 & 2,000 & 8,000 \\
        CIFAR-100 & 50,000 & 2,000 & 8,000 \\
        Tiny ImageNet & 100,000 & 2,000 & 8,000 \\
        \Xhline{2\arrayrulewidth}
    \end{tabular}
    \label{table:dataset}
\end{table}

\begin{figure*}[tp]
	\centering
	\subfigure[VGG-5 (MNIST)]{
	    \includegraphics[width=0.3\textwidth]{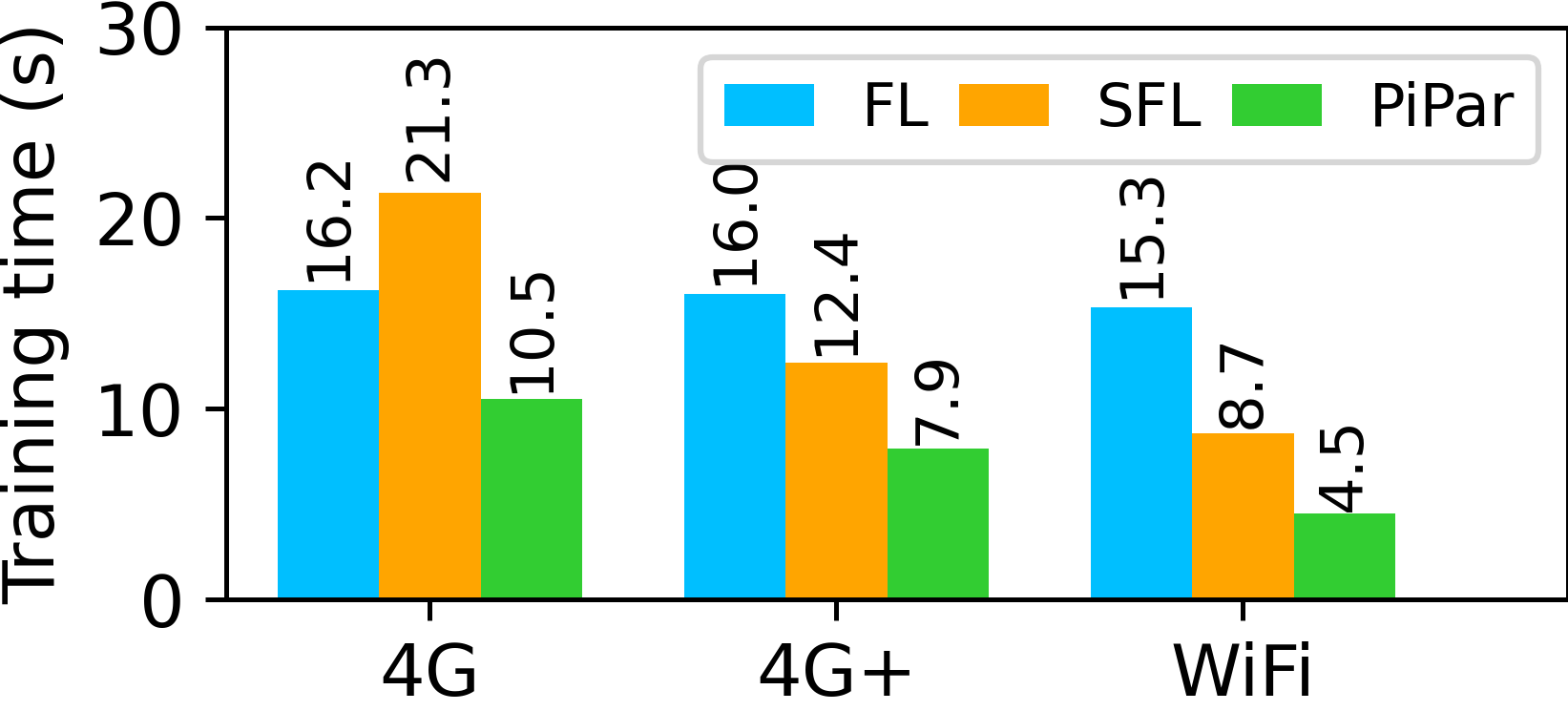}
	    \label{fig:vgg5-mnist-time}
	    }
	\hfill
	\subfigure[ResNet-18 (MNIST)]{
	    \includegraphics[width=0.3\textwidth]{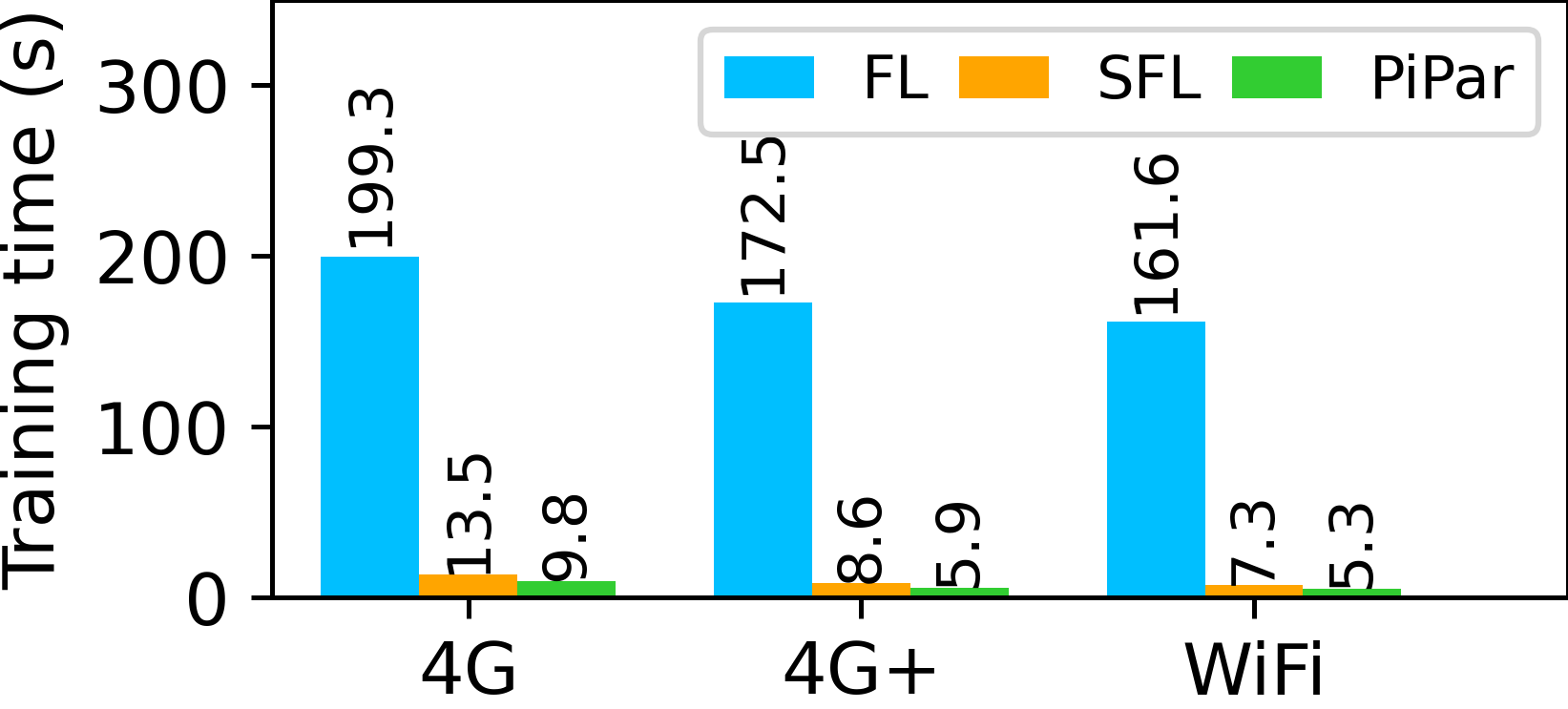}
	    \label{fig:resnet18-mnist-time}
	    }
        \hfill
        \subfigure[MobileNetV3-Small (MNIST)]{
	    \includegraphics[width=0.3\textwidth]{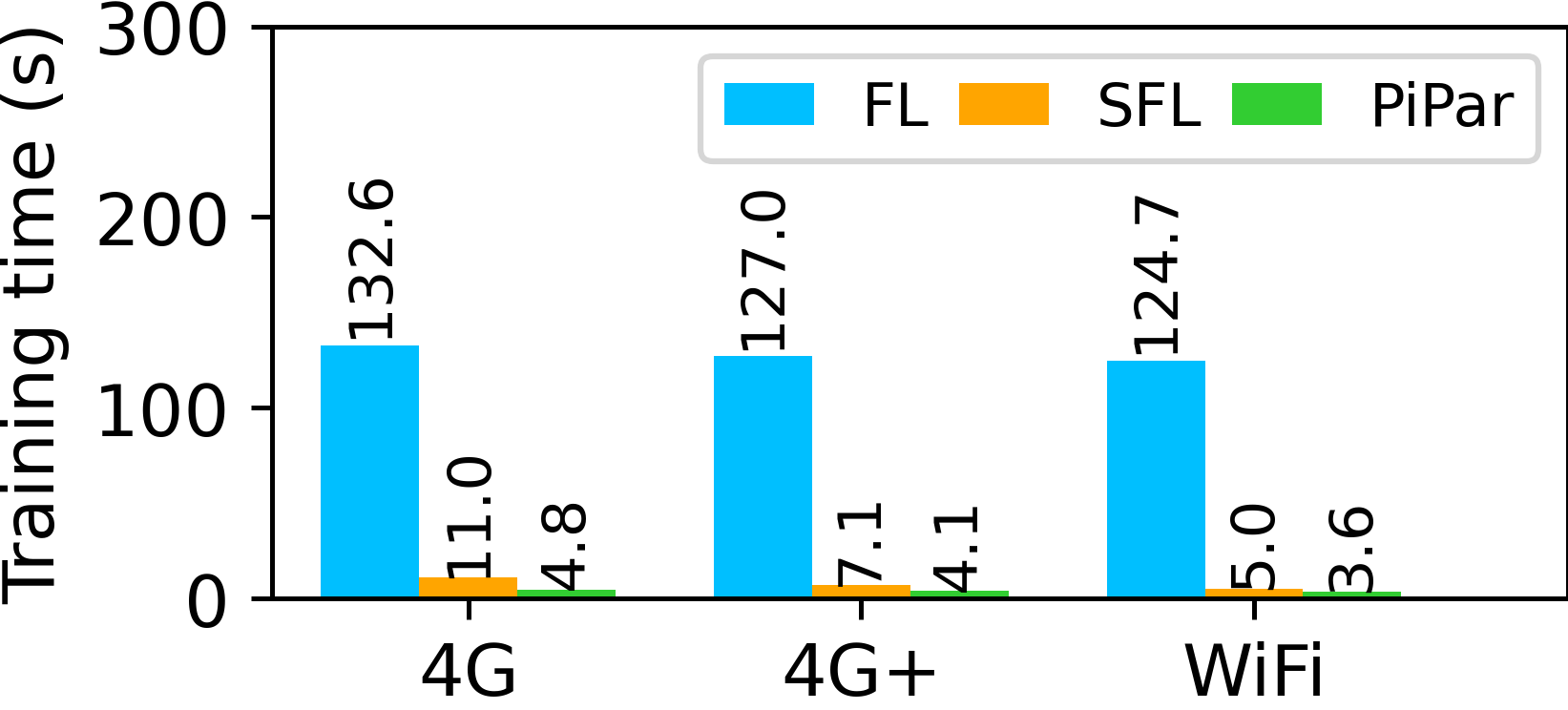}
	    \label{fig:mobilesmall-mnist-time}
	    }
        \subfigure[VGG-5 (CIFAR-10)]{
	    \includegraphics[width=0.3\textwidth]{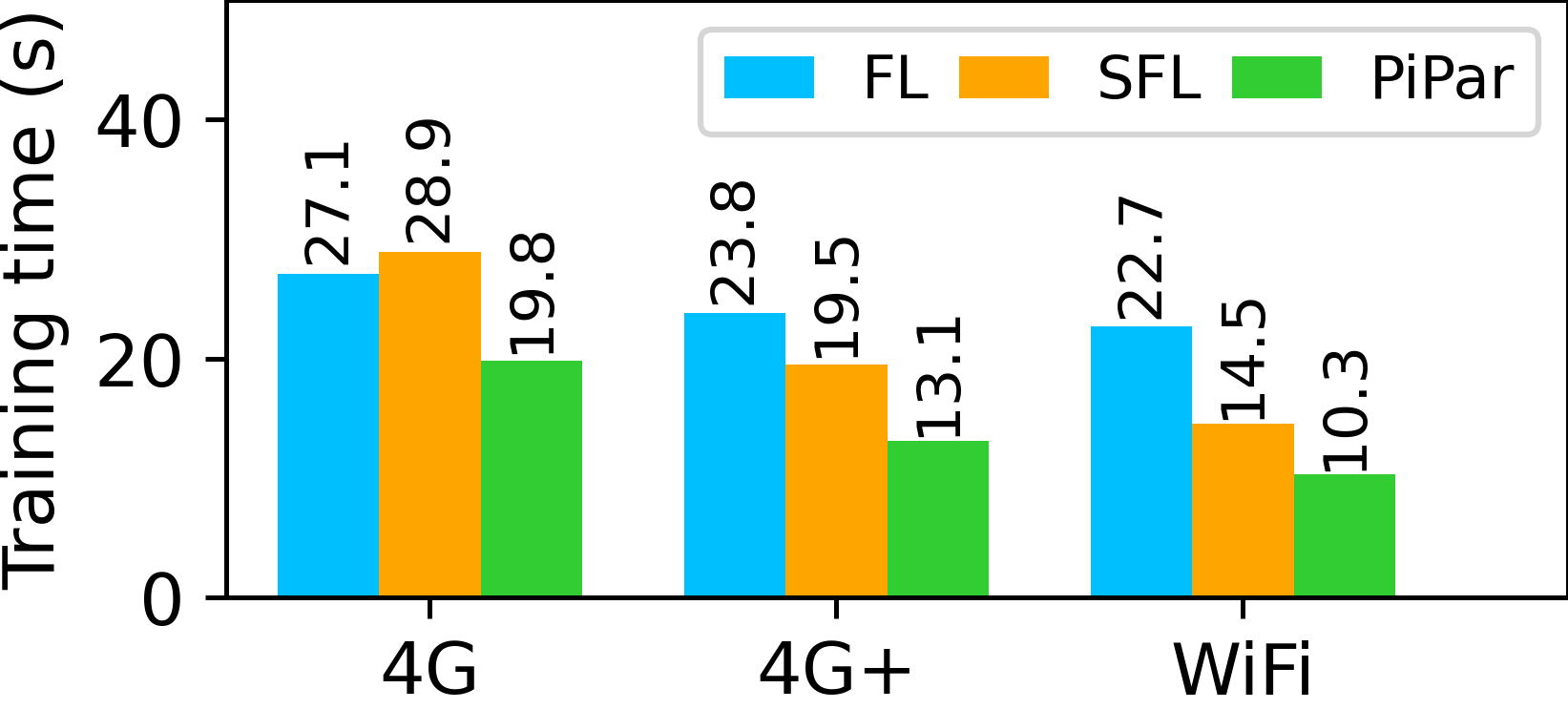}
	    \label{fig:vgg5-cifar10-time}
	    }
	\hfill
	\subfigure[ResNet-18 (CIFAR-10)]{
	    \includegraphics[width=0.3\textwidth]{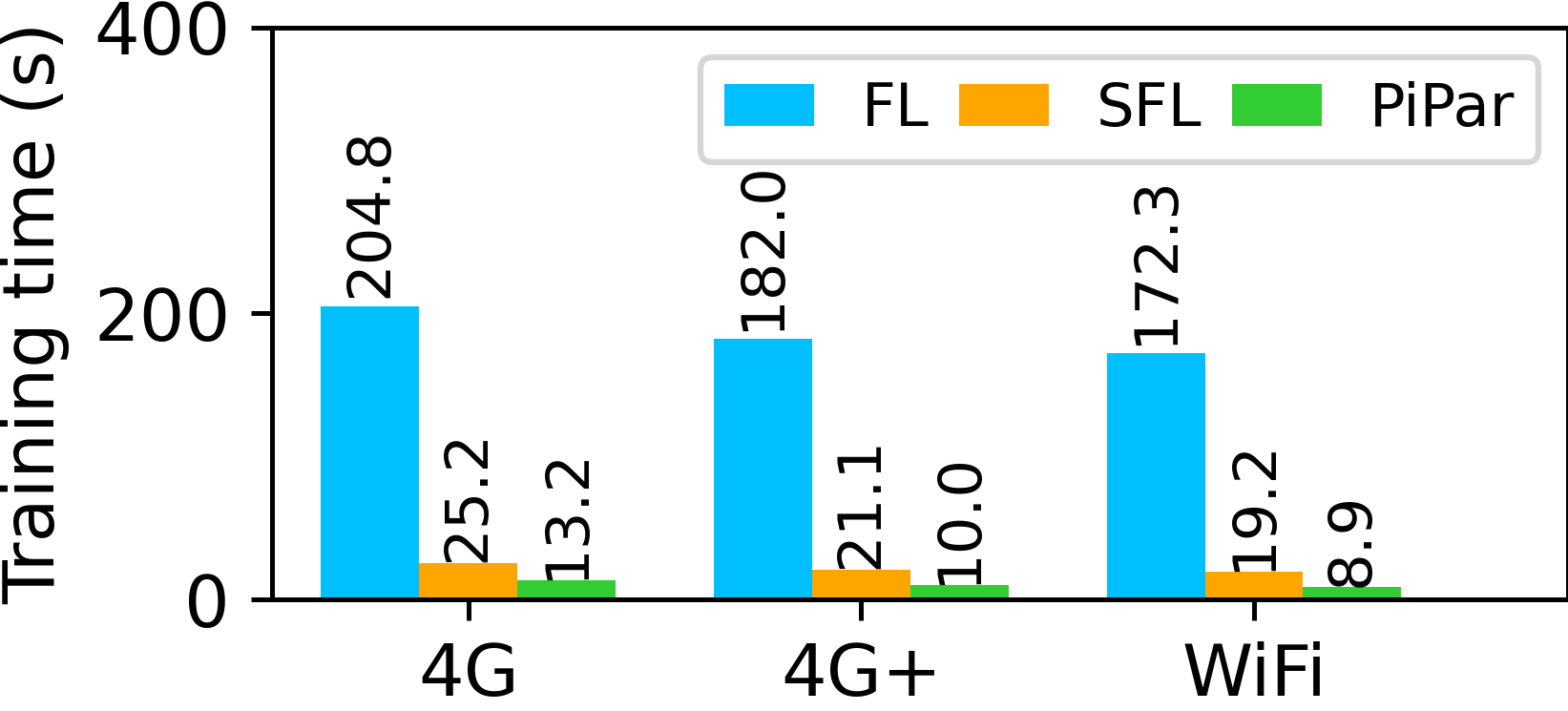}
	    \label{fig:resnet18-cifar10-time}
	    }
        \hfill
        \subfigure[MobileNetV3-Small (CIFAR-10)]{
	    \includegraphics[width=0.3\textwidth]{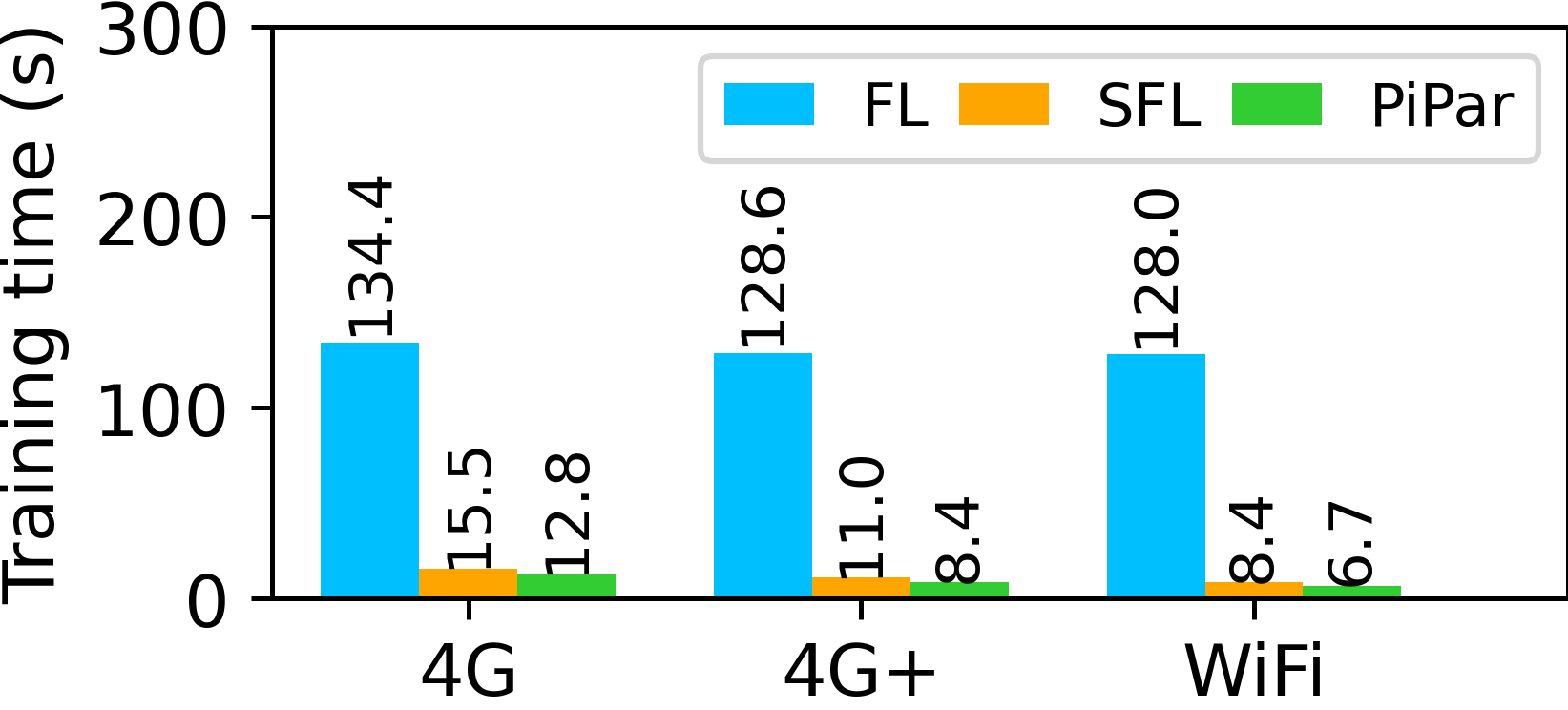}
	    \label{fig:mobilesmall-cifar10-time}
	    }
	\caption{Training time per epoch for FL, SFL and \PiPar\ under different network conditions for small DNNs.}
	\label{fig:training-time}
\end{figure*}

\begin{figure*}[tp]
	\centering
	\subfigure[VGG-16 (CIFAR-100)]{
	    \includegraphics[width=0.3\textwidth]{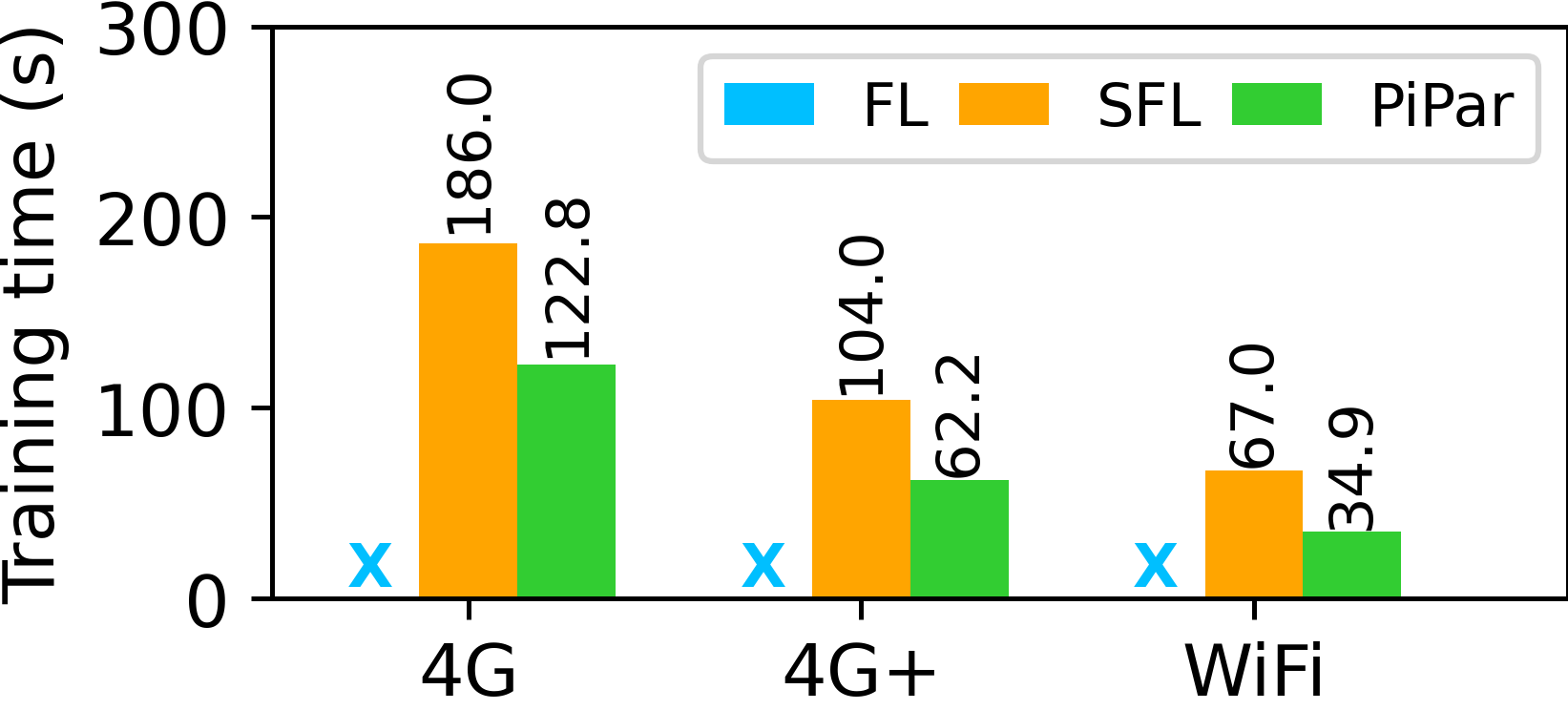}
	    \label{fig:vgg16-cifar100-time}
	    }
	\hfill
	\subfigure[ResNet-101 (CIFAR-100)]{
	    \includegraphics[width=0.3\textwidth]{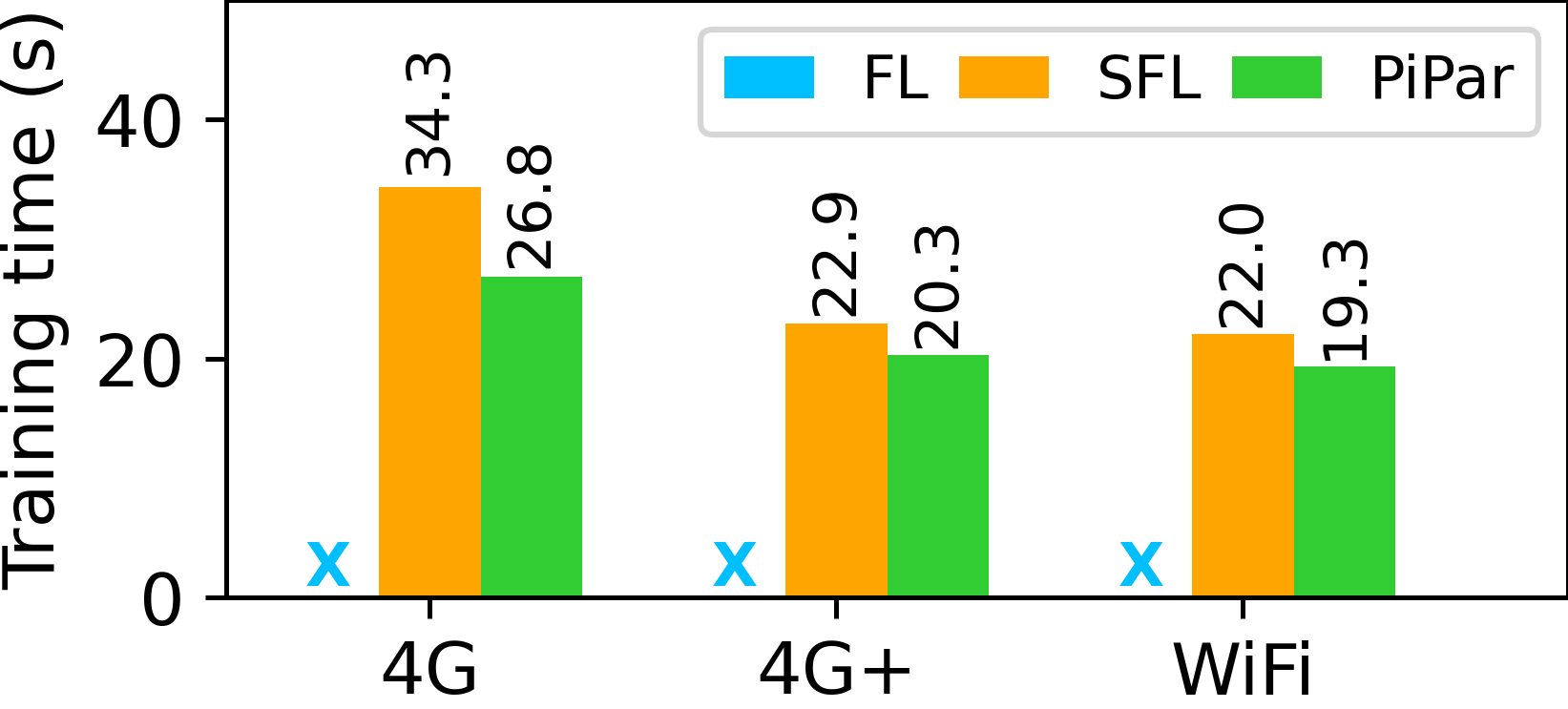}
	    \label{fig:resnet101-cifar100-time}
	    }
        \hfill
        \subfigure[MobileNetV3-Large (CIFAR-100)]{
	    \includegraphics[width=0.3\textwidth]{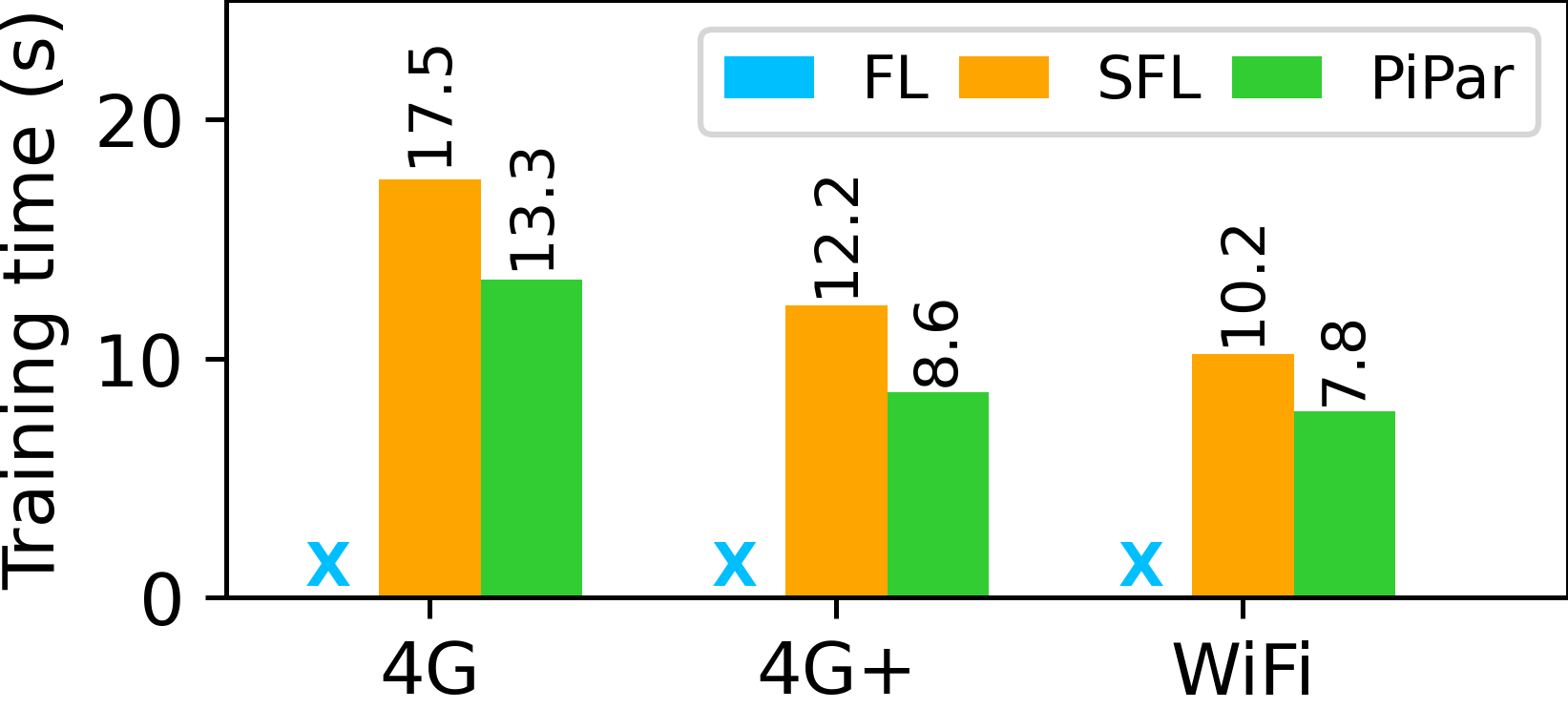}
	    \label{fig:mobilelarge-cifar100-time}
	    }
        \subfigure[VGG-16 (Tiny ImageNet)]{
	    \includegraphics[width=0.3\textwidth]{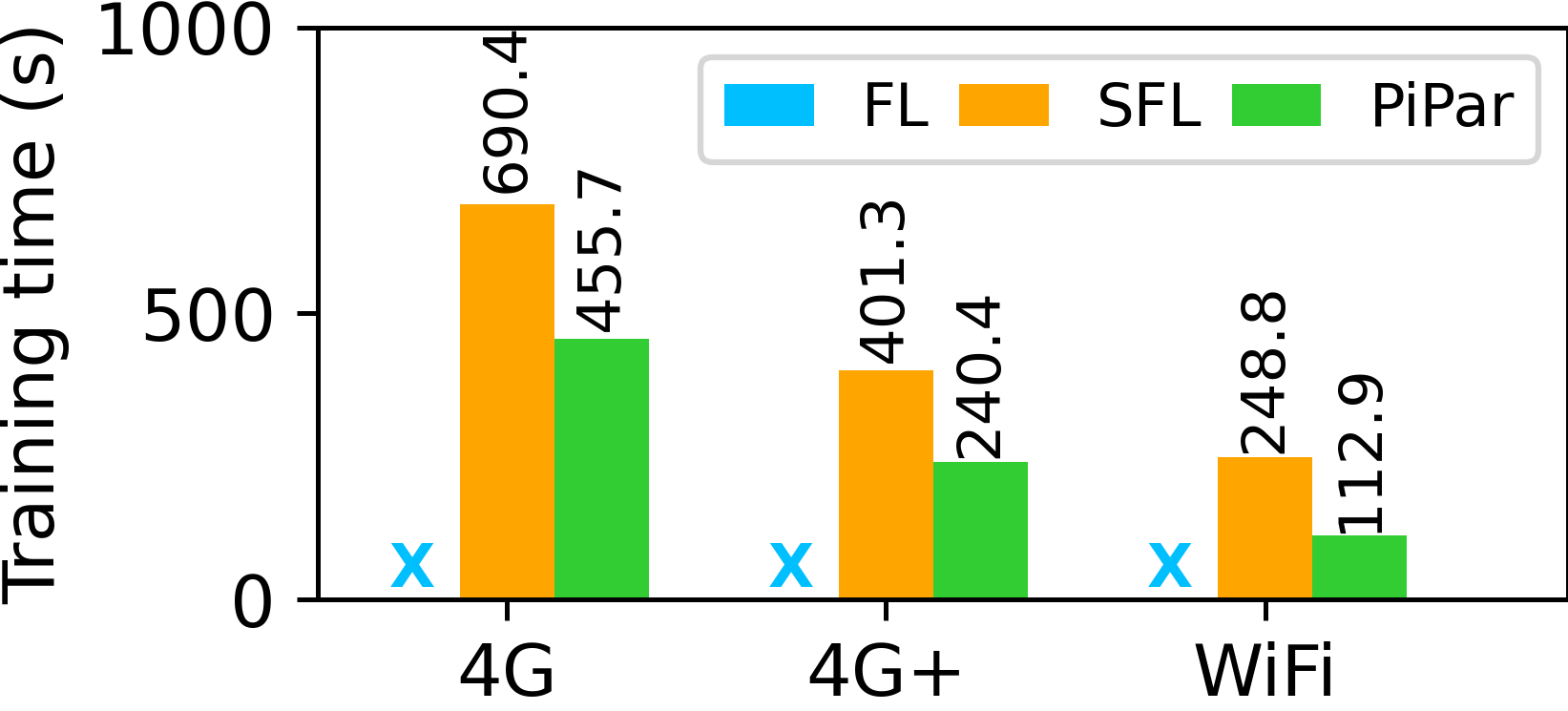}
	    \label{fig:vgg16-tinyimagenet-time}
	    }
	\hfill
	\subfigure[ResNet-101 (Tiny ImageNet)]{
	    \includegraphics[width=0.3\textwidth]{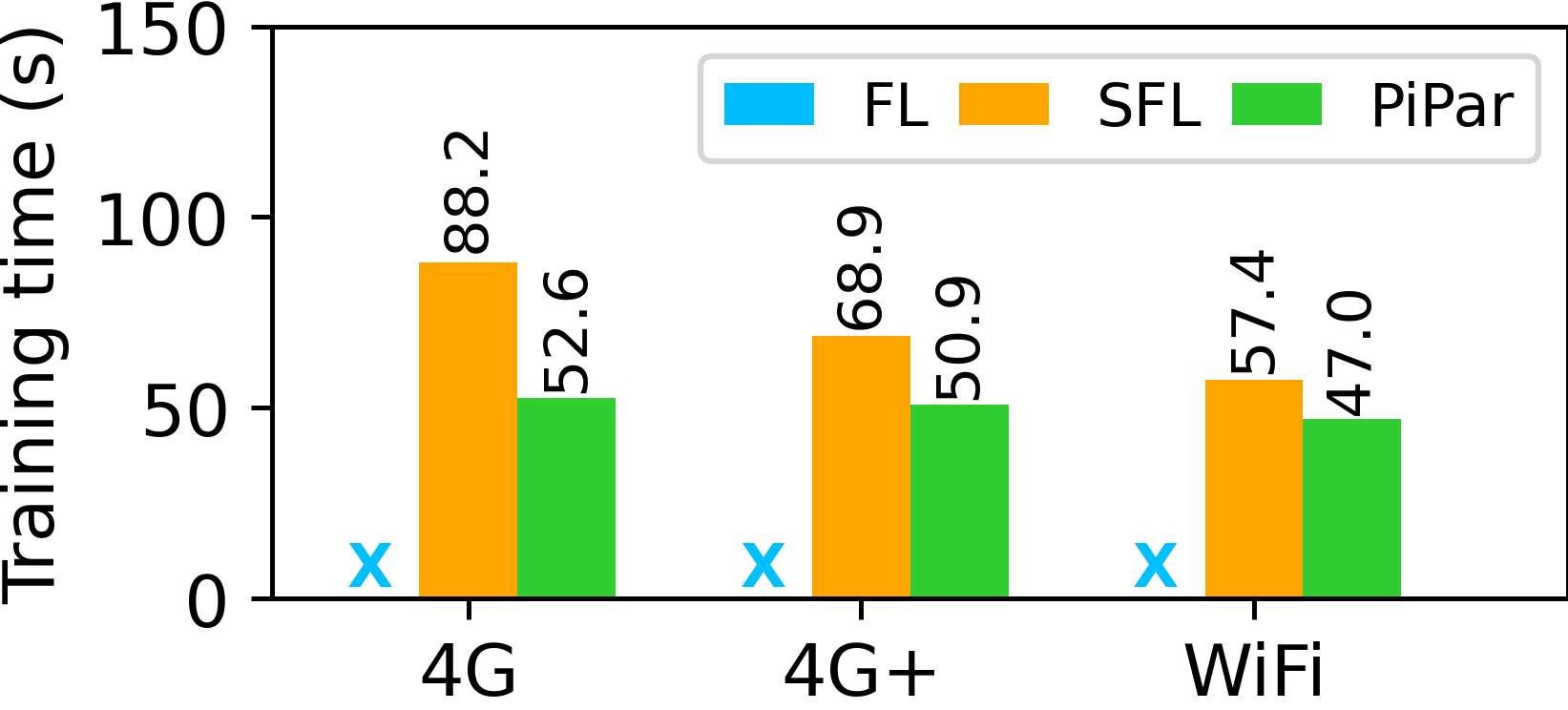}
	    \label{fig:resnet101-tinyimagenet-time}
	    }
        \hfill
        \subfigure[MobileNetV3-Large (Tiny ImageNet)]{
	    \includegraphics[width=0.3\textwidth]{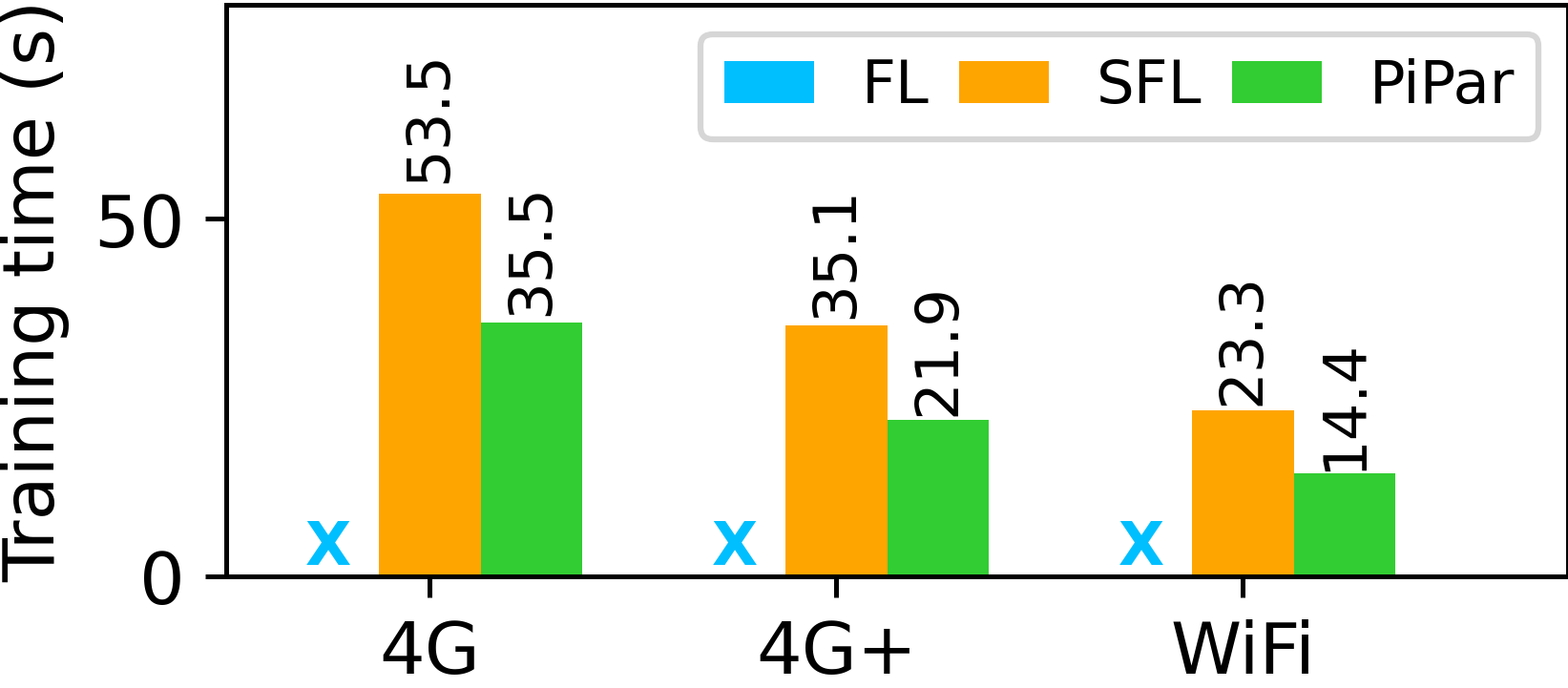}
	    \label{fig:mobilelarge-tinyimagenet-time}
	    }
	\caption{Training time per epoch for SFL and \PiPar\ under different network conditions for large DNNs. FL results are not shown as the entire DNN does not fit on the device memory.}
	\label{fig:training-time-large}
\end{figure*}

\begin{figure*}[tp]
	\centering
	\subfigure[VGG-5 (MNIST)]{
	    \includegraphics[width=0.3\textwidth]{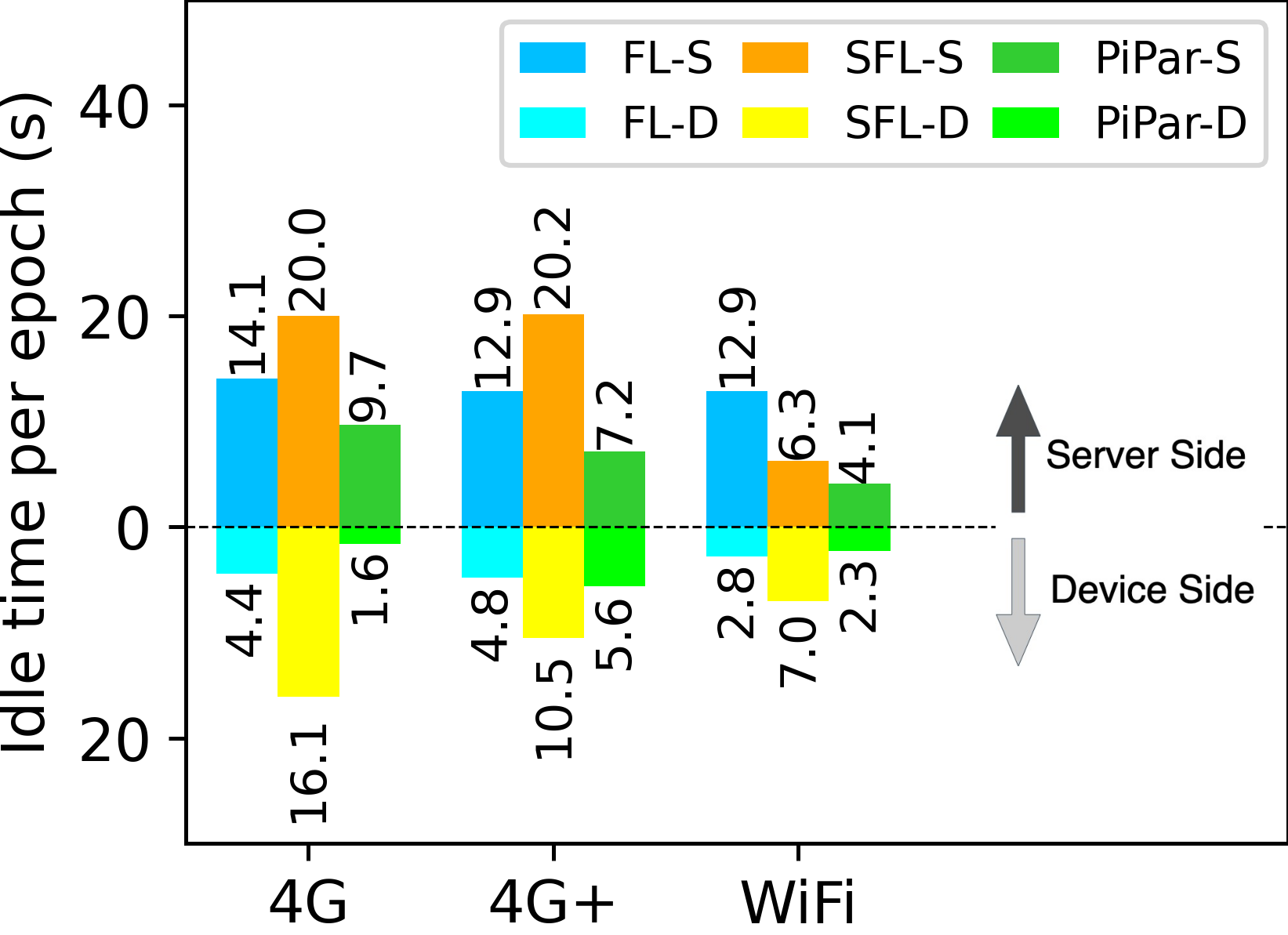}
	    \label{fig:vgg5-mnist-idle}
	    }
	\hfill
	\subfigure[ResNet-18 (MNIST)]{
	    \includegraphics[width=0.3\textwidth]{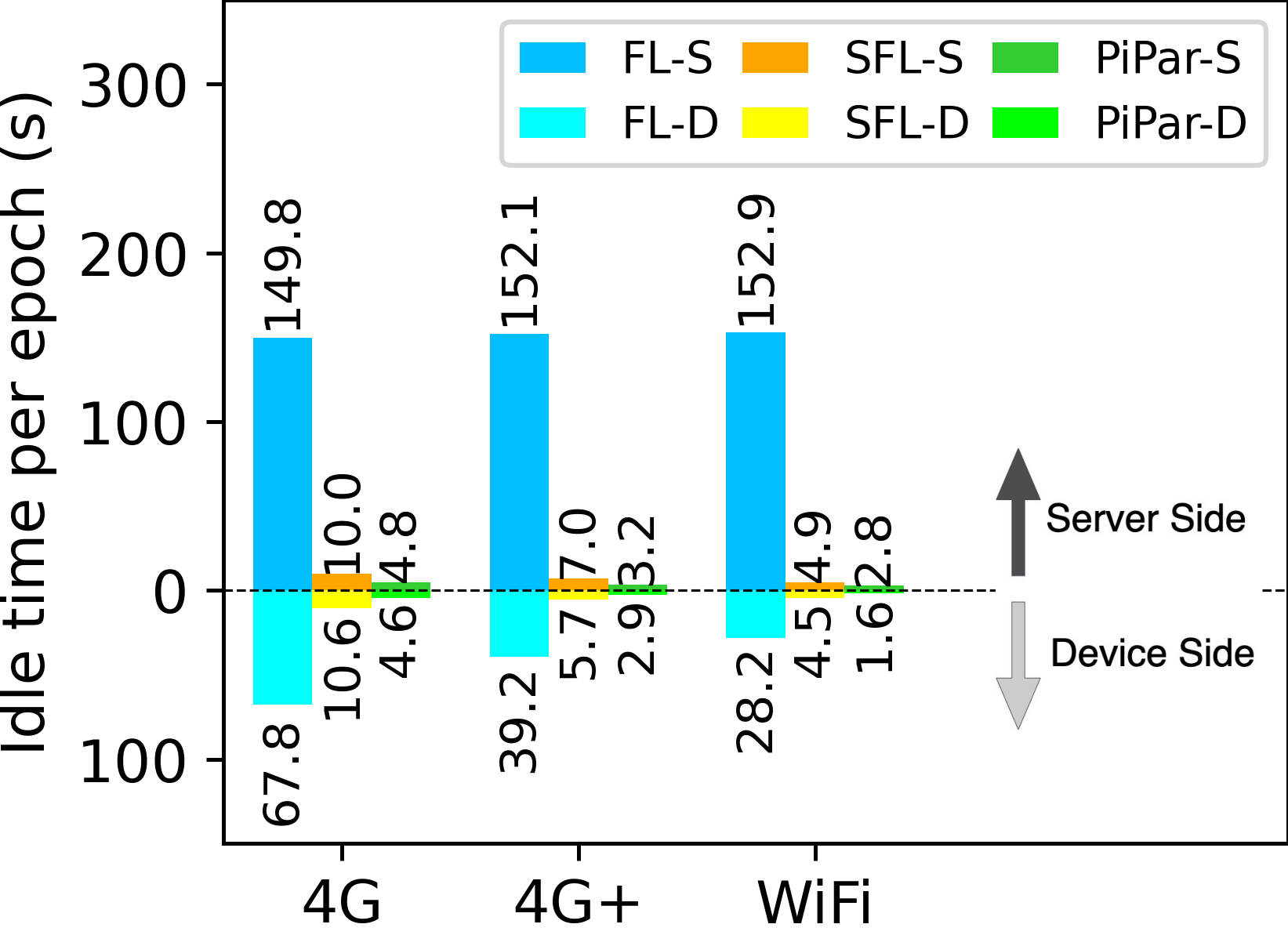}
	    \label{fig:resnet18-mnist-idle}
	    }
        \hfill
	\subfigure[MobileNetV3-Small (MNIST)]{
	    \includegraphics[width=0.3\textwidth]{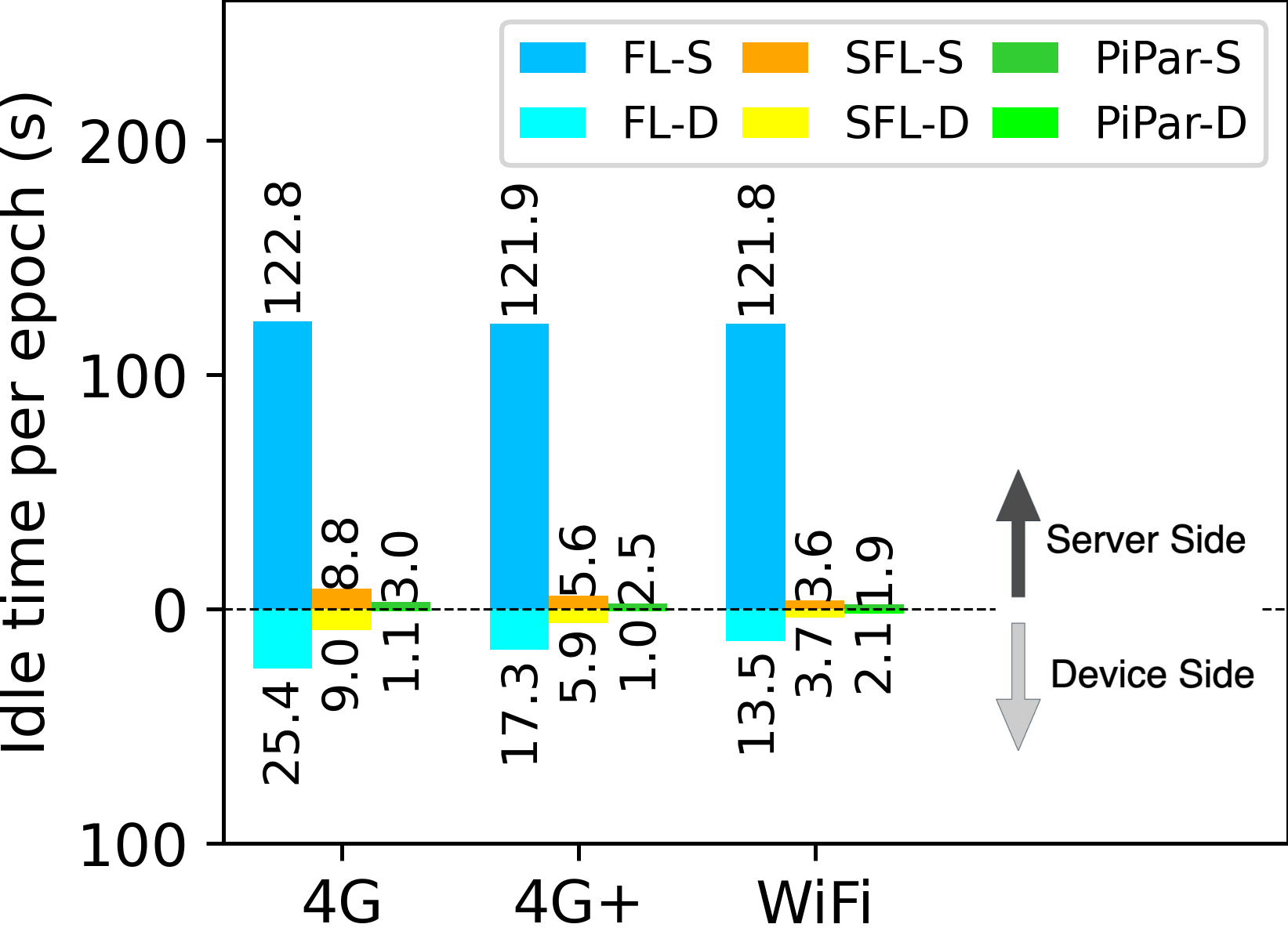}
	    \label{fig:mobilesmall-mnist-idle}
	    }
        \subfigure[VGG-5 (CIFAR-10)]{
	    \includegraphics[width=0.3\textwidth]{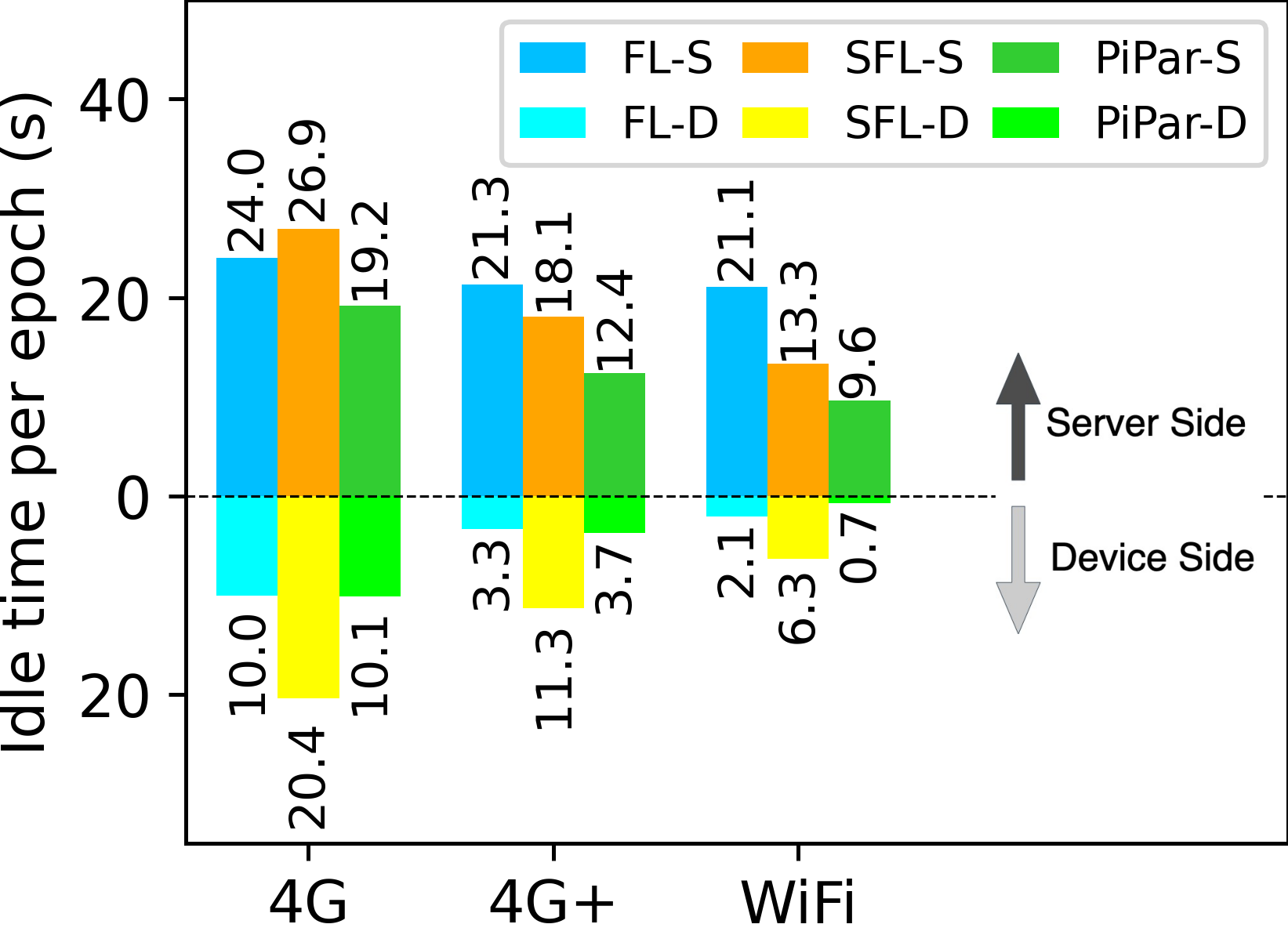}
	    \label{fig:vgg5-cifar10-idle}
	    }
	\hfill
	\subfigure[ResNet-18 (CIFAR-10)]{
	    \includegraphics[width=0.3\textwidth]{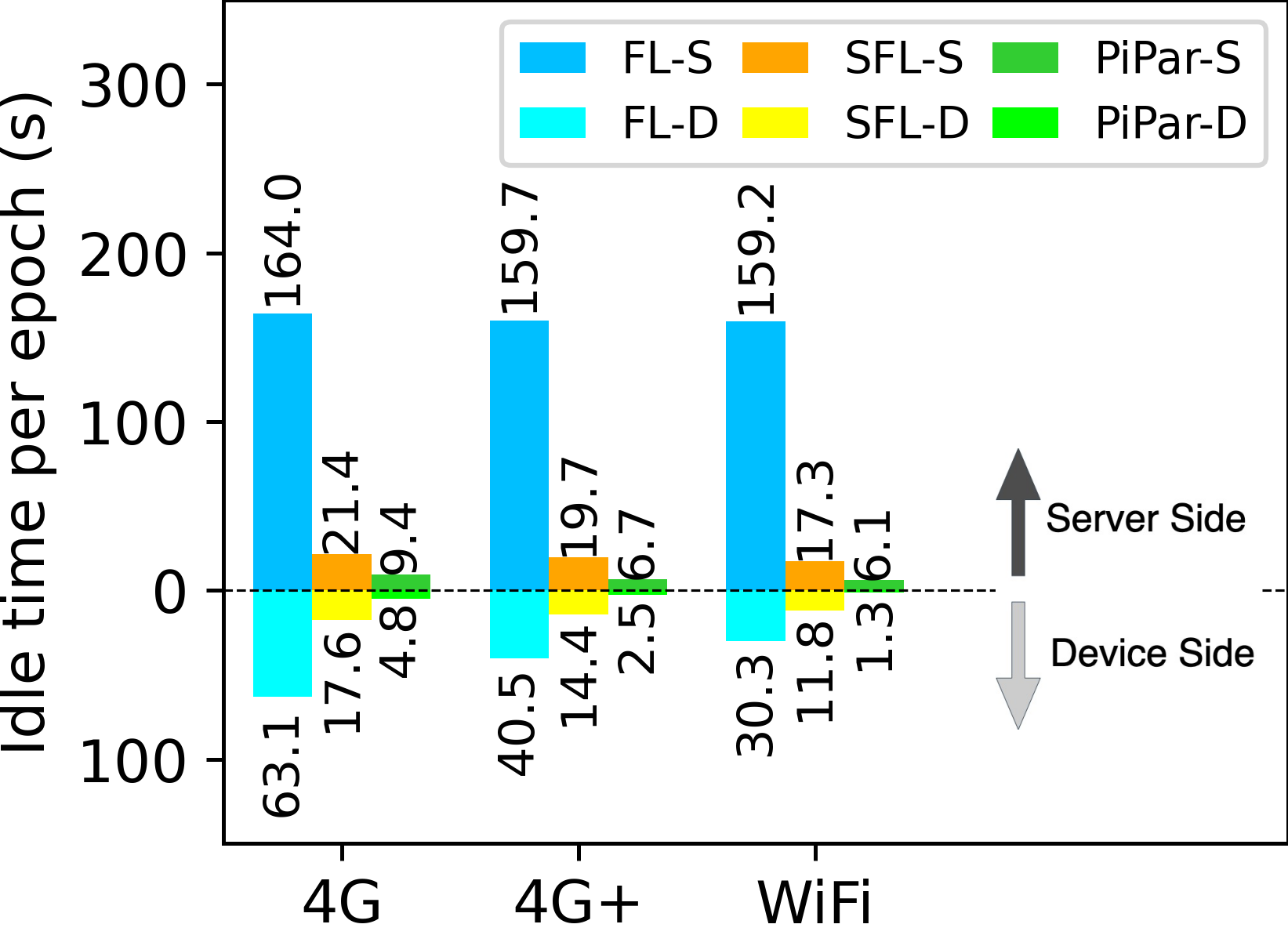}
	    \label{fig:resnet18-cifar10-idle}
	    }
        \hfill
	\subfigure[MobileNetV3-Small (CIFAR-10)]{
	    \includegraphics[width=0.3\textwidth]{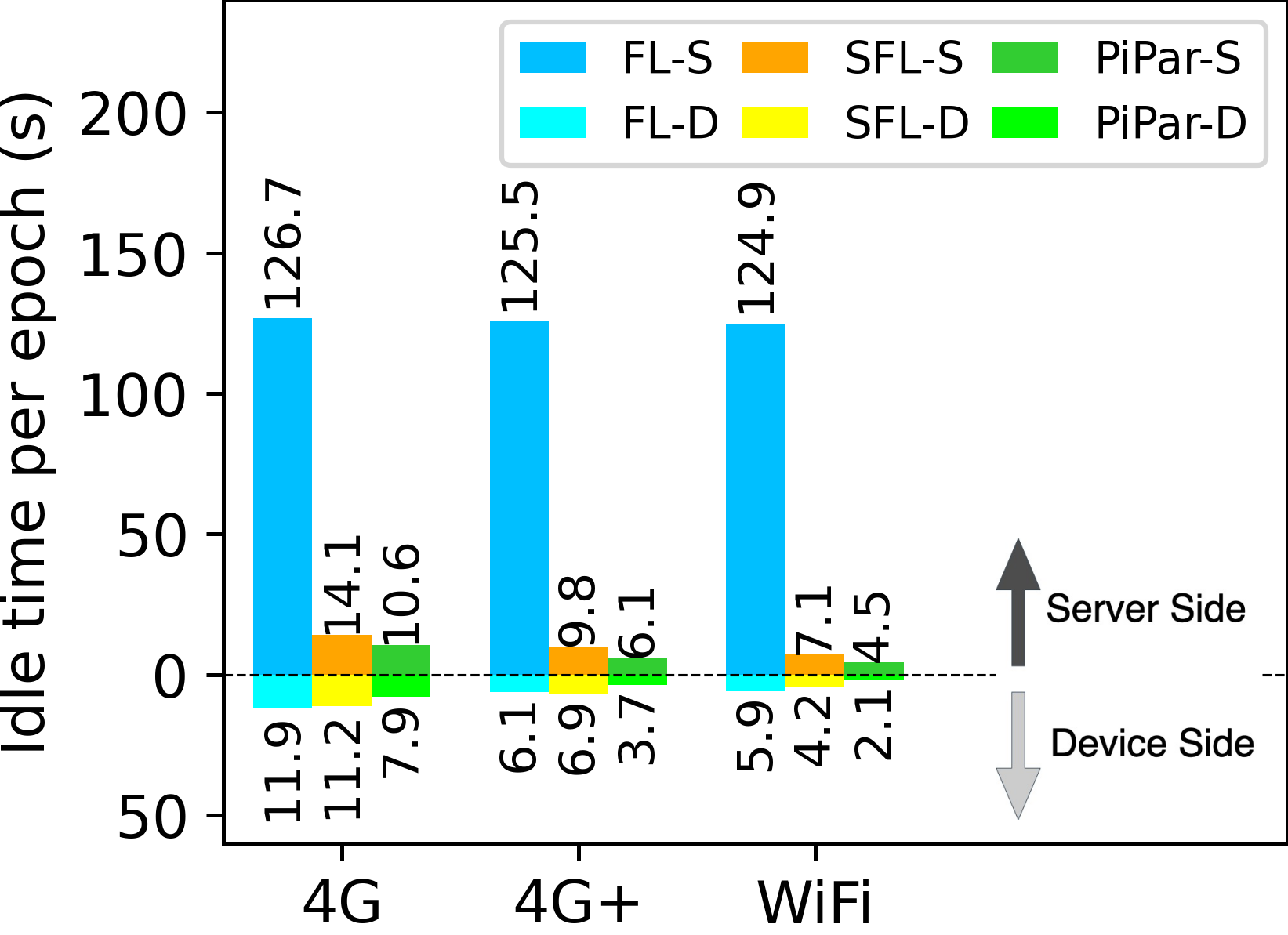}
	    \label{fig:mobilesmall-cifar10-idle}
	    }
	\caption{Idle time per epoch on the server and devices in FL, SFL and \PiPar\ under different network conditions for small DNNs. `S' and `D' in the legend represent server-side and device-side idle time, respectively. They are shown in the upward and downward bars.}
	\label{fig:idle-time}
\end{figure*}

\begin{figure*}[tp]
	\centering
	\subfigure[VGG-16 (CIFAR-100)]{
	    \includegraphics[width=0.3\textwidth]{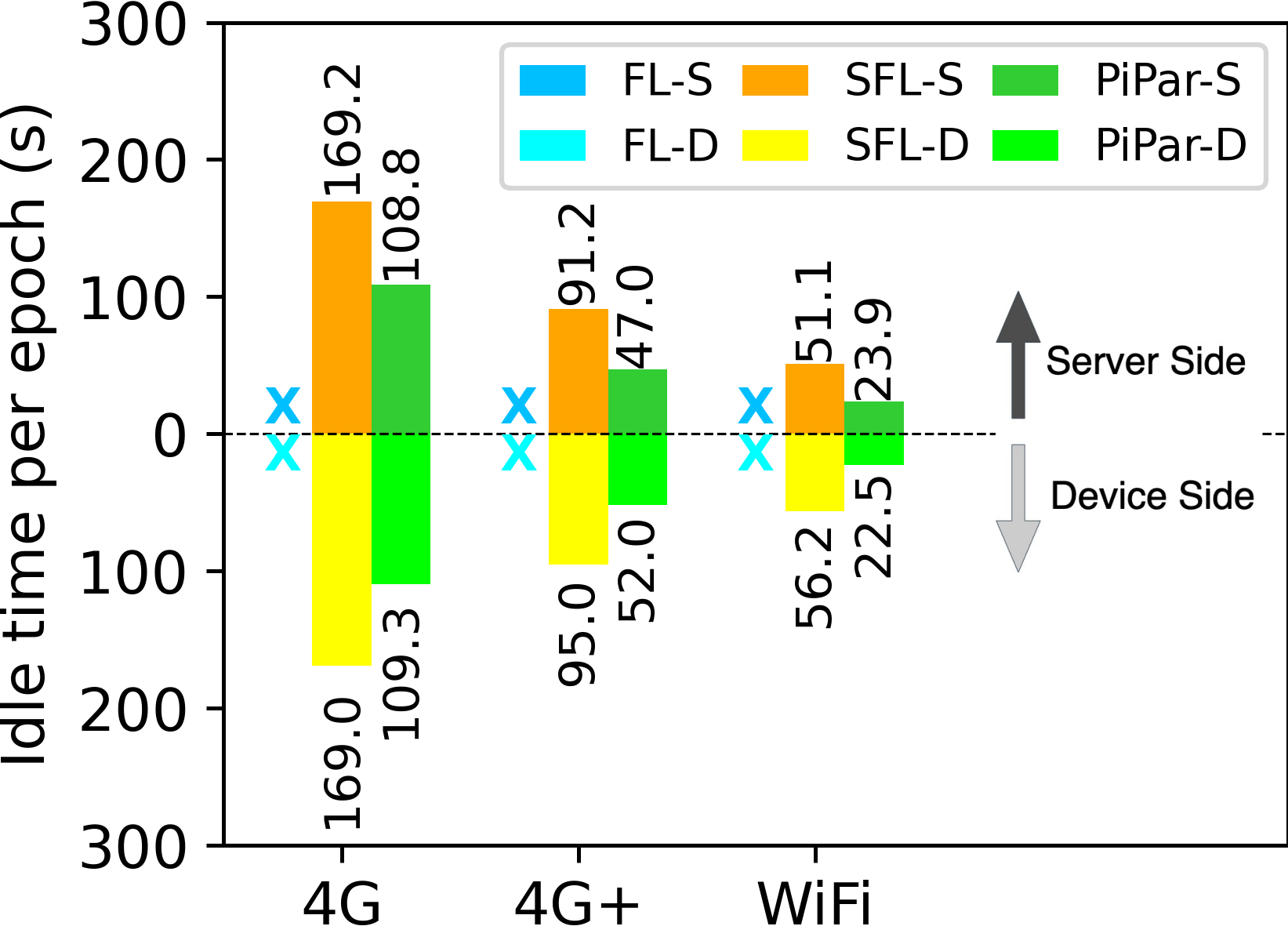}
	    \label{fig:vgg16-cifar100-idle}
	    }
	\hfill
	\subfigure[ResNet-101 (CIFAR-100)]{
	    \includegraphics[width=0.3\textwidth]{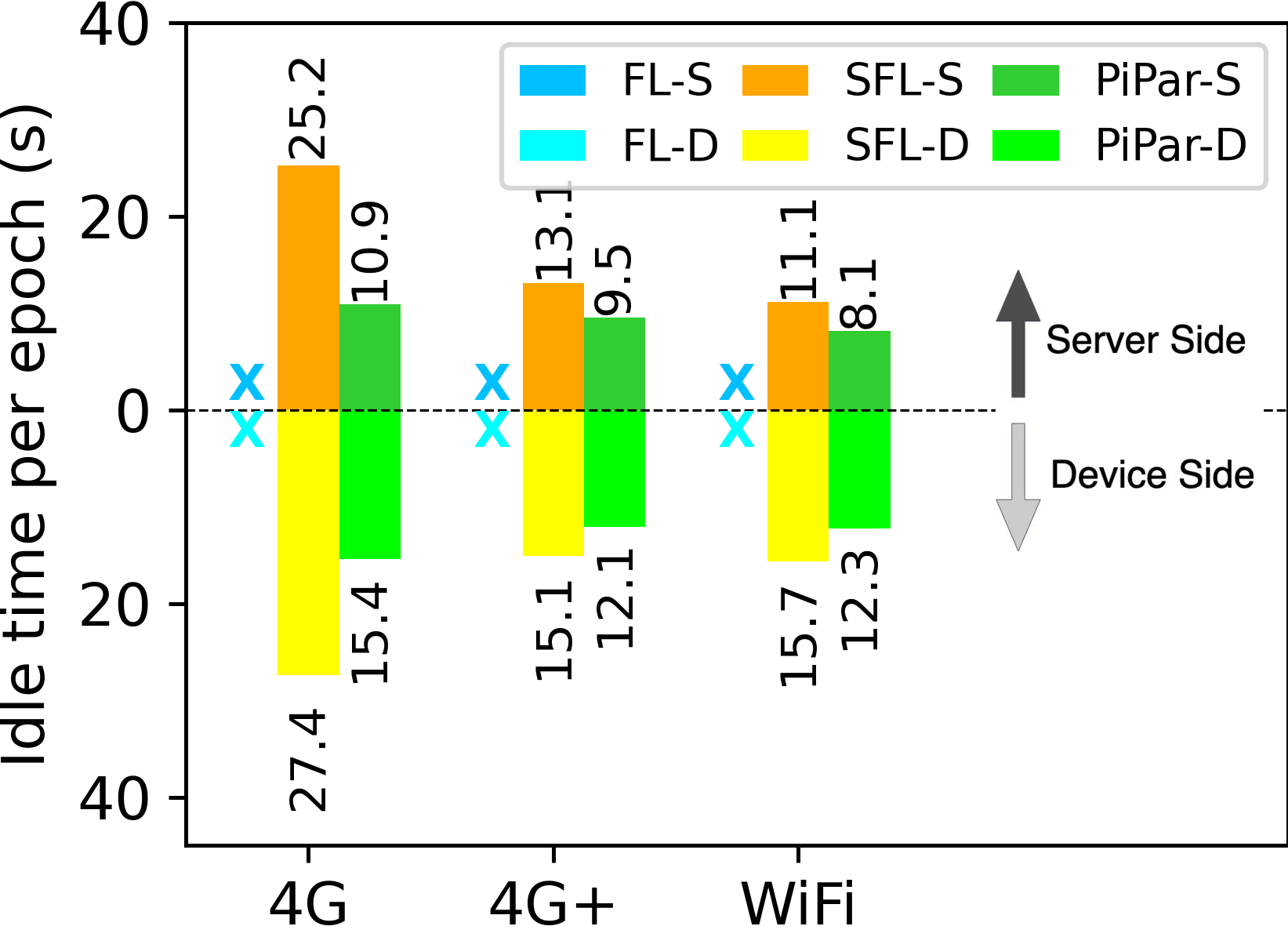}
	    \label{fig:resnet101-cifar100-idle}
	    }
        \hfill
	\subfigure[MobileNetV3-Large (CIFAR-100)]{
	    \includegraphics[width=0.3\textwidth]{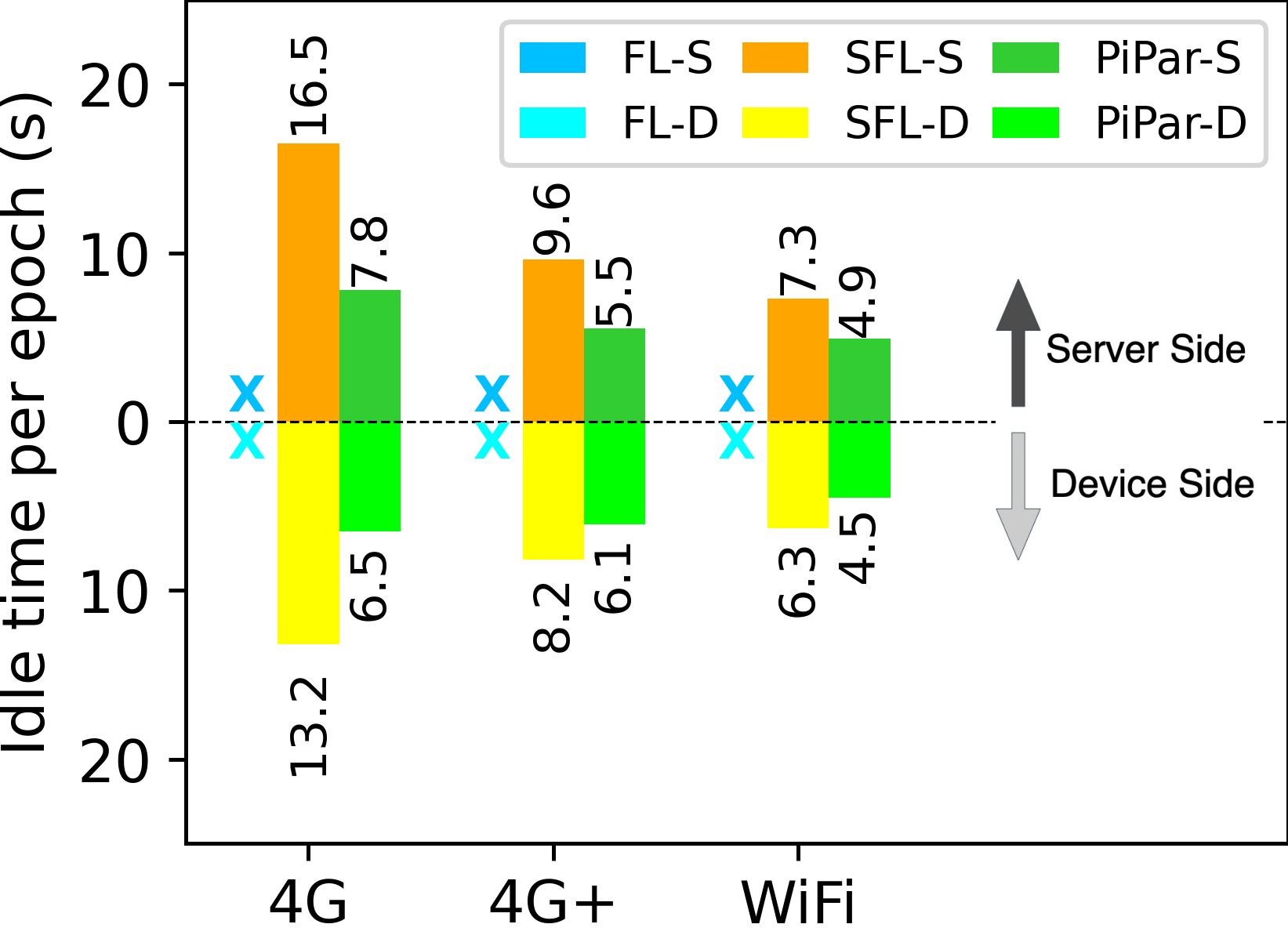}
	    \label{fig:mobilelarge-cifar100-idle}
	    }
        \subfigure[VGG-16 (Tiny ImageNet)]{
	    \includegraphics[width=0.3\textwidth]{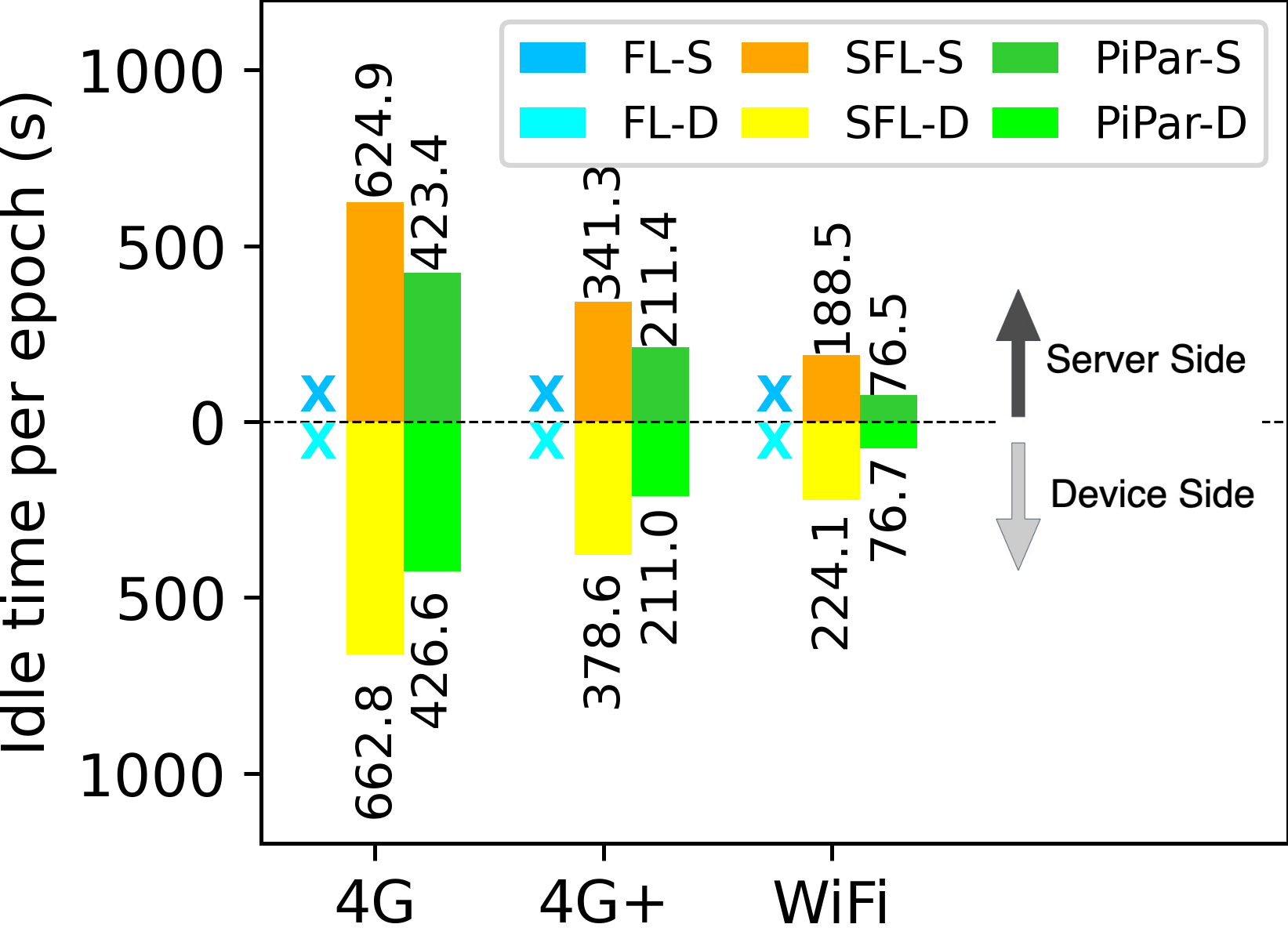}
	    \label{fig:vgg16-tinyimagenet-idle}
	    }
	\hfill
	\subfigure[ResNet-101 (Tiny ImageNet)]{
	    \includegraphics[width=0.3\textwidth]{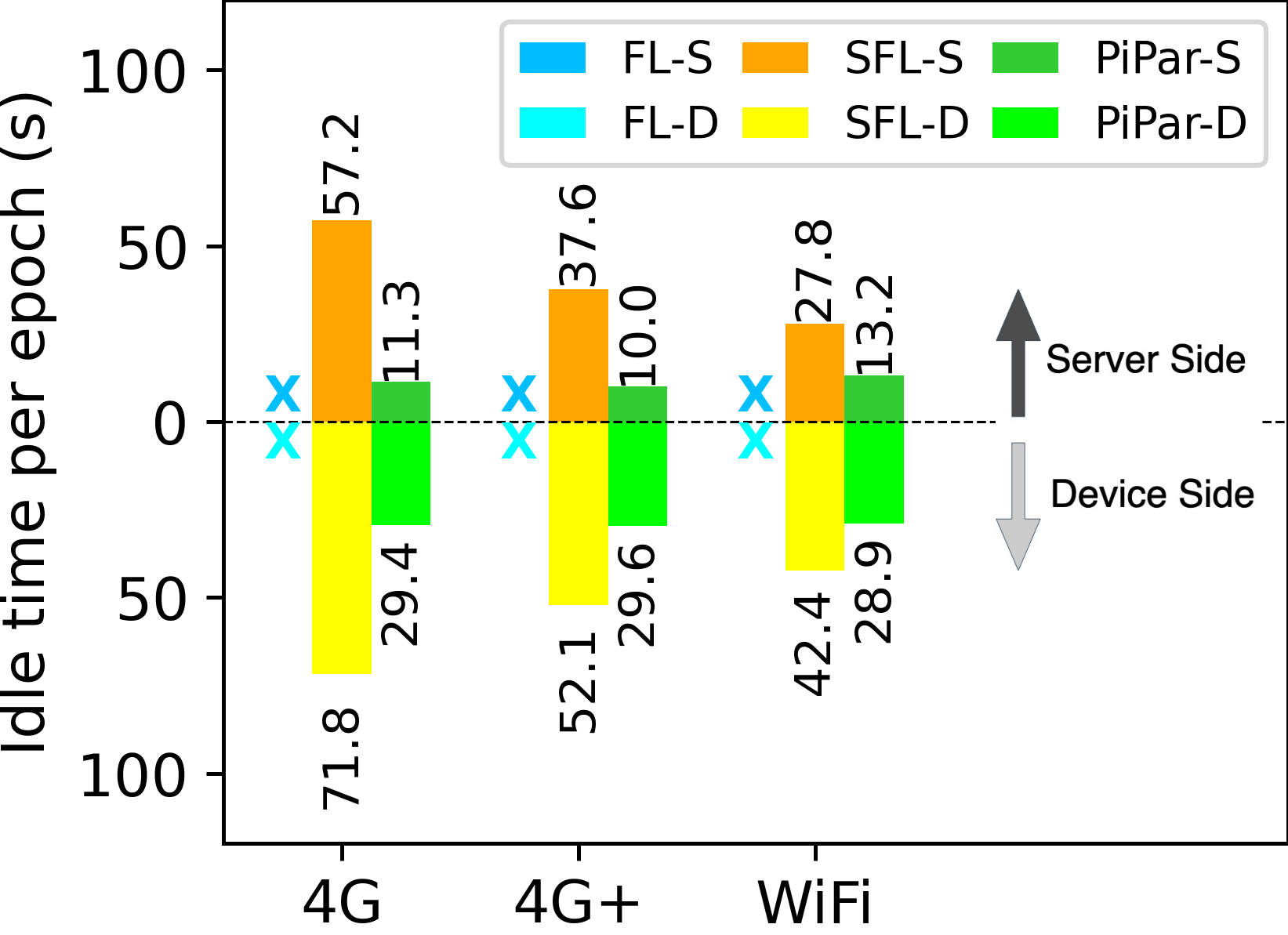}
	    \label{fig:resnet101-tinyimagenet-idle}
	    }
        \hfill
	\subfigure[MobileNetV3-Large (Tiny ImageNet)]{
	    \includegraphics[width=0.3\textwidth]{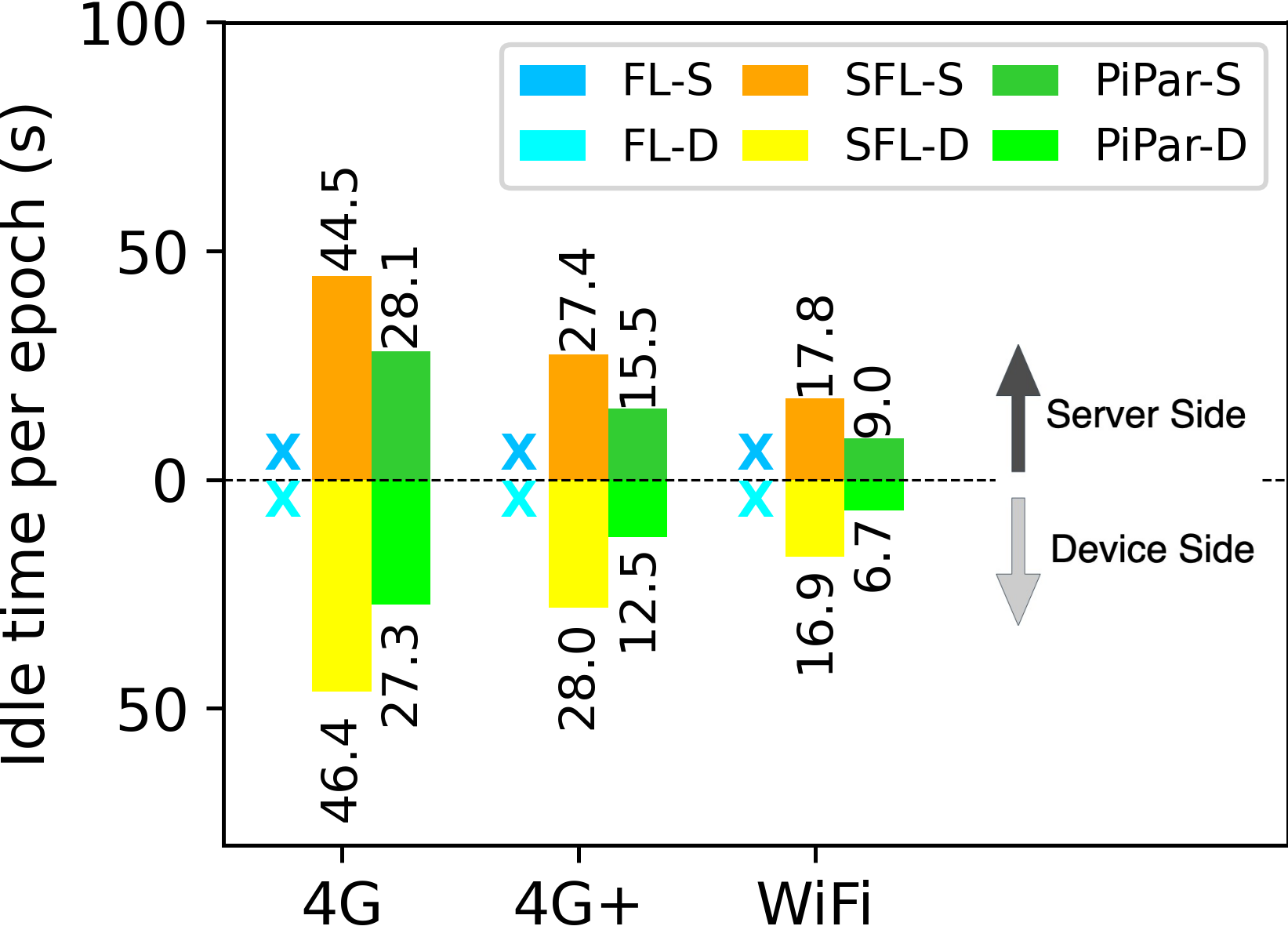}
	    \label{fig:mobilelarge-tinyimagenet-idle}
	    }
	\caption{Idle time per epoch on the server and devices in SFL and \PiPar\ under different network conditions for large DNNs. `S' and `D' in the legend represent server-side and device-side idle time, respectively. They are shown in the upward and downward bars. FL results are not shown as the entire DNN does not fit on the device memory.}
	\label{fig:idle-time-large}
\end{figure*}

\begin{figure*}[tp]
	\centering
	\subfigure[VGG-5 (MNIST)]{
	    \includegraphics[width=0.3\textwidth]{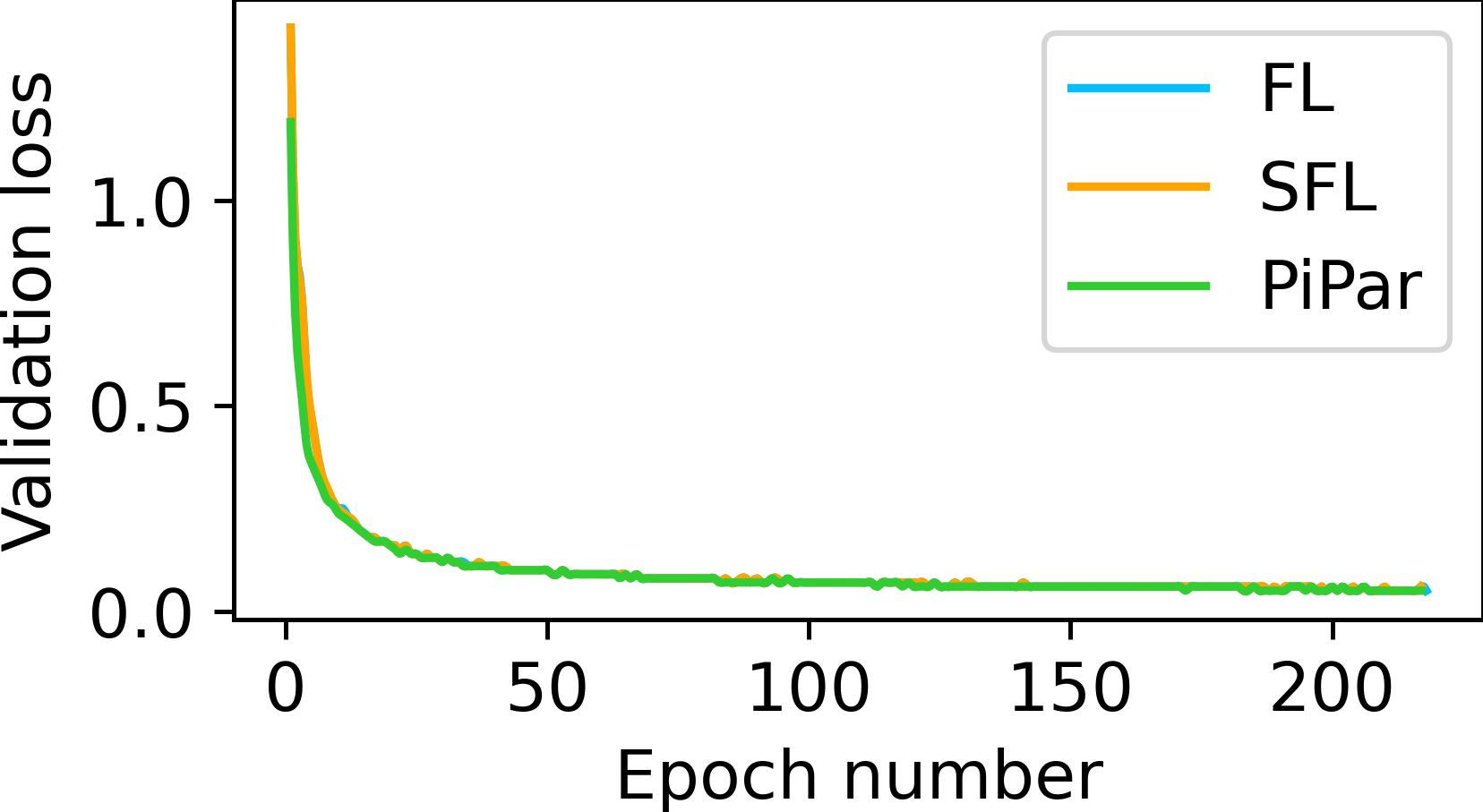}
	    \label{fig:vgg5-mnist-loss}
	    }
	\hfill
        \subfigure[ResNet-18 (MNIST)]{
	    \includegraphics[width=0.3\textwidth]{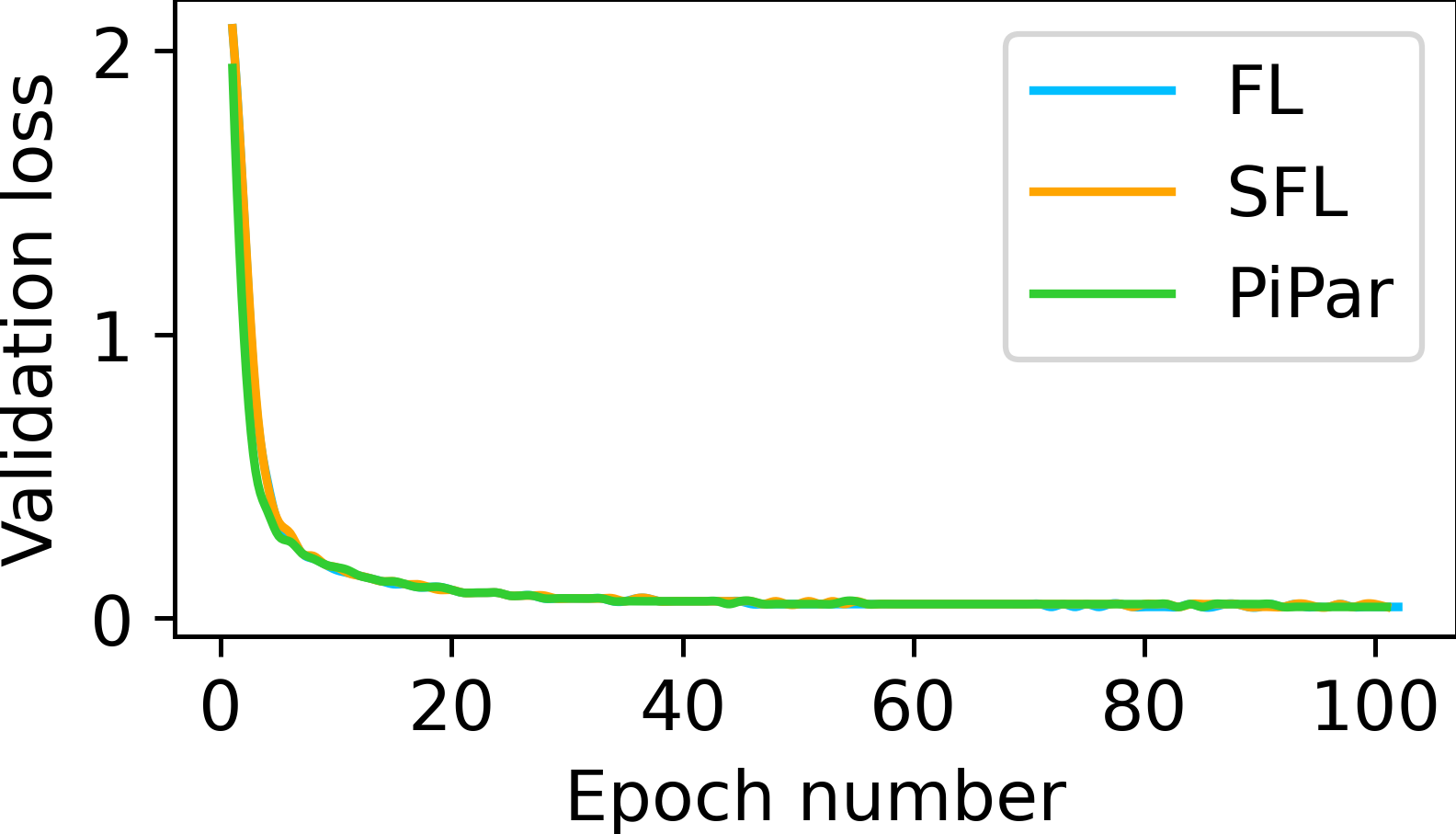}
	    \label{fig:resnet18-mnist-loss}
	    }
	\hfill
        \subfigure[MobileNetV3-Small (MNIST)]{
	    \includegraphics[width=0.3\textwidth]{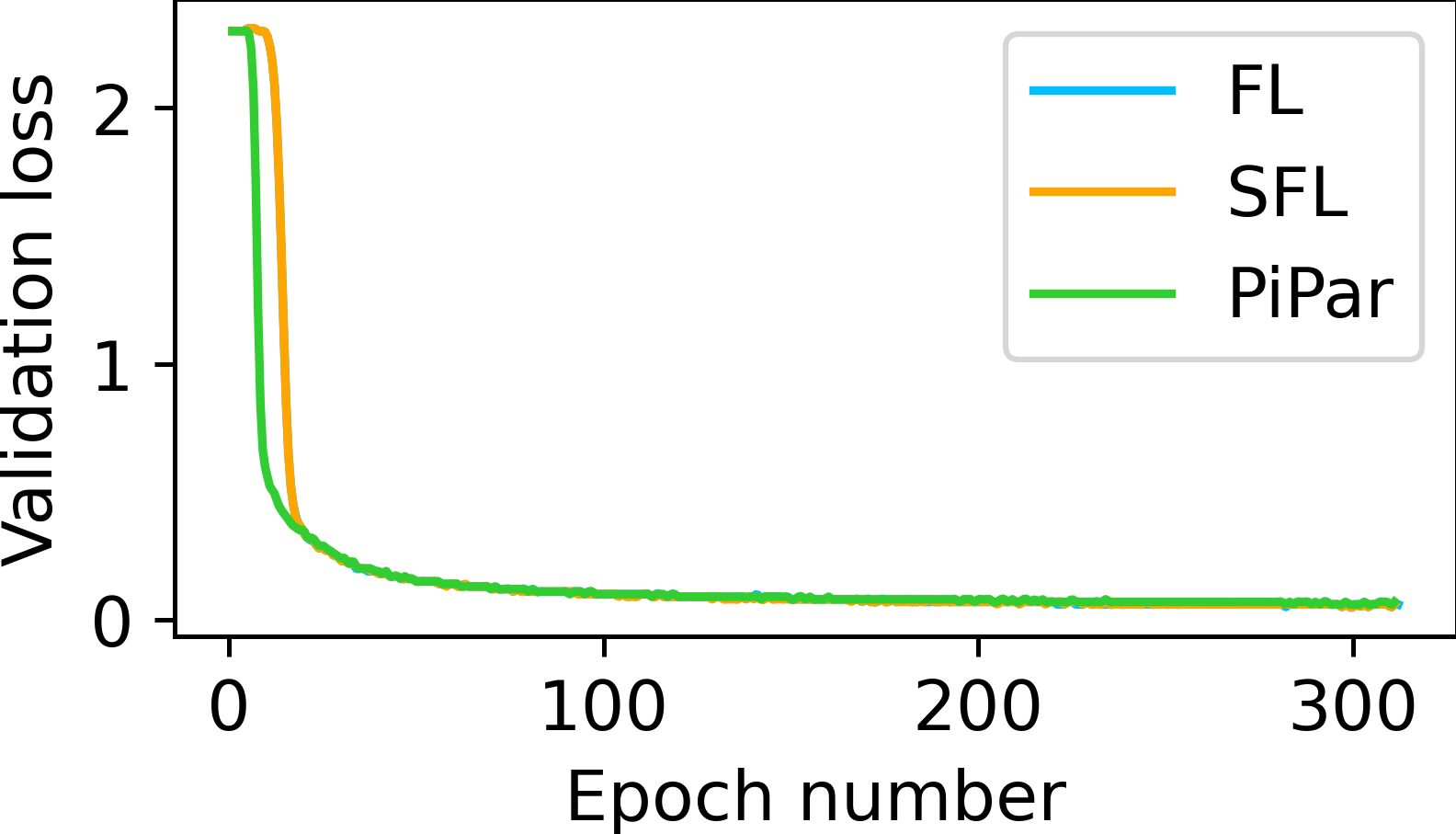}
	    \label{fig:mobilesmall-mnist-loss}
	    }
	\hfill
	\subfigure[VGG-5 (CIFAR-10)]{
	    \includegraphics[width=0.3\textwidth]{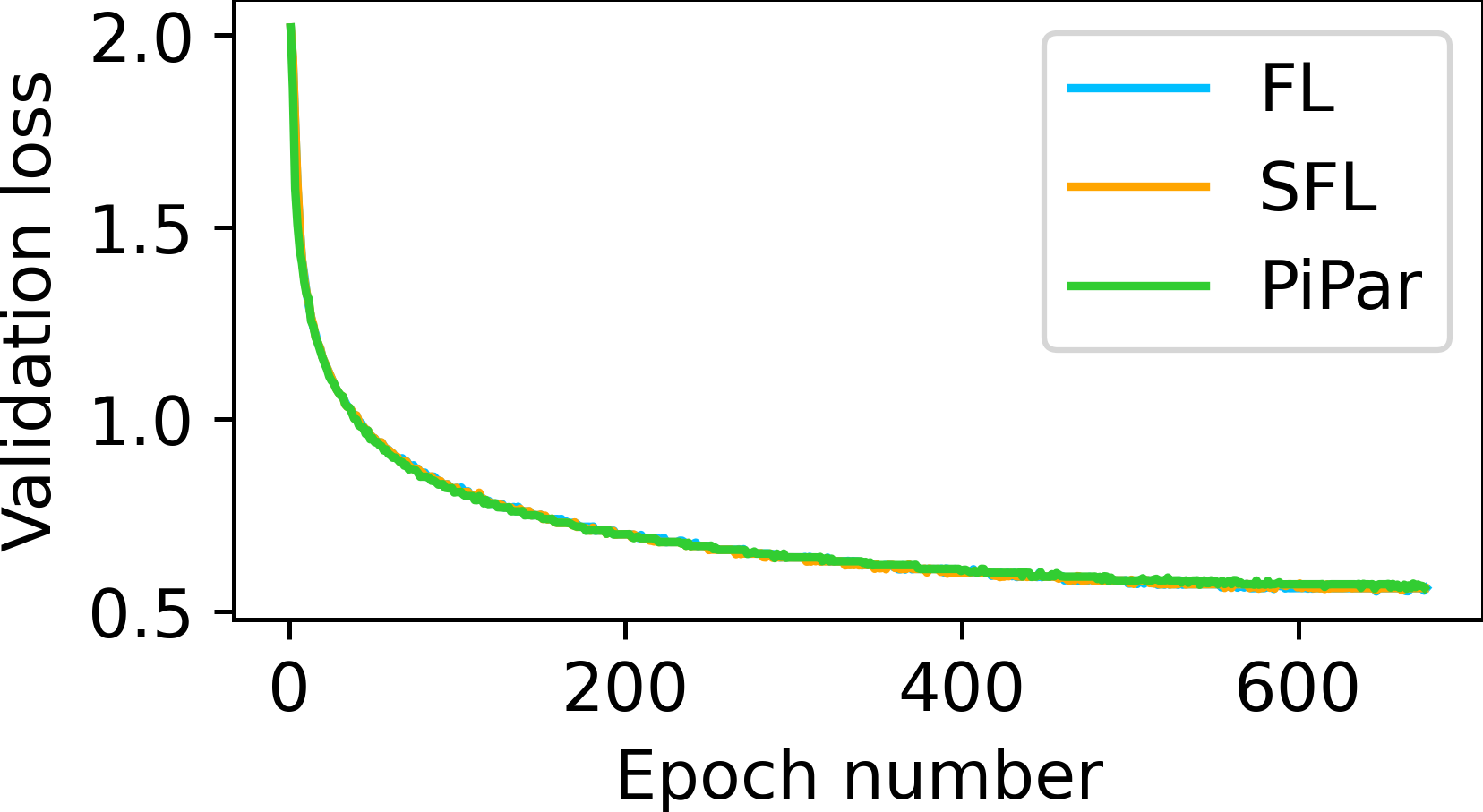}
	    \label{fig:vgg5-cifar10-loss}
	    }
	\hfill
        \subfigure[ResNet-18 (CIFAR-10)]{
	    \includegraphics[width=0.3\textwidth]{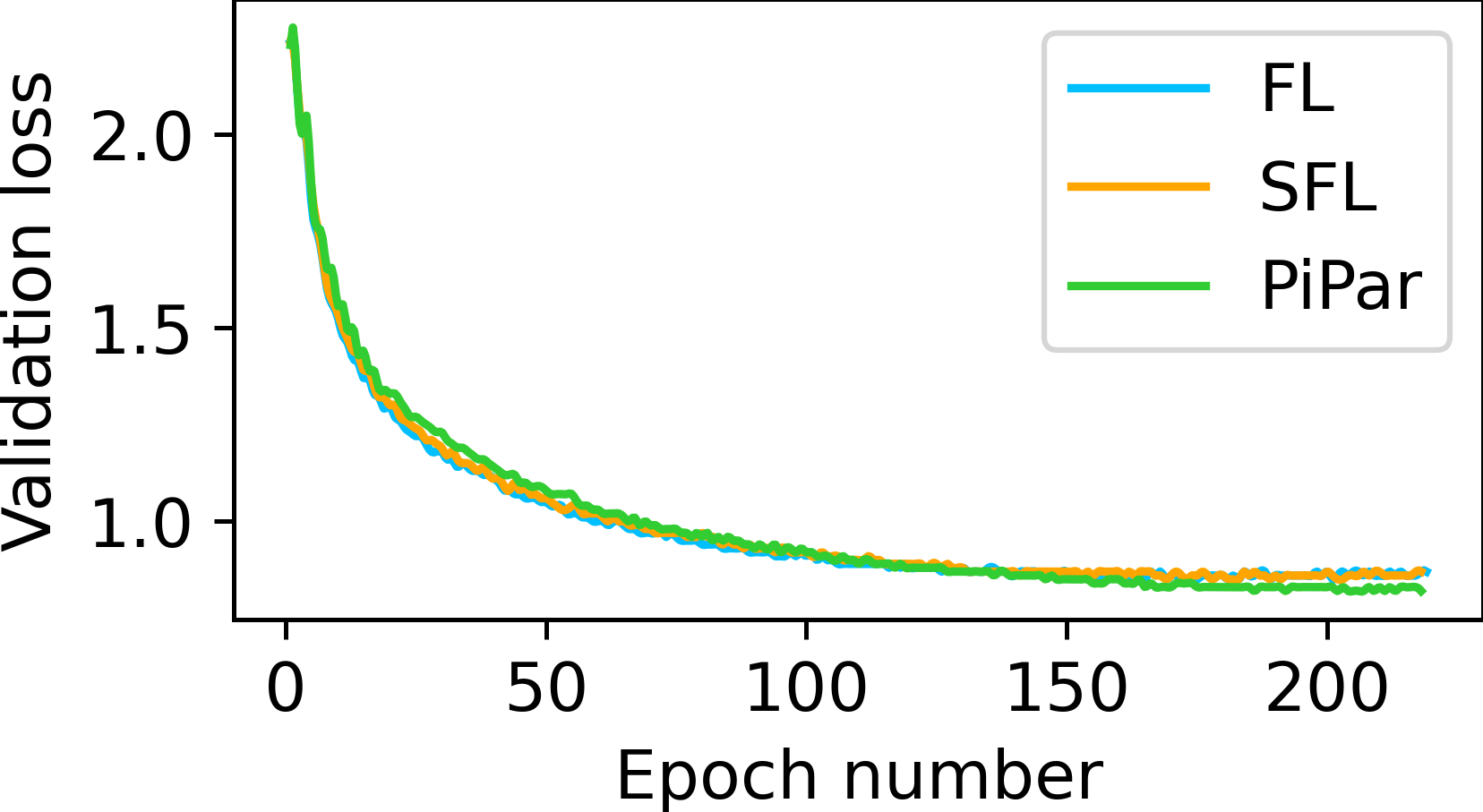}
	    \label{fig:resnet18-cifar10-loss}
	    }
	\hfill
        \subfigure[MobileNetV3-Small (CIFAR-10)]{
	    \includegraphics[width=0.3\textwidth]{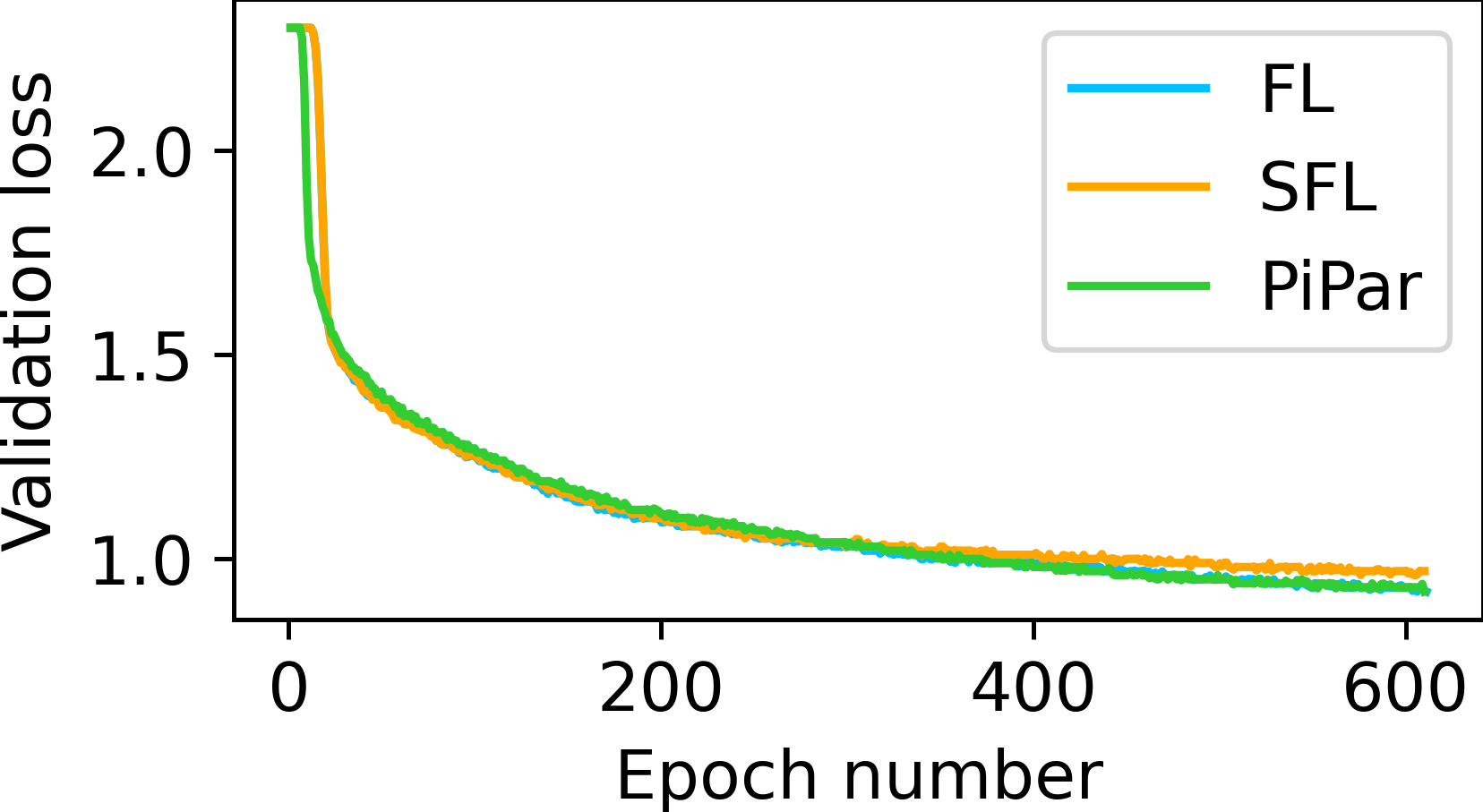}
	    \label{fig:mobilesmall-cifar10-loss}
	    }
	\hfill
	\caption{Validation loss for FL, SFL and \PiPar\ using small DNNs.}
	\label{fig:loss-small}
\end{figure*}

\begin{figure*}[tp]
	\centering
	\subfigure[VGG-16 (CIFAR-100)]{
	    \includegraphics[width=0.3\textwidth]{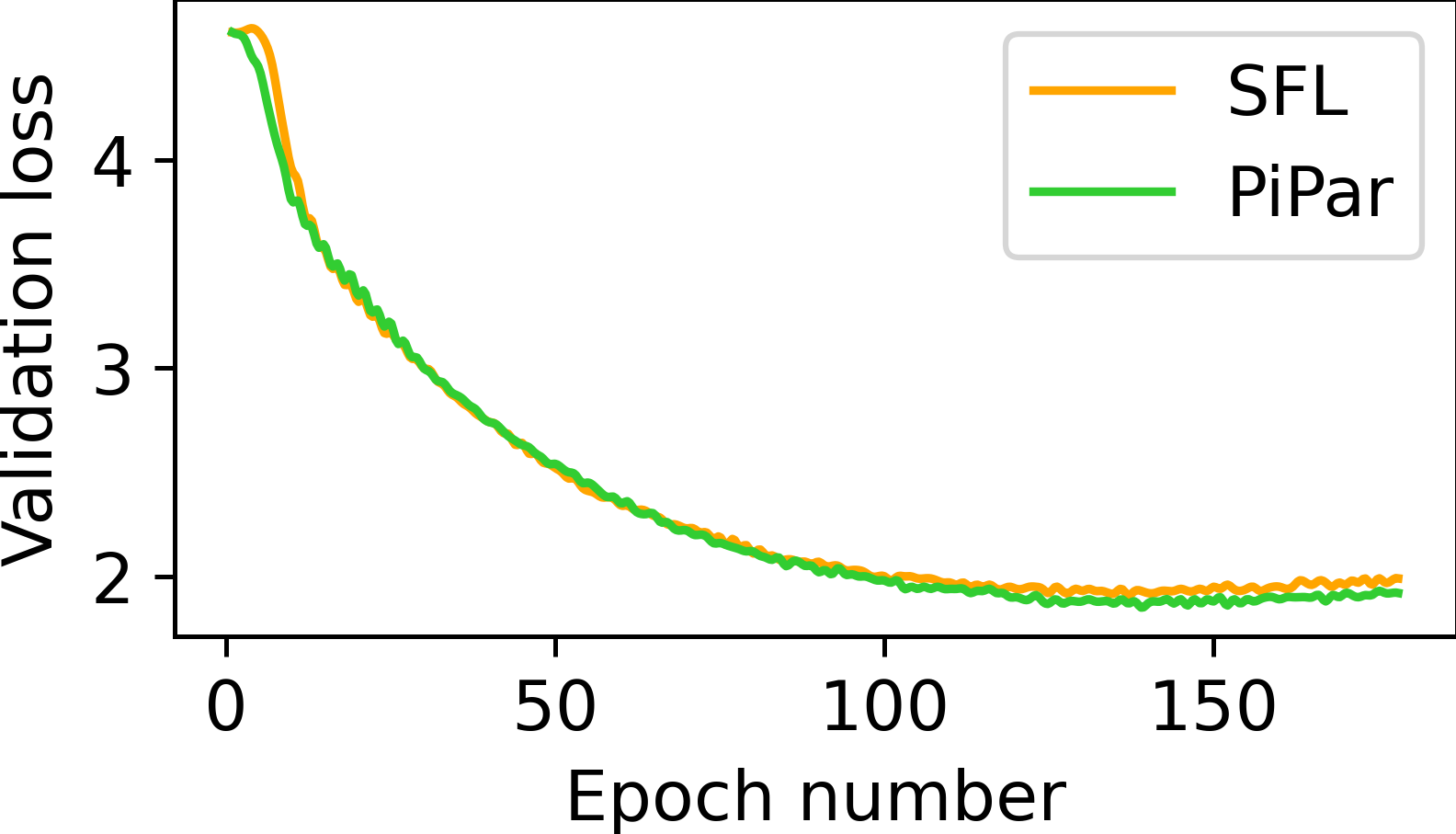}
	    \label{fig:vgg16-cifar100-loss}
	    }
	\hfill
        \subfigure[ResNet-101 (CIFAR-100)]{
	    \includegraphics[width=0.3\textwidth]{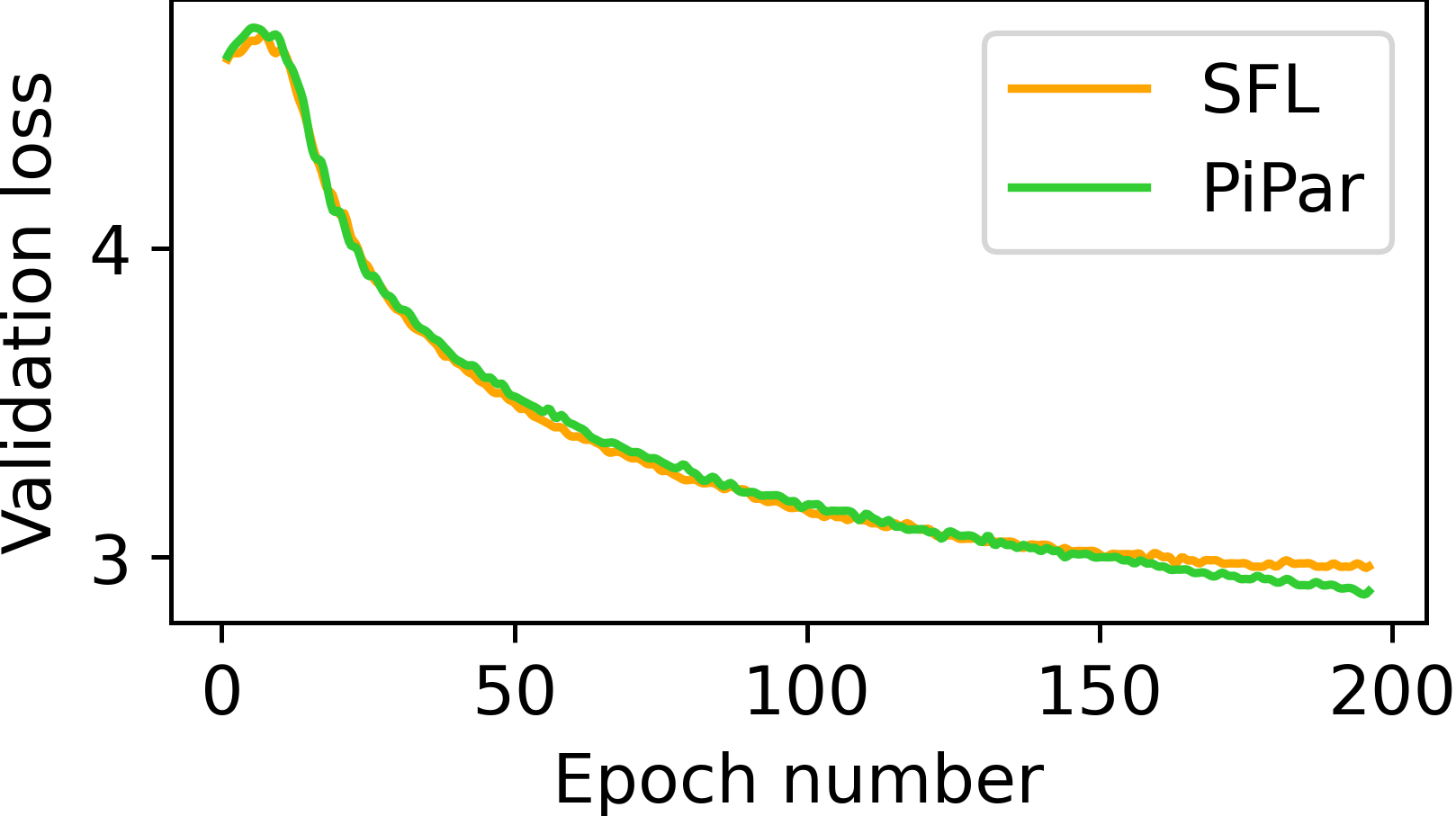}
	    \label{fig:resnet101-cifar100-loss}
	    }
	\hfill
        \subfigure[MobileNetV3-Large (CIFAR-100)]{
	    \includegraphics[width=0.3\textwidth]{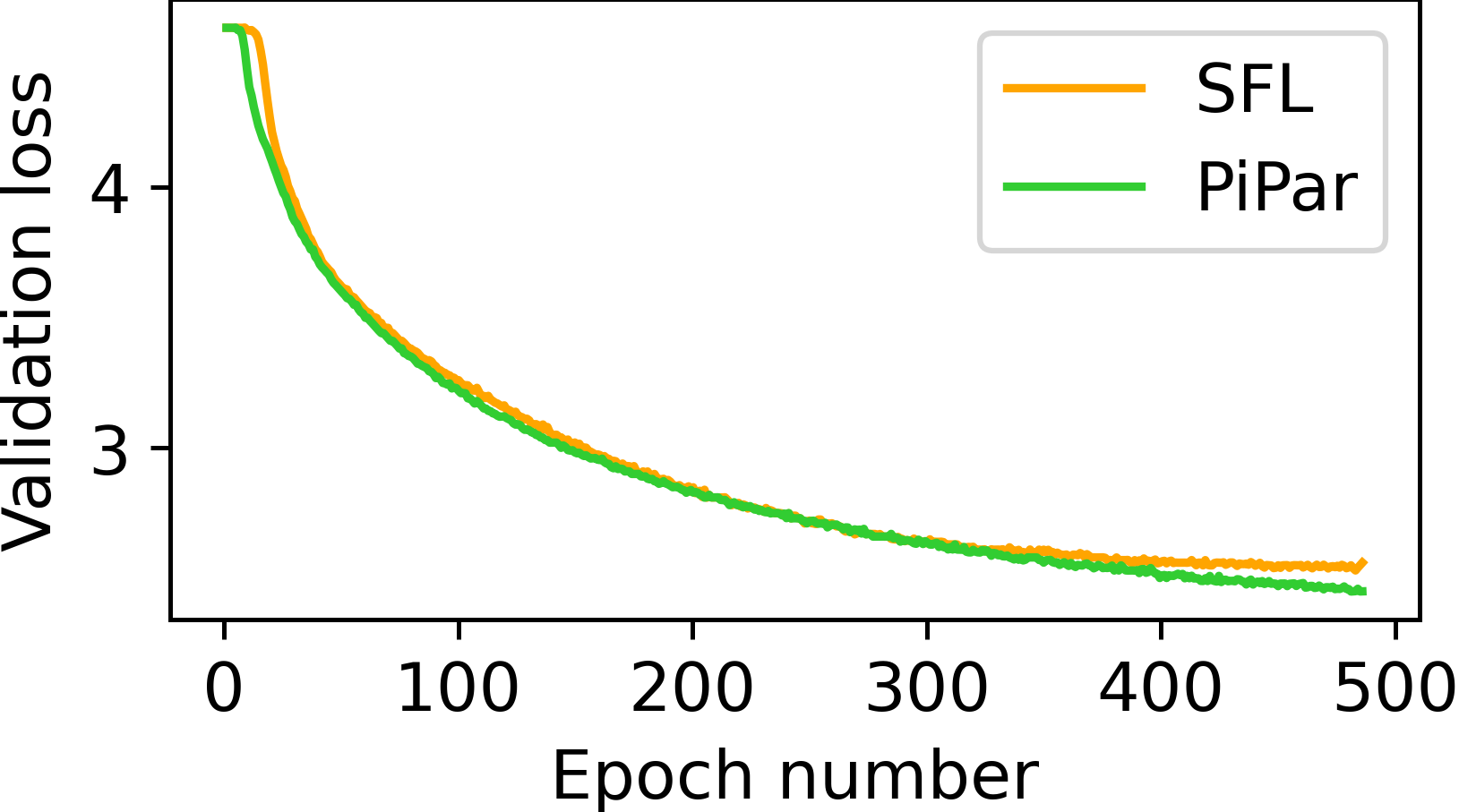}
	    \label{fig:mobilelarge-cifar100-loss}
	    }
	\hfill
	\subfigure[VGG-16 (Tiny ImageNet)]{
	    \includegraphics[width=0.3\textwidth]{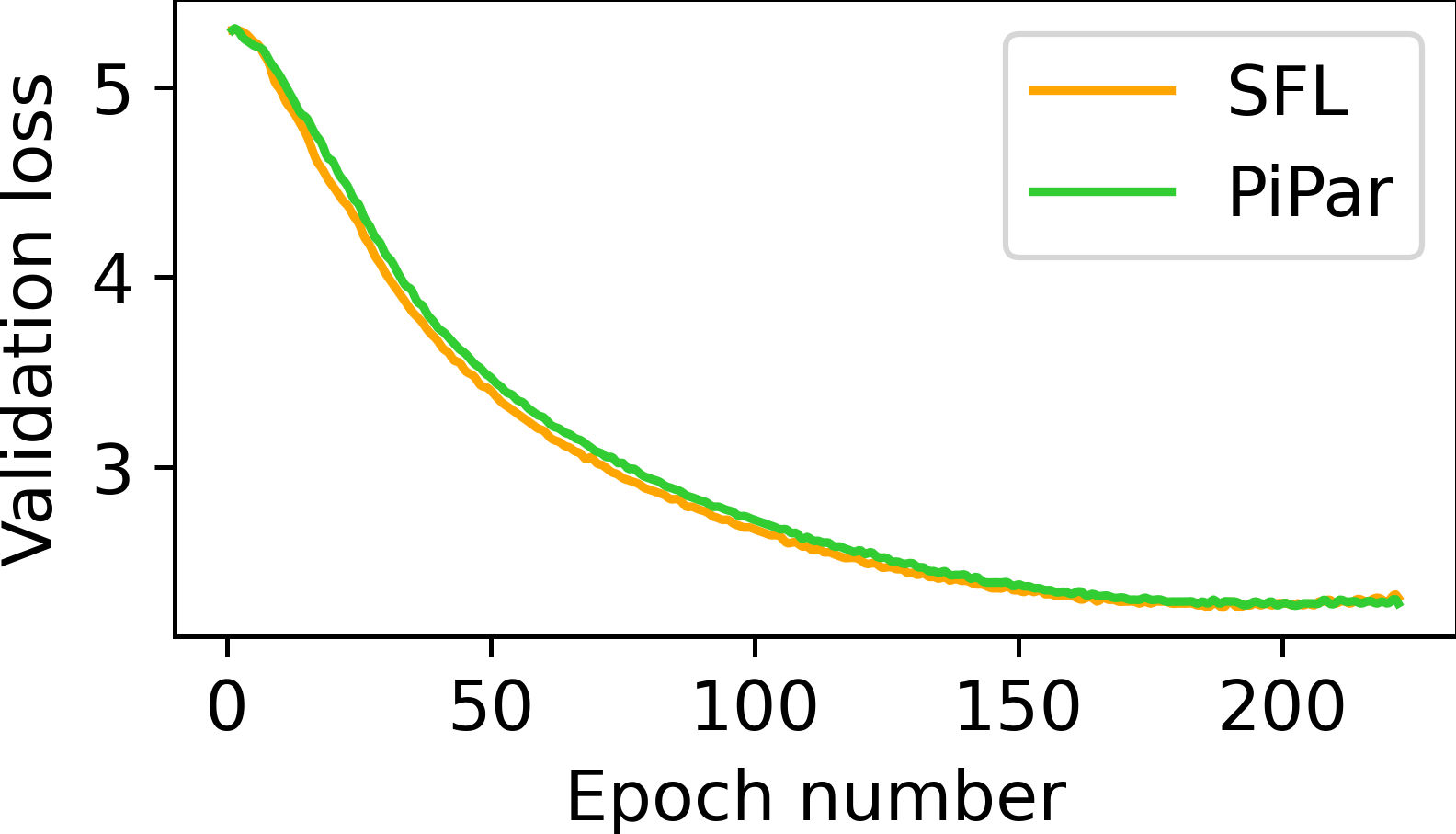}
	    \label{fig:vgg16-tinyimagenet-loss}
	    }
	\hfill
        \subfigure[ResNet-101 (Tiny ImageNet)]{
	    \includegraphics[width=0.3\textwidth]{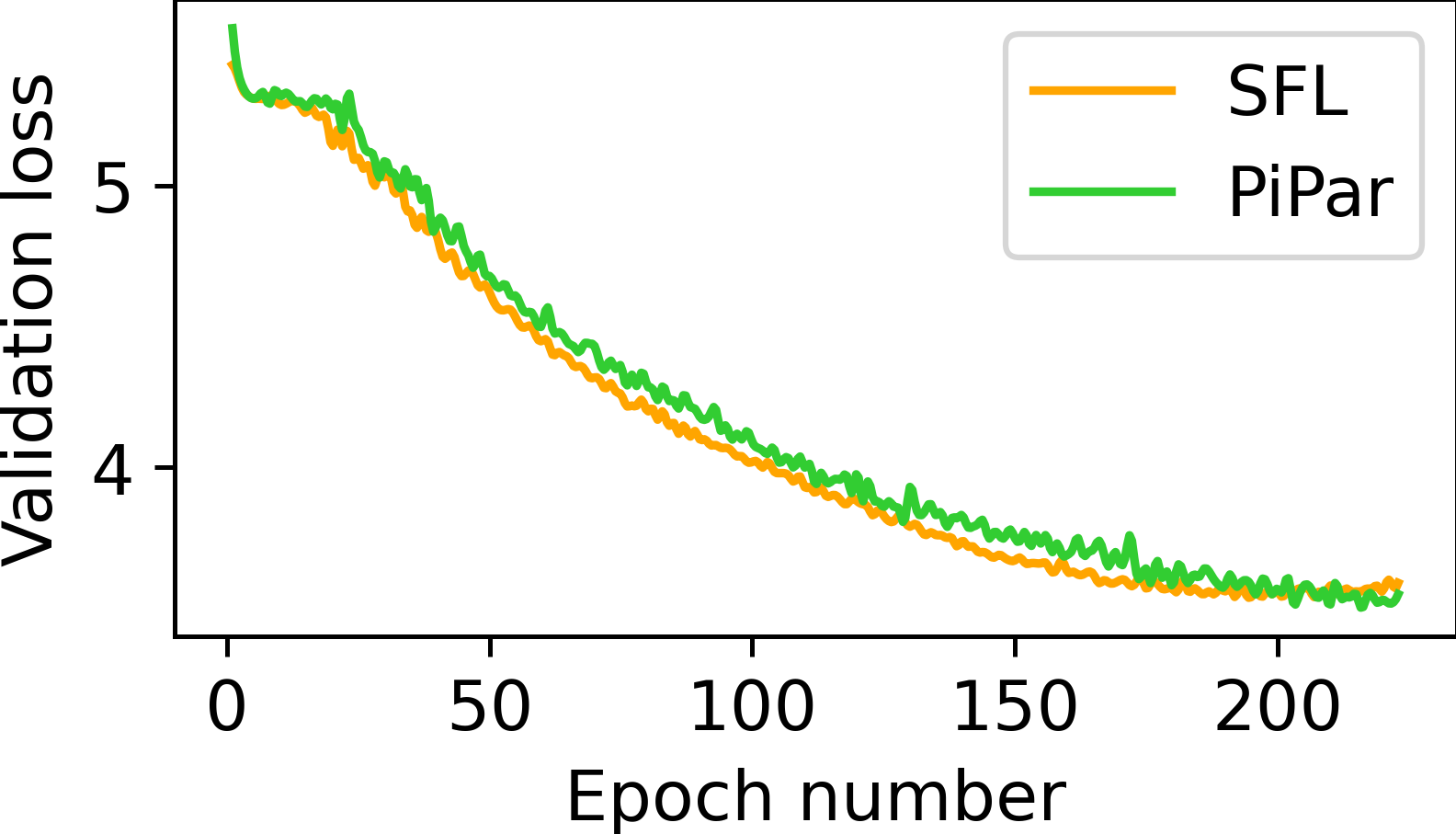}
	    \label{fig:resnet101-tinyimagenet-loss}
	    }
	\hfill
        \subfigure[MobileNetV3-Large (Tiny ImageNet)]{
	    \includegraphics[width=0.3\textwidth]{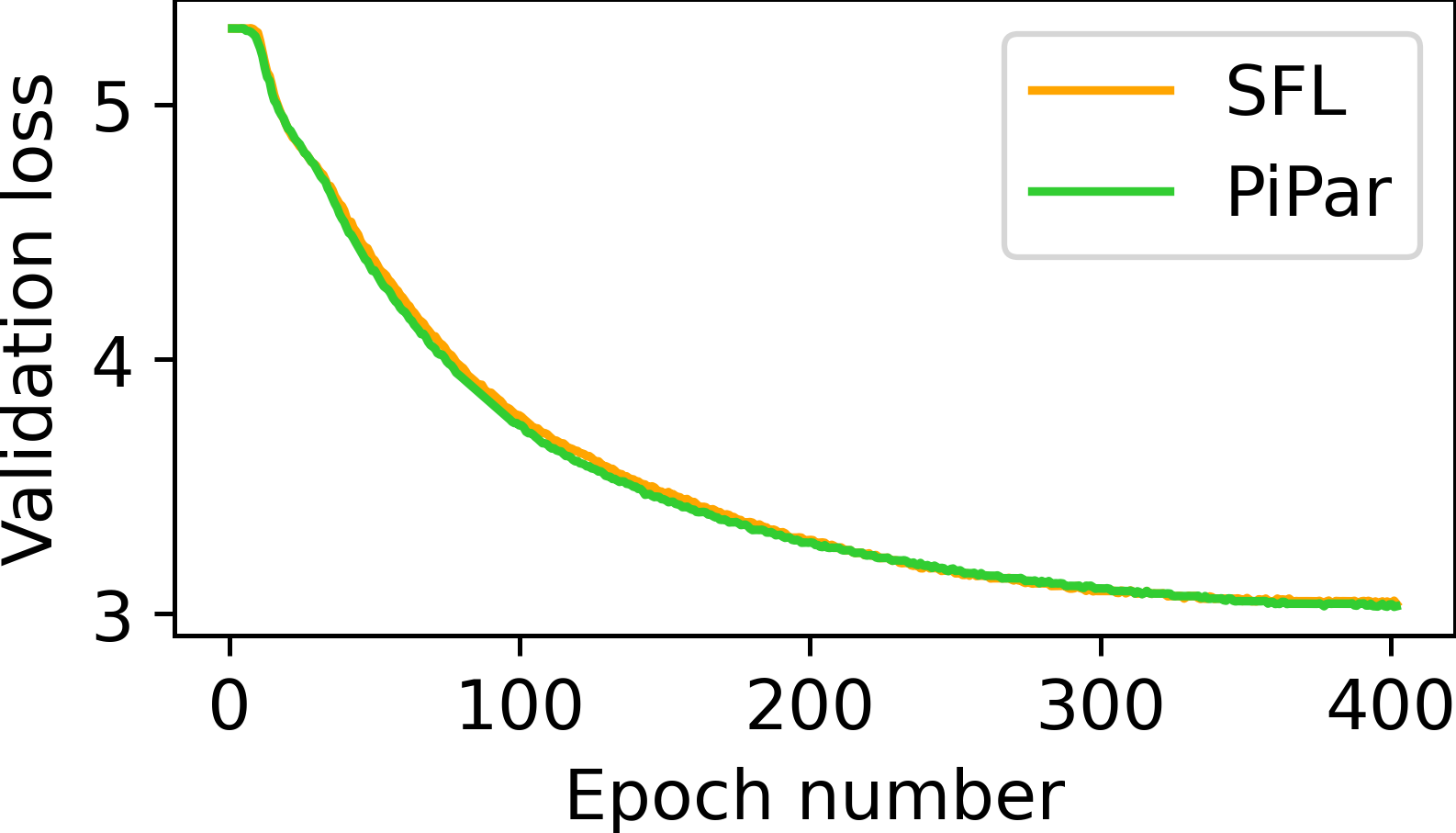}
	    \label{fig:mobilelarge-tinyimagenet-loss}
	    }
	\hfill
	\caption{Validation loss for SFL and \PiPar\ using large DNNs. FL results are not shown as the entire DNN does not fit on the device memory.}
	\label{fig:loss-large}
\end{figure*}

\begin{figure*}[tp]
	\centering
	\subfigure[4G]{
	    \includegraphics[width=0.3\textwidth]{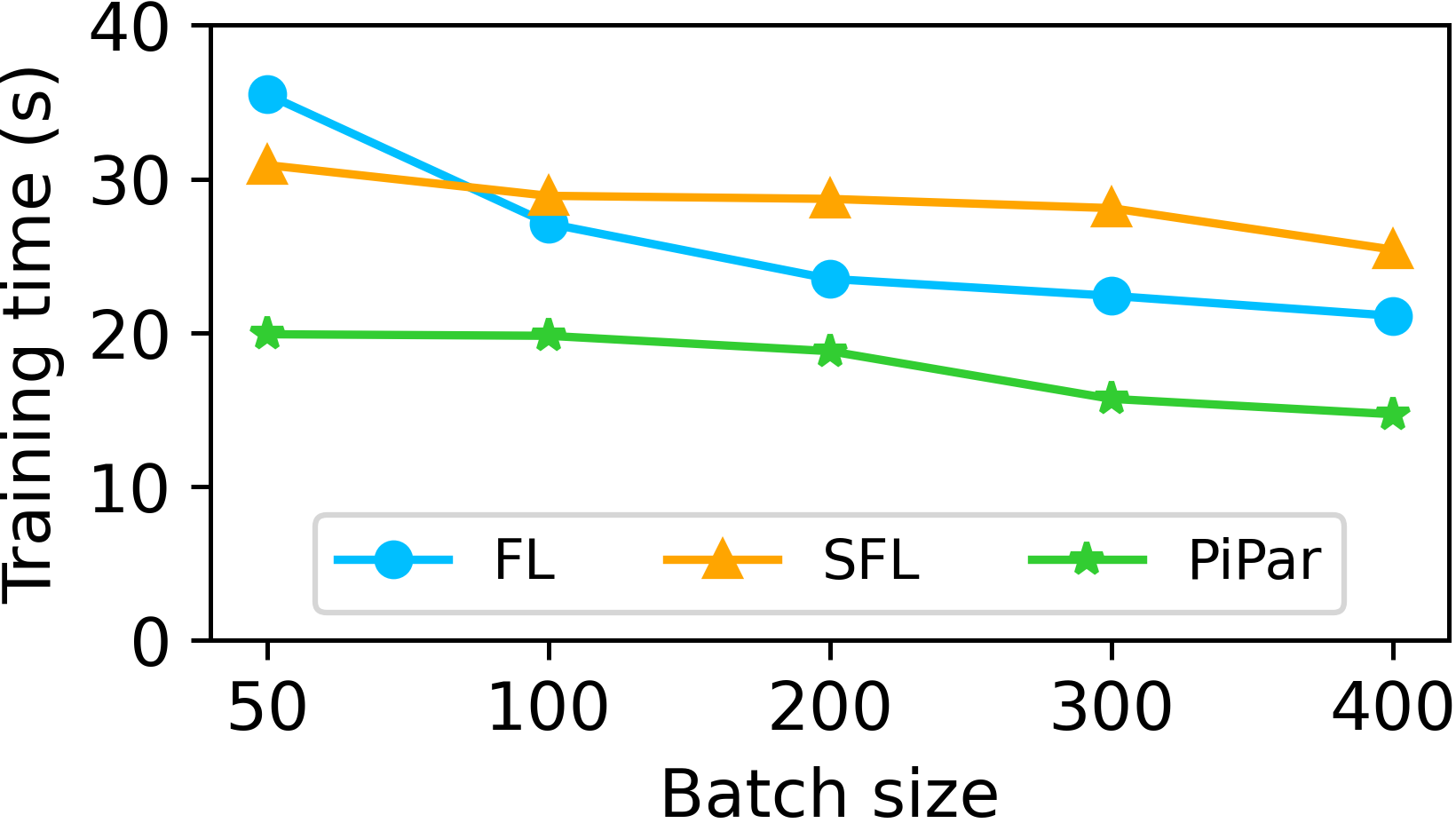}
	    \label{fig:bs-vgg5-4g}
	    }
	\hfill
	\subfigure[4G+]{
	    \includegraphics[width=0.3\textwidth]{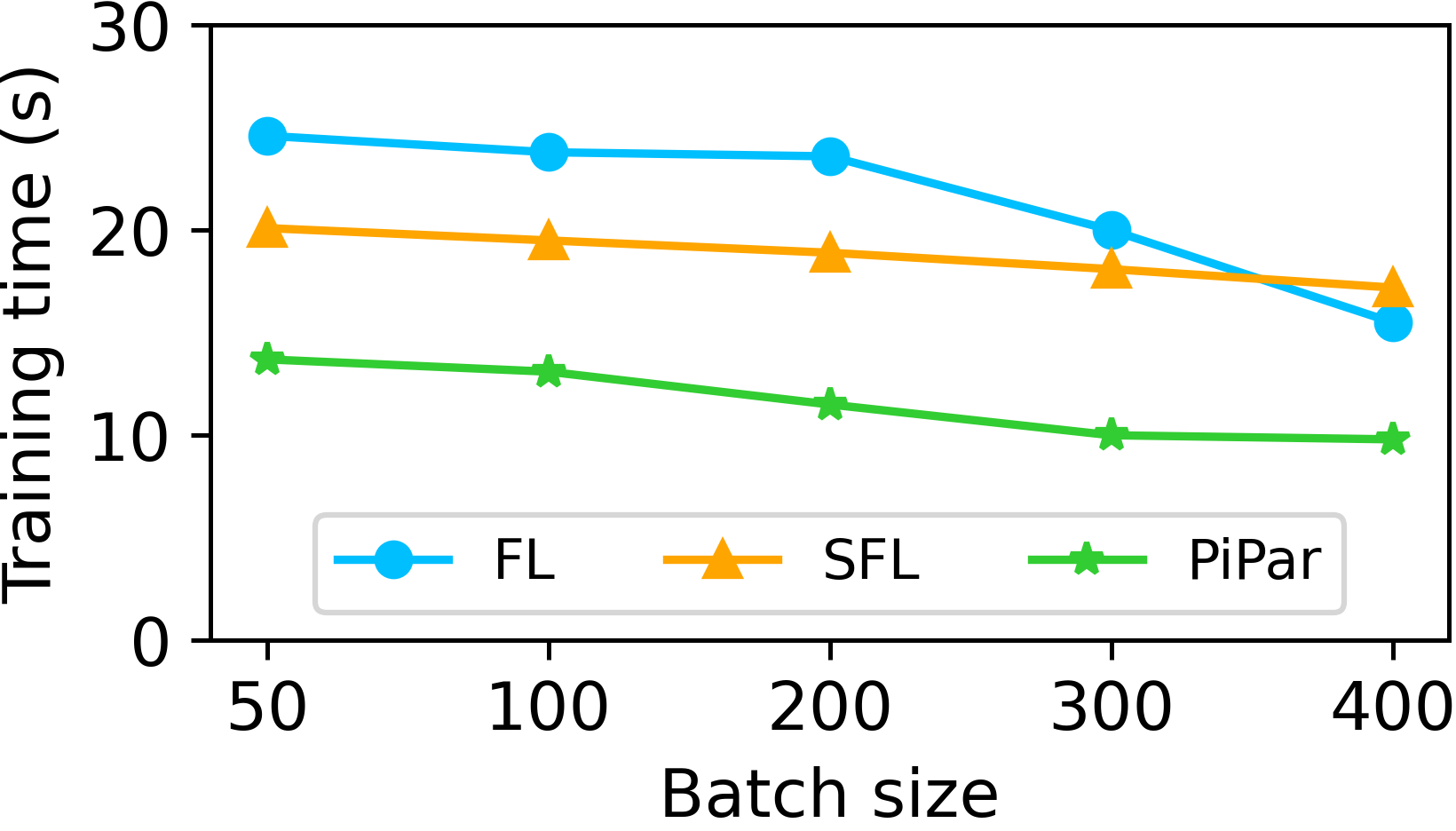}
	    \label{fig:bs-vgg5-4gp}
	    }
        \hfill
	\subfigure[WiFi]{
	    \includegraphics[width=0.3\textwidth]{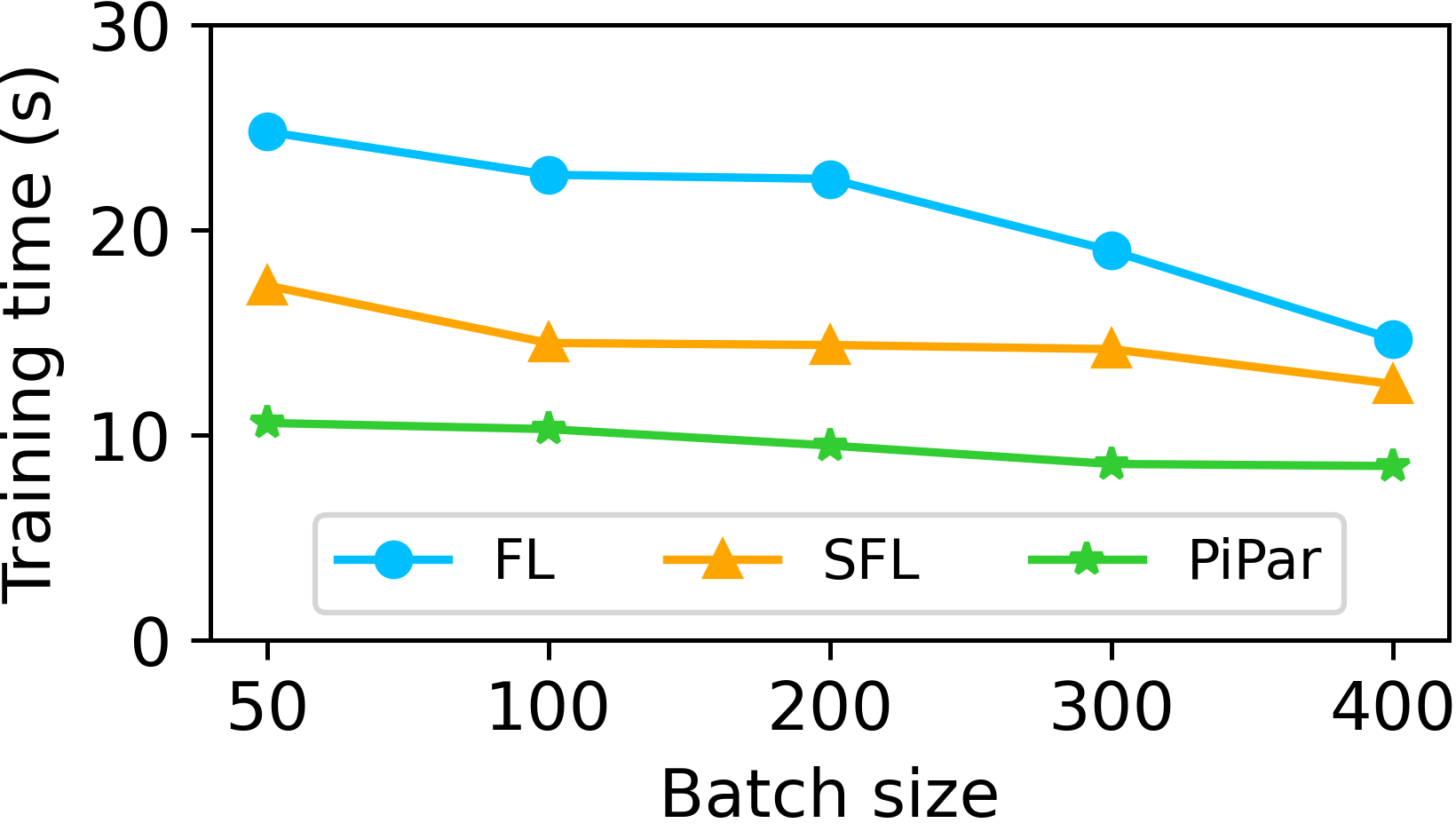}
	    \label{fig:bs-vgg5-wifi}
	    }
	\centering
	\caption{Training time per epoch for FL, SFL and \PiPar\ using VGG-5 and the CIFAR-10 dataset with different batch sizes $B$.}
	\label{fig:batch-size}
\end{figure*}

\begin{figure*}[tp]
	\centering
	\subfigure[4G]{
	    \includegraphics[width=0.3\textwidth]{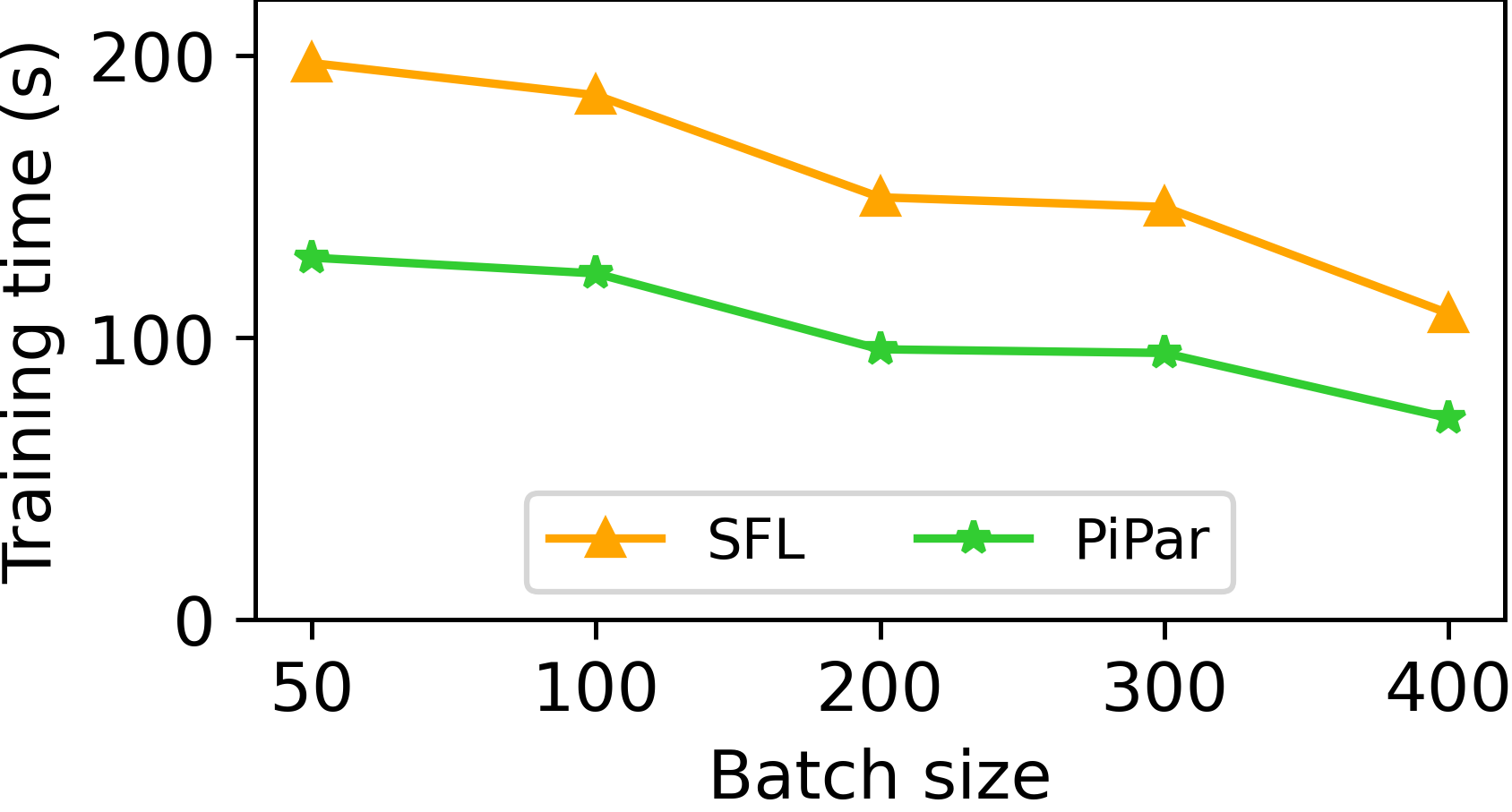}
	    \label{fig:bs-vgg16-4g}
	    }
	\hfill
	\subfigure[4G+]{
	    \includegraphics[width=0.3\textwidth]{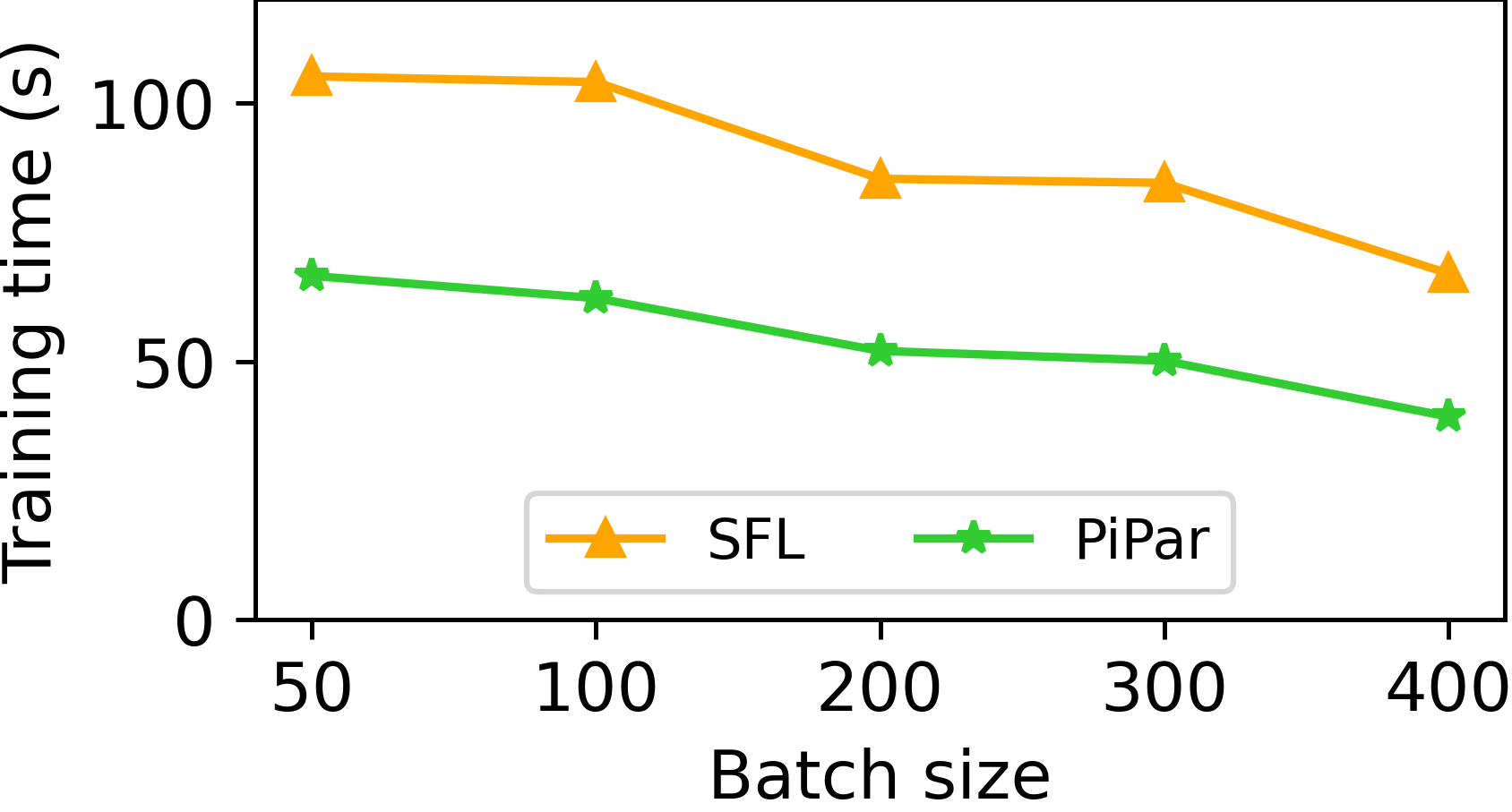}
	    \label{fig:bs-vgg16-4gp}
	    }
        \hfill
	\subfigure[WiFi]{
	    \includegraphics[width=0.3\textwidth]{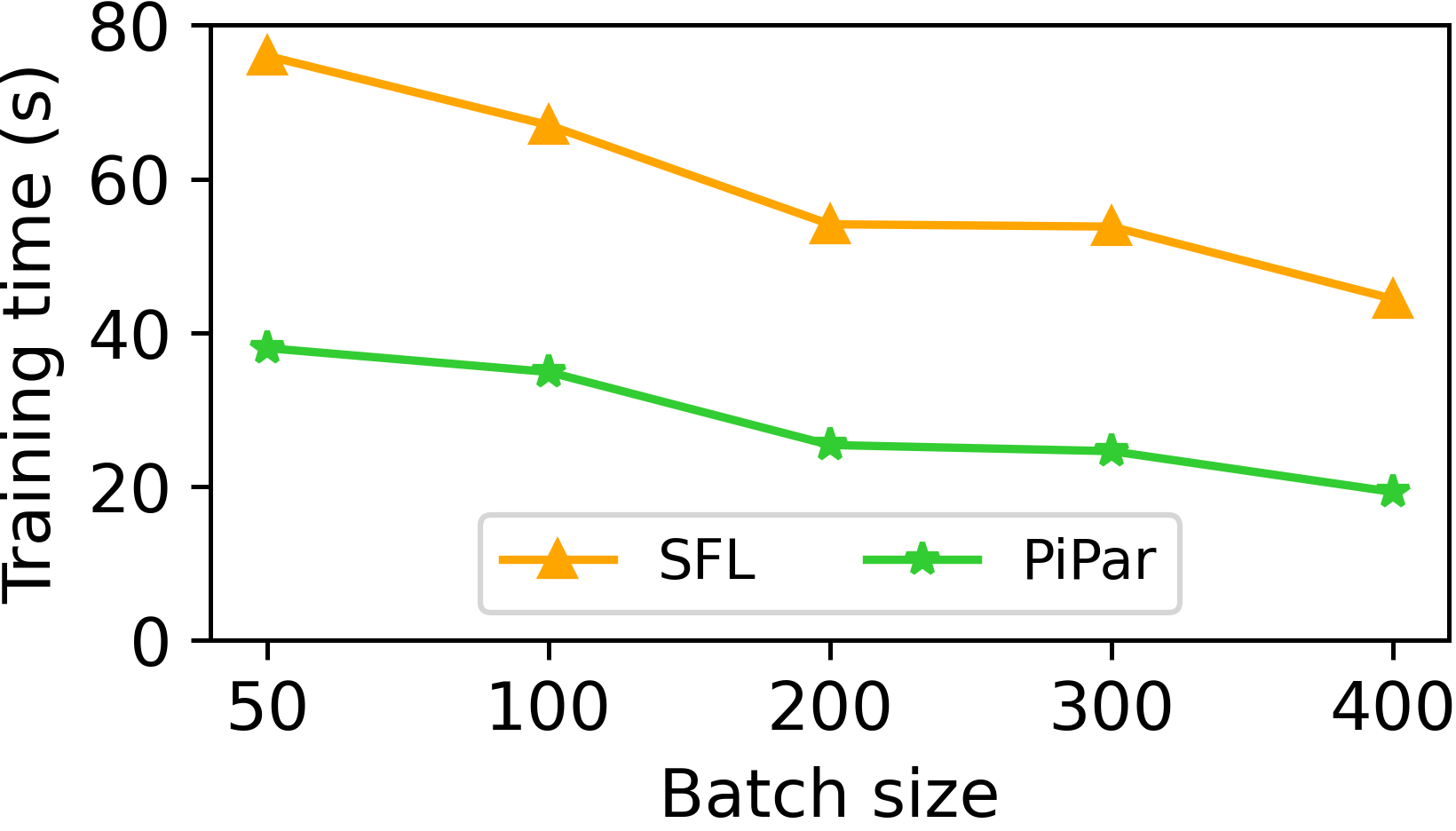}
	    \label{fig:bs-vgg16-wifi}
	    }
	\centering
	\caption{Training time per epoch for SFL and \PiPar\ using VGG-16 and the CIFAR-100 dataset with different batch sizes $B$. FL results are not shown as the entire DNN does not fit on the device memory.}
	\label{fig:batch-size-large}
\end{figure*}

\subsection{Efficiency results}
\label{subsec:eff-exp}

The experiments in this section compare the efficiency of \PiPar\ with FL and SFL. Although SL is a popular CML technique, it is significantly slower than SFL since each device operates sequentially. Hence, SL is not considered in these experiments. All possible split points for SFL are benchmarked (based on the benchmarking method adopted in Scission~\cite{DBLP:journals/corr/abs-2008-03523}), and the efficiency of SFL with the best split point is reported. The split point and parallel batch number for \PiPar\ are selected by the approach proposed in Section~\ref{subsec:optimise}.

\subsubsection{Comparing efficiency}
\label{subsubsec:exp-eff}

The efficiency of the CML techniques is measured by \textit{training time per epoch}. Section~\ref{subsubsec:convergence} will highlight that the loss curves of FL, SFL and \PiPar\ overlap, so the same number of epochs are required for model convergence using the three techniques. Hence, if \PiPar\ reduces the training time per epoch, it reduces the overall training time.

Figure~\ref{fig:training-time} shows the training time per epoch of six DNNs (three small and three large DNNs) for FL, SFL and \PiPar\ under 4G, 4G+ and WiFi network conditions. The three large DNNs, namely VGG-16, ResNet-101 and MobileNetV3-Large cannot be trained using FL as the entire model needs to be trained on the device but the device memory is limited and does not support the size of these models. It is immediately evident that the training time per epoch for \PiPar\ is lower than FL and SFL in all cases. 

When training VGG-5 models on MNIST (Figure~\ref{fig:vgg5-mnist-time}) and CIFAR-10 (Figure~\ref{fig:vgg5-cifar10-time}), the difference of FL training time under three network conditions (5.6\%) is smaller than that of SFL (59\%) and \PiPar\ (57\%), because the devices only upload and download once at the end of the training epoch (it requires less communication compared to SFL and \PiPar). However, FL trains the entire model on each device, which requires longer computational time. When the bandwidth is low (for example, 4G), FL outperforms SFL, because the latter requires more communication time. However, under 4G+ and WiFi, SFL has shorter training times because of fewer device-side computations. Under all network conditions, \PiPar\ outperforms FL and SFL. It is noteworthy that the benefits of \PiPar\ are evident when training needs to occur in a limited bandwidth environment since more computations can be overlapped with communication (communication takes more time under limited bandwidth). \PiPar\ accelerates FL by 1.4$\times$~-~3.4$\times$ and SFL by 1.4$\times$~-~2.0$\times$. 

FL is slow when training ResNet-18 (Figure~\ref{fig:resnet18-mnist-time} and Figure~\ref{fig:resnet18-cifar10-time}) and MobileNet-Small (Figure~\ref{fig:mobilesmall-mnist-time} and Figure~\ref{fig:mobilesmall-cifar10-time}) because they are deeper networks with more layers. Both SFL and \PiPar\ outperform FL. \PiPar\ has the shortest training time per epoch under all network conditions. In training ResNet-18, \PiPar\ accelerates FL by 15.5$\times$~-~30.5$\times$ and SFL by 1.4$\times$~-~2.2$\times$; in training MobileNetV3-Small, \PiPar\ accelerates FL by 10.5$\times$~-~34.6$\times$ and SFL by 1.2$\times$~-~2.3$\times$.
 
Although FL cannot be executed on the devices when training large DNNs, namely VGG-16 (Figure~\ref{fig:vgg16-cifar100-time} and Figure~\ref{fig:vgg16-tinyimagenet-time}), ResNet-101 (Figure~\ref{fig:resnet101-cifar100-time} and Figure~\ref{fig:resnet101-tinyimagenet-time}) and MobileNetV3-Large (Figure~\ref{fig:mobilelarge-cifar100-time} and Figure~\ref{fig:mobilelarge-tinyimagenet-time}), \PiPar\ accelerates the training process by 1.12$\times$~-~2.2$\times$ when compared to SFL.

\subsubsection{Comparing resources utilization}
\label{subsubsec:exp-ru}

The metric used to compare the utilization of hardware resources is the idle time of the server and devices, which is the total time that the server/device does not contribute to training models in an epoch. The device-side idle time is the average idle time for all devices. A lower idle time corresponds to a higher hardware resource utilization. Since the devices are homogeneous, it is assumed that there is a negligible impact of stragglers.

As shown in Figure~\ref{fig:idle-time}, \PiPar\ reduces the server-side idle time under all network conditions when training VGG-5, ResNet-18 and MobileNetV3-Small on MNIST and CIFAR-10. Since the server has more computing resources than the devices, model training is faster on the server. Hence, reducing the server-side idle time takes precedence over reducing the device-side idle time. Since FL trains complete models on the devices, the devices are rarely idle. However, the server is idle for a large proportion of the time when the model is trained. Compared to FL, SFL utilizes more resources on the server because the server trains multiple layers. \PiPar\ reduces the server-side idle time by overlapping the server-side computations, device-side computations and communication between the server and the devices. Compared to FL and SFL, the server-side idle time using \PiPar\ is reduced up to 64.1$\times$ and 2.9$\times$, respectively. \PiPar\ also reduces the device-side idle time of SFL up to 23.1$\times$ in all cases.

Figure~\ref{fig:idle-time-large} highlights that, compared to SFL, \PiPar\ also reduces idle time when training VGG-16, ResNet-101 and MobileNetV3-Large on CIFAR-100 and Tiny ImageNet. Server-side idle time and device-side idle time of SFL are reduced up to 2.3$\times$ and 2.5$\times$, respectively.

\subsection{Convergence and model accuracy results}
\label{subsec:acc-exp}

It is theoretically proven in Section~\ref{sec:convergence} that \PiPar\ achieves comparable model accuracy and convergence as FL. It will be empirically demonstrated that \PiPar\ does not adversely impact the convergence and accuracy of models. 

The convergence curves and test accuracy of the small and large DNNs using FL, SFL and \PiPar\ are reported. Note that due to the limited memory of devices, the large DNNs could not be executed using FL. Since network conditions do not affect model convergence and accuracy in FL and SFL, only the results for WiFi are reported.

\subsubsection{Comparing convergence}
\label{subsubsec:convergence}

Figure~\ref{fig:loss-small} and Figure~\ref{fig:loss-large} report the loss curves of FL, SFL and \PiPar\ on the validation datasets using the small and large DNNs, respectively. The results highlight that for all combinations of DNNs and datasets, the loss curves of \PiPar\ generally overlaps those of FL and SFL. Therefore, \PiPar\ does not affect model convergence.

It is noted that regardless of the DNN and dataset choice, \PiPar\ converges within the same number of epochs as FL and SFL. Since \PiPar\ reduces the training time per epoch, as presented in Section~\ref{subsubsec:exp-eff}, the overall training time is therefore reduced.

\subsubsection{Comparing accuracy}

In Table~\ref{table:acc-small}, the test accuracy of the small DNNs using FL, SFL and \PiPar\ are reported. The last row shows the difference between the model accuracy of \PiPar\ and the higher one of FL and SFL, denoted as $\Delta$. The results for FL\footnote{Since FL cannot be run on the devices due to limited memory, a high-performance testbed is used to train large DNNs using FL. The high-performance testbed comprising 128 CPUs, two A6000 GPUs and 256 GB memory. Note that the compute capability of devices does not affect the accuracy but only training time.}, SFL and \PiPar\ on the large DNNs are shown in Table~\ref{table:acc-large}. As seen in both tables, in all cases \PiPar\ achieves comparable accuracy as FL and SFL on the test dataset, where the difference in accuracy ranges from -0.2\% to +2.07\%. Specifically, in the worst case, the test accuracy of training MobileNetV3-Small on CIFAR-10 achieved by \PiPar\ is 0.2 lower than FL but still 0.7 higher than SFL.

These results empirically demonstrate that splitting a DNN and reordering the training stages in \PiPar\ does not sacrifice model accuracy while obtaining a higher training efficiency.

\begin{table*}[tp]
    \centering
    \caption{Model accuracy (percentage) for FL, SFL and \PiPar\ using small DNNs.}
    \begin{tabular}{cccc|ccc}
        \Xhline{2\arrayrulewidth}
         \multirow{2}{*}{\textbf{Technique}} & \multicolumn{3}{c}{\textbf{MNIST}} & \multicolumn{3}{c}{{\textbf{CIFAR-10}}}  \\
        \cline{2-7}
          & VGG-5 & ResNet-18 & MobileNetV3-Small & VGG-5 & ResNet-18 & MobileNetV3-Small \\
        \Xhline{2\arrayrulewidth}
          FL & 97.96 & 98.11 & 97.71 & 81.39 & 71.79 & 67.35 \\
          SFL & 97.96 & 98.23 & 97.71 & 81.34 & 71.1 & 66.45 \\
          \PiPar\ & 97.94 & 98.49 & 97.73 & 81.31 & 72.39 & 67.15 \\
        \cline{1-7}
         $\Delta$ & -0.02 & +0.26 & +0.02 & -0.08 & +0.6 & -0.2 \\
        \Xhline{2\arrayrulewidth}
    \end{tabular}
    \label{table:acc-small}
\end{table*}

\begin{table*}[tp]
    \centering
    \caption{Model accuracy (percentage) for FL, SFL and \PiPar\ using large DNNs  }
    \begin{tabular}{cccc|ccc}
        \Xhline{2\arrayrulewidth}
         \multirow{2}{*}{\textbf{Technique}} & \multicolumn{3}{c}{\textbf{CIFAR-100}} & \multicolumn{3}{c}{{\textbf{Tiny ImageNet}}}  \\
        \cline{2-7}
          & VGG-16 & ResNet-101 & MobileNetV3-Large & VGG-16 & ResNet-101 & MobileNetV3-Large \\
        \Xhline{2\arrayrulewidth}
        FL & 52.79 & 26.16 & 37.01 & 44.94 & 23.49 & 29.3 \\
          SFL & 51.89 & 27.11 & 37.3 & 44.93 & 23.49 & 29.53 \\
          \PiPar\ & 52.98 & 27.64 & 37.64 & 45.78 & 25.56 & 29.6 \\
        \cline{1-7}
         $\Delta$ & +0.19 & +0.53 & +0.34 & +0.84 & +2.07 & +0.07 \\
        \Xhline{2\arrayrulewidth}
    \end{tabular}
    \label{table:acc-large}
\end{table*}

\subsection{Evaluation of automated parameter selection}
\label{subsec:op-exp}

The results presented here demonstrate the effectiveness of the automated parameter selection approach in \PiPar. Initially, we exhaustively benchmarked all possible parameters to obtain the optimal parameters. We then show that \PiPar\ selects parameters that are obtained in less time than an exhaustive search, but achieves optimal or near-optimal training time.


The control parameters of \PiPar, namely split point $P^k$ and parallel batch number $N^k$ for device $k$, where $k=1,2,...,K$, affects training efficiency (Section~\ref{subsec:optimise}). $P^k$ and $N^k$ for all devices are the same in our experiments since we consider homogeneous devices; so we use $P$ and $N$. 

The optimal split point $P_{opt}$ and parallel batch number $N_{opt}$ can be found by exhaustively searching given a finite search space. 
As shown in Table~\ref{table:structure} and Table~\ref{table:large-structure}, VGG-5, ResNet-18, MobileNetV3-Small, VGG-16, ResNet-101 and MobileNetV3-Large consist of 5, 10, 15, 16, 35 and 19 sequential layers, respectively. We have $P\in [1,5]$ for VGG-5, $P\in [1,10]$ for ResNet-18, $P\in [1,15]$ for MobileNetV3-Small, $P\in [1,16]$ for VGG-16, $P\in [1,35]$ for ResNet-101 and $P\in [1,16]$ for MobileNetV3-Large. Note that DNNs, such as ResNet-18 and ResNet-101, have parallel branches that cannot be split. In this case, only connections between sequential layers can be selected as split points.
Assuming batch size for FL is $B$, since the batch size for \PiPar\ $\lfloor B/N \rfloor$ is no less than 1, we have $N\in [1,B]$. 
To exhaustively search for the optimal pair $\{P_{opt}, N_{opt}\}$, the DNNs are trained for one iteration using all possible $\{P, N\}$ pairs in \PiPar, and the pair with the shortest training time is considered optimal.

The proposed method is to select $P$ and $N$ for each experiment. There is only one training iteration in the profiling stage. We compare $\{P, N\}$ selected by our approach against $\{P_{opt}, N_{opt}\}$ determined by the exhaustive search in terms of training time and search time. $T_{P,N}$ denotes the training time for each epoch given $P$ and $N$. We have $T_{P,N} \geq T_{P_{opt}, N_{opt}}$. The score in Equation~\ref{eq:score} measures how close $T_{P,N}$ is to $T_{P_{opt}, N_{opt}}$, which is between 0 and 1. The higher the score, the better the $\{P, N\}$ values perform in terms of training time.

\begin{equation}
\label{eq:score}
    score = \frac{T_{P_{opt}, N_{opt}}}{T_{P,N}}
\end{equation}

The results for small and large DNNs are shown in Table~\ref{table:opt-exp} and Table~\ref{table:opt-exp-large}, respectively. $S_{P,N}$ is the search time to obtain $P$ and $N$ for the automated parameter selection approach or the exhaustive search; smaller is better. For both small and large DNNs, the proposed method selects near optimal parameters in all cases (with a $Score \geq 0.96$) and optimal parameters in 83.3\% cases. In addition, our approach selects optimal split point $P$ in all cases. 

The results highlight that the time to exhaustively search is substantially high to be practical in the real-world. The average cost of our approach is 27\% of one training epoch, which is 6957$\times$ faster than exhaustively searching. The approach is only executed once before training. Since raining consists of hundreds of epochs or more, the overhead of executing this algorithm is negligible (less than 0.3\%). Therefore, our approach provides a practical approach to determine the parameters of \PiPar.

\begin{table*}[tp]
    \centering
    \caption{Parameters selected by the approach in \PiPar\ in contrast to the optimal parameters for small DNNs.}
    \begin{tabu}{ccccccc|ccccc}
    \Xhline{2\arrayrulewidth}
       \multirow{2}{*}{\textbf{Model}} & \multirow{2}{*}{\textbf{Dataset}} & \multirow{2}{*}{\textbf{Network}} & \multicolumn{4}{c}{\textbf{Proposed Approach}} & \multicolumn{4}{c}{\textbf{Exhaustive Search}} & \multirow{2}{*}{\makecell{$Score$\\(Equation~\ref{eq:score})}} \\
       \cline{4-11}
       & & & $P$ & $N$ & $T_{P,N}$ & $S_{P,N}$ & $P_{opt}$ & $N_{opt}$ & $T_{P_{opt}, N_{opt}}$ & $S_{P_{opt}, N_{opt}}$ &  \\
      \Xhline{2\arrayrulewidth}
       \multirow{6}{*}{VGG-5} & \multirow{3}{*}{MNIST} & 4G & 2 & 4 & 10.5 & 1.7 & 2 & 4 & 10.5 & 481.8 & 1 \\
        & & 4G+ & 1 & 10 & 7.9 & 1.4 & 1 & 11 & 7.7 & 477.0 & 0.97 \\
        & & WiFi & 1 & 6 & 4.5 & 1.3 & 1 & 6 & 4.5 & 474.2 & 1 \\
        \cline{2-12}
         & \multirow{3}{*}{CIFAR-10} & 4G & 1 & 12 & 19.8 & 6.2 & 1 & 12 & 19.8 & 2034.2 & 1 \\
        & & 4G+ & 1 & 6 & 13.1 & 4.5 & 1 & 8 & 12.9 & 1818.4 & 0.98 \\
        & & WiFi & 1 & 3 & 10.3 & 4.2 & 1 & 3 & 10.3 & 1816.2 & 1 \\
        \hline
       \multirow{6}{*}{ResNet-18} & \multirow{3}{*}{MNIST} & 4G & 1 & 7 & 9.8 & 1.6 & 1 & 7 & 9.8 & 3393.9 & 1 \\
        & & 4G+ & 1 & 4 & 5.9 & 1.5 & 1 & 4 & 5.9 & 3130.5 & 1 \\
        & & WiFi & 1 & 3 & 5.3 & 1.4 & 1 & 3 & 5.3 & 3018.4 & 1 \\
        \cline{2-12}
         & \multirow{3}{*}{CIFAR-10} & 4G & 1 & 8 & 13.2 & 4.2 & 1 & 8 & 13.2 & 13753.4 & 1 \\
        & & 4G+ & 1 & 6 & 10.0 & 4.0 & 1 & 6 & 10.0 & 13634.0 & 1 \\
        & & WiFi & 1 & 3 & 8.9 & 3.8 & 1 & 3 & 8.9 & 13535.9 & 1 \\
        \hline
        \multirow{6}{*}{MobileNetV3-Small} & \multirow{3}{*}{MNIST} & 4G & 2 & 5 & 4.8 & 1.2 & 2 & 5 & 4.8 & 4955.0 & 1 \\
        & & 4G+ & 2 & 4 & 4.1 & 1.1 & 2 & 4 & 4.1 & 4847.5 & 1 \\
        & & WiFi & 1 & 4 & 3.6 & 1.1 & 1 & 4 & 3.6 & 4816.7 & 1 \\
        \cline{2-12}
         & \multirow{3}{*}{CIFAR-10} & 4G & 2 & 8 & 12.8 & 4.1 & 2 & 8 & 12.8 & 33549.2 & 1 \\
        & & 4G+ & 1 & 12 & 8.4 & 3.7 & 1 & 12 & 8.4 & 33382.1 & 1 \\
        & & WiFi & 1 & 9 & 6.7 & 3.1 & 1 & 9 & 6.7 & 32780.5 & 1 \\
    \Xhline{2\arrayrulewidth}
    \end{tabu}
    \label{table:opt-exp}
\end{table*}

\begin{table*}[tp]
    \centering
    \caption{Parameters selected by the approach in \PiPar\ in contrast to the optimal parameters for large DNNs.}
    \begin{tabu}{ccccccc|ccccc}
    \Xhline{2\arrayrulewidth}
       \multirow{2}{*}{\textbf{Model}} & \multirow{2}{*}{\textbf{Dataset}} & \multirow{2}{*}{\textbf{Network}} & \multicolumn{4}{c}{\textbf{Proposed Approach}} & \multicolumn{4}{c}{\textbf{Exhaustive Search}} & \multirow{2}{*}{\makecell{$Score$\\(Equation~\ref{eq:score})}} \\
       \cline{4-11}
       & & & $P$ & $N$ & $T_{P,N}$ & $S_{P,N}$ & $P_{opt}$ & $N_{opt}$ & $T_{P_{opt}, N_{opt}}$ & $S_{P_{opt}, N_{opt}}$ &  \\
      \Xhline{2\arrayrulewidth}
        \multirow{6}{*}{VGG-16} & \multirow{3}{*}{CIFAR-100} & 4G & 1 & 20 & 122.8 & 13.6 & 1 & 16 & 117.6 & 34274.4 & 0.96 \\
        & & 4G+ & 1 & 16 & 62.2 & 6.1 & 1 & 16 & 62.2 & 19753.2 & 1 \\
        & & WiFi & 1 & 6 & 34.9 & 5.2 & 1 & 6 & 34.9 & 12535.2 & 1 \\
        \cline{2-12}
         & \multirow{3}{*}{Tiny ImageNet} & 4G & 1 & 16 & 455.7 & 22.4 & 1 & 16 & 455.7 & 824884.0 & 0.96 \\
        & & 4G+ & 1 & 14 & 240.4 & 18.7 & 1 & 16 & 233.7 & 452372.8 & 0.97 \\
        & & WiFi & 1 & 12 & 112.9 & 16.8 & 1 & 16 & 110.2 & 245357.0 & 1 \\
        \hline
       \multirow{6}{*}{ResNet-101} & \multirow{3}{*}{CIFAR-100} & 4G & 1 & 4 & 26.8 & 5.9 & 1 & 4 & 26.8 & 29004.6 & 1 \\
        & & 4G+ & 1 & 2 & 20.3 & 4.3 & 1 & 2 & 20.3 & 28408.6 & 1 \\
        & & WiFi & 1 & 2 & 19.3 & 3.8 & 1 & 2 & 19.3 & 28347.6 & 1 \\
        \cline{2-12}
        & \multirow{3}{*}{Tiny ImageNet} & 4G & 1 & 8 & 52.6 & 11.1 & 1 & 8 & 52.6 & 122736.22 & 1 \\
        & & 4G+ & 1 & 6 & 50.9 & 10.3 & 1 & 4 & 49.0 & 119846.0 & 0.96 \\
        & & WiFi & 1 & 3 & 47.0 & 9.9 & 1 & 3 & 47.0 & 117718.5 & 1 \\
        \hline
        \multirow{6}{*}{MobileNetV3-Large} & \multirow{3}{*}{CIFAR-100} & 4G & 1 & 16 & 13.3 & 4.1 & 1 & 16 & 13.3 & 41844.5 & 1 \\
        & & 4G+ & 1 & 6 & 8.6 & 3.6 & 1 & 6 & 8.6 & 36807.8 & 1 \\
        & & WiFi & 1 & 4 & 7.8 & 3.4 & 1 & 4 & 7.8 & 35651.0 & 1 \\
        \cline{2-12}
        & \multirow{3}{*}{Tiny ImageNet} & 4G & 1 & 16 & 35.5 & 8.6 & 1 & 16 & 35.5 & 79864.6 & 1 \\
        & & 4G+ & 1 & 12 & 21.9 & 7.8 & 1 & 12 & 21.9 & 57710.0 & 1 \\
        & & WiFi & 1 & 8 & 14.4 & 7.7 & 1 & 8 & 14.4 & 46406.2 & 1 \\
    \Xhline{2\arrayrulewidth}
    \end{tabu}
    \label{table:opt-exp-large}
\end{table*}

\subsection{Batch size analysis}
\label{subsec:batch-size}

Compared to FL, \PiPar\ in Phase 2 increases the number of mini-batches involved in each training iteration and reduces the batch size. Assume $B$ is the FL batch size. The batch size in \PiPar\ is $B^\prime = \lfloor B/N \rfloor$, where $N$ is the parallel batch number (Equation~\ref{eq:batch-size}). The DNNs are trained using FL and SFL for different $B$ and \PiPar\ for corresponding $B^\prime$ under different network conditions.

Figure~\ref{fig:batch-size} shows the training time per epoch for FL, SFL and \PiPar\ using VGG-5 and CIFAR-10, while Figure~\ref{fig:batch-size-large} shows the training time per epoch for SFL and \PiPar\ using VGG-16 and CIFAR-100. The training time of FL/SFL/\PiPar\ decreases as the batch size increases because intra-batch parallelisation can be leveraged for matrix multiplication operations when a larger batch is trained. However, increasing batch sizes is not an effective way to speed up training because it requires more memory and reduces model accuracy~\cite{DBLP:journals/corr/KeskarMNST16}. The results highlight that \PiPar\ is consistently faster than FL and SFL for VGG-5 and faster than SFL for VGG-16 under all network conditions, regardless of batch sizes. The same trend is seen when training other four DNNs and two datasets.

\subsection{Robustness Analysis}
\label{subsec:robust}
We explore the robustness of \PiPar\ to more complex environments. In Section~\ref{subsec:hetero}, heterogeneous devices are used to evaluate the performance of \PiPar\ against FL and SFL. The impact of using differential privacy methods is considered in Section~\ref{subsec:privacy}. Finally, the impact of changing network bandwidth on the overhead of the automated parameter selection approach is considered in Section~\ref{subsec:bandwidth}. In this section, only representative results are shown that are obtained from an evaluation using VGG-5, ResNet-18 and MobileNetV3-Small on CIFAR-10 under different network conditions. A similar trend is noted for other datasets. 

\subsubsection{Impact on Performance with Heterogeneous Devices}
\label{subsec:hetero}

The impact on the performance using a homogeneous and heterogeneous testbed is considered. The setup of the homogeneous testbed was presented in Section~\ref{subsec:setup}. In the heterogeneous testbed, the same number of devices are used but the CPU frequency of half of the devices is reduced from 1.2 GHz to 600 MHz to create an environment with different compute capabilities of devices. 

\begin{figure*}[tp]
	\centering
	\subfigure[VGG-5 (4G)]{
	    \includegraphics[width=0.3\textwidth]{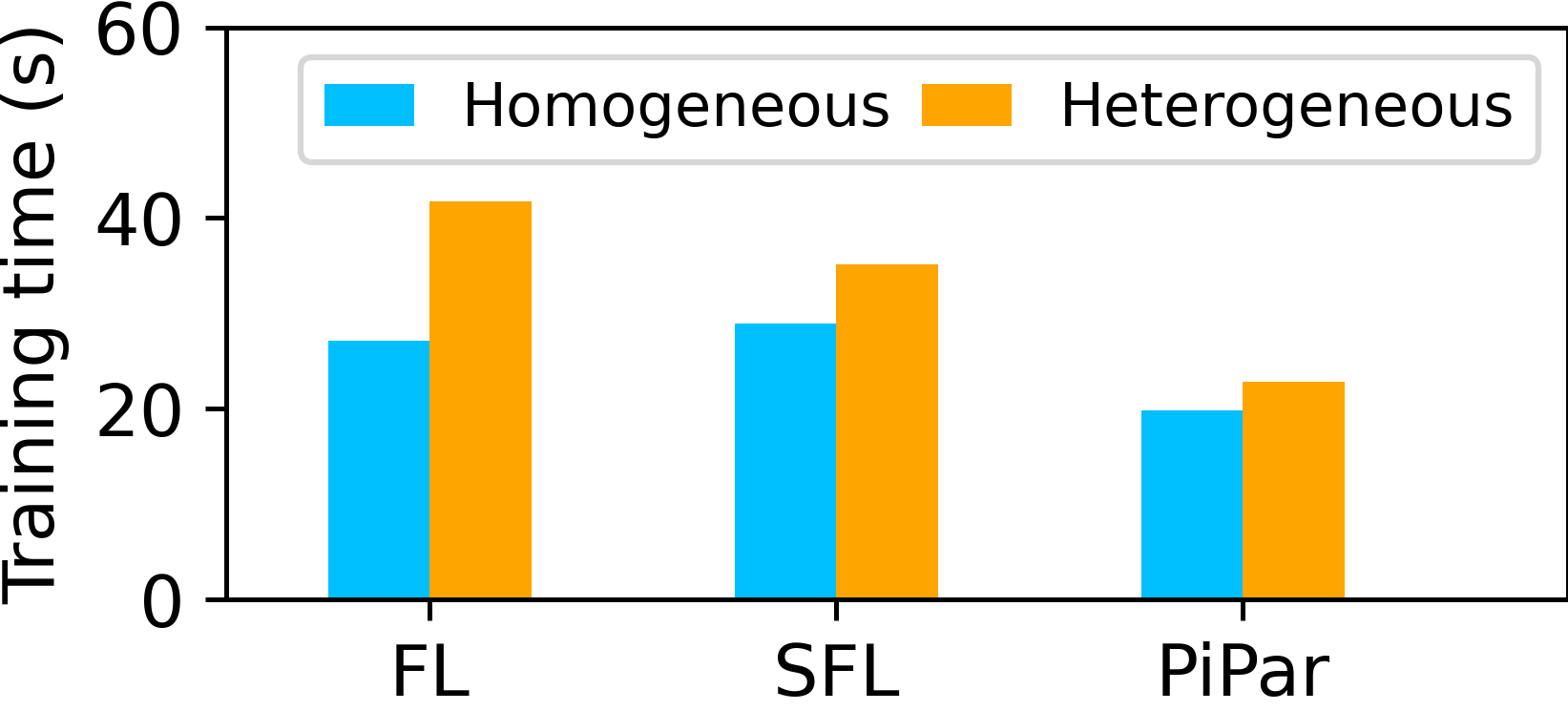}
	    \label{fig:vgg5-hetero-4g}
	    }
	\hfill
        \subfigure[VGG-5 (4G+)]{
	    \includegraphics[width=0.3\textwidth]{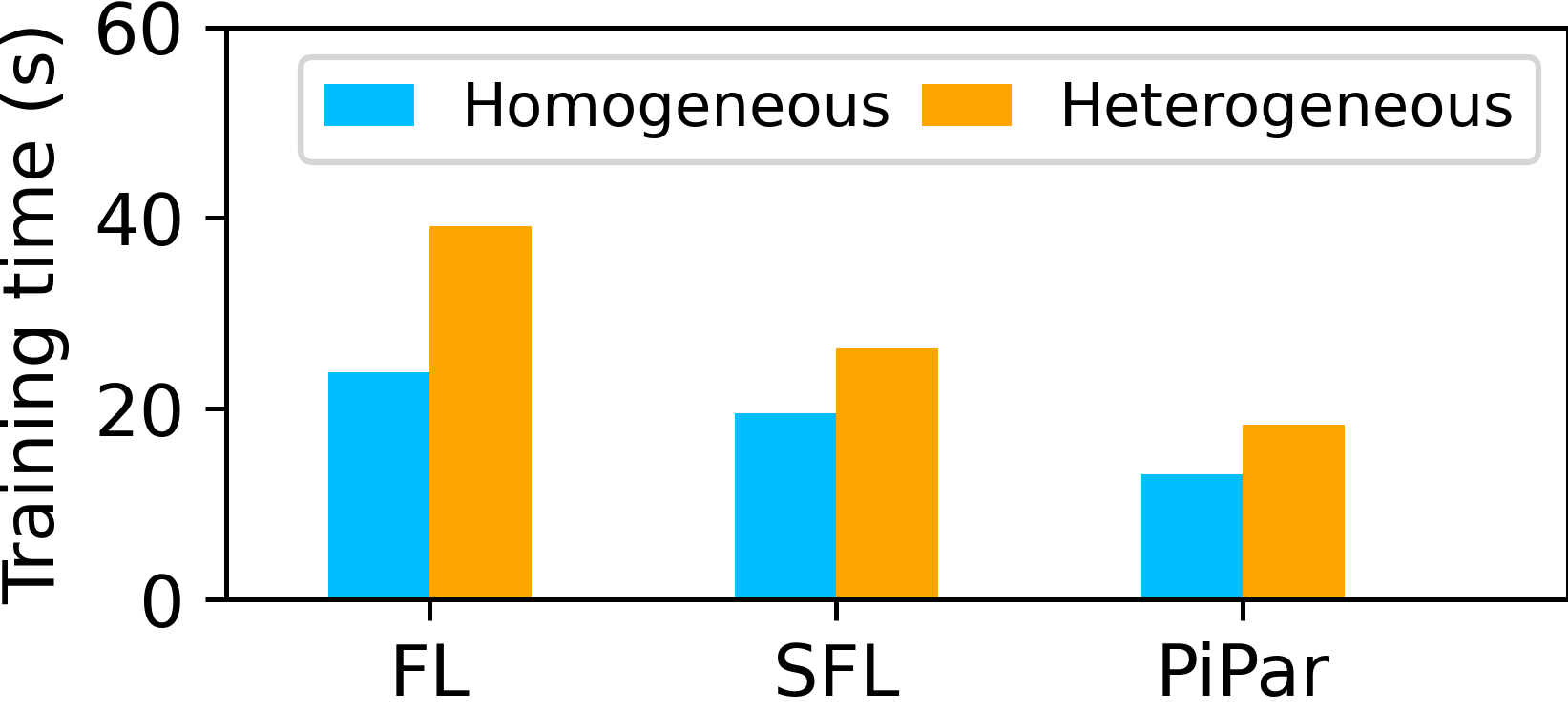}
	    \label{fig:vgg5-hetero-4gp}
	    }
	\hfill
        \subfigure[VGG-5 (WiFi)]{
	    \includegraphics[width=0.3\textwidth]{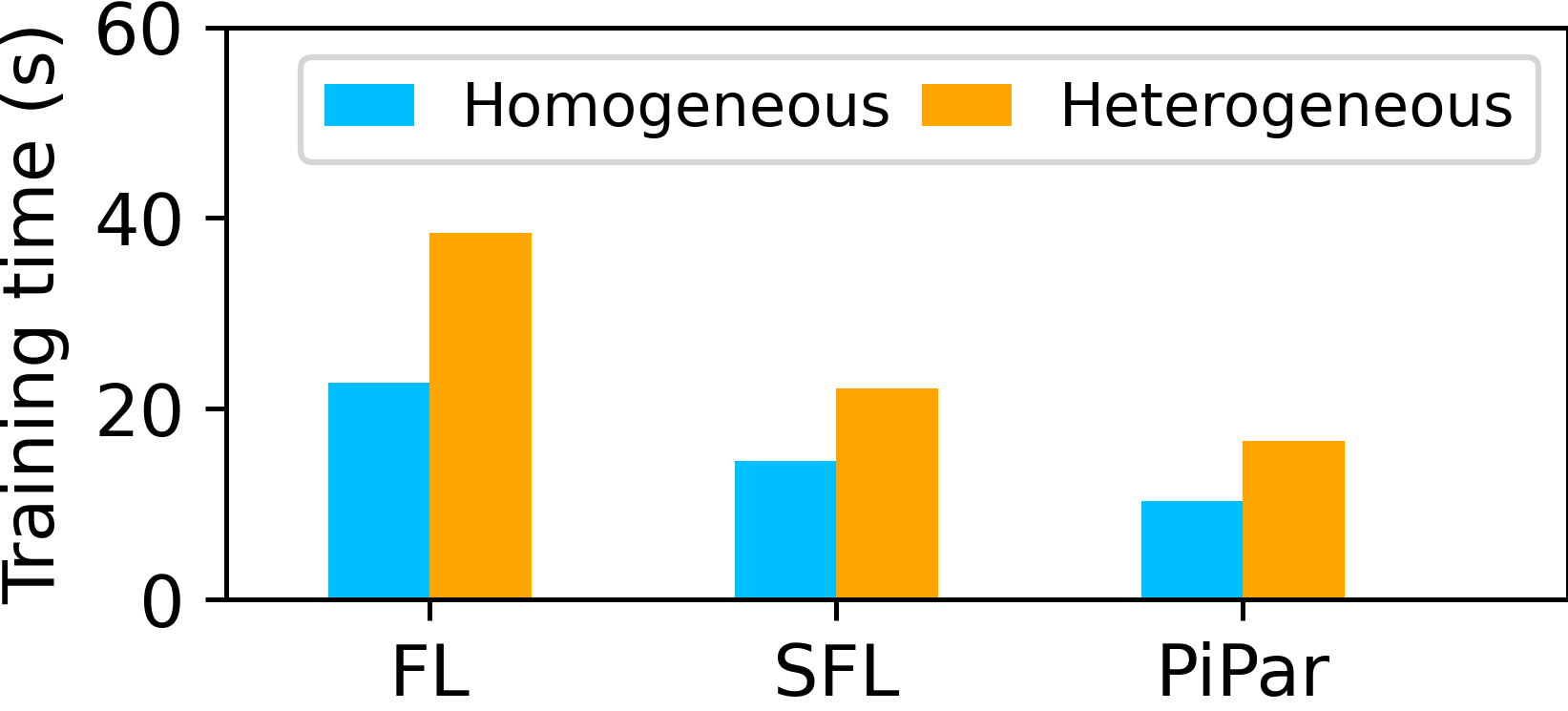}
	    \label{fig:vgg5-hetero-wifi}
	    }
	\hfill
	\subfigure[ResNet-18 (4G)]{
	    \includegraphics[width=0.3\textwidth]{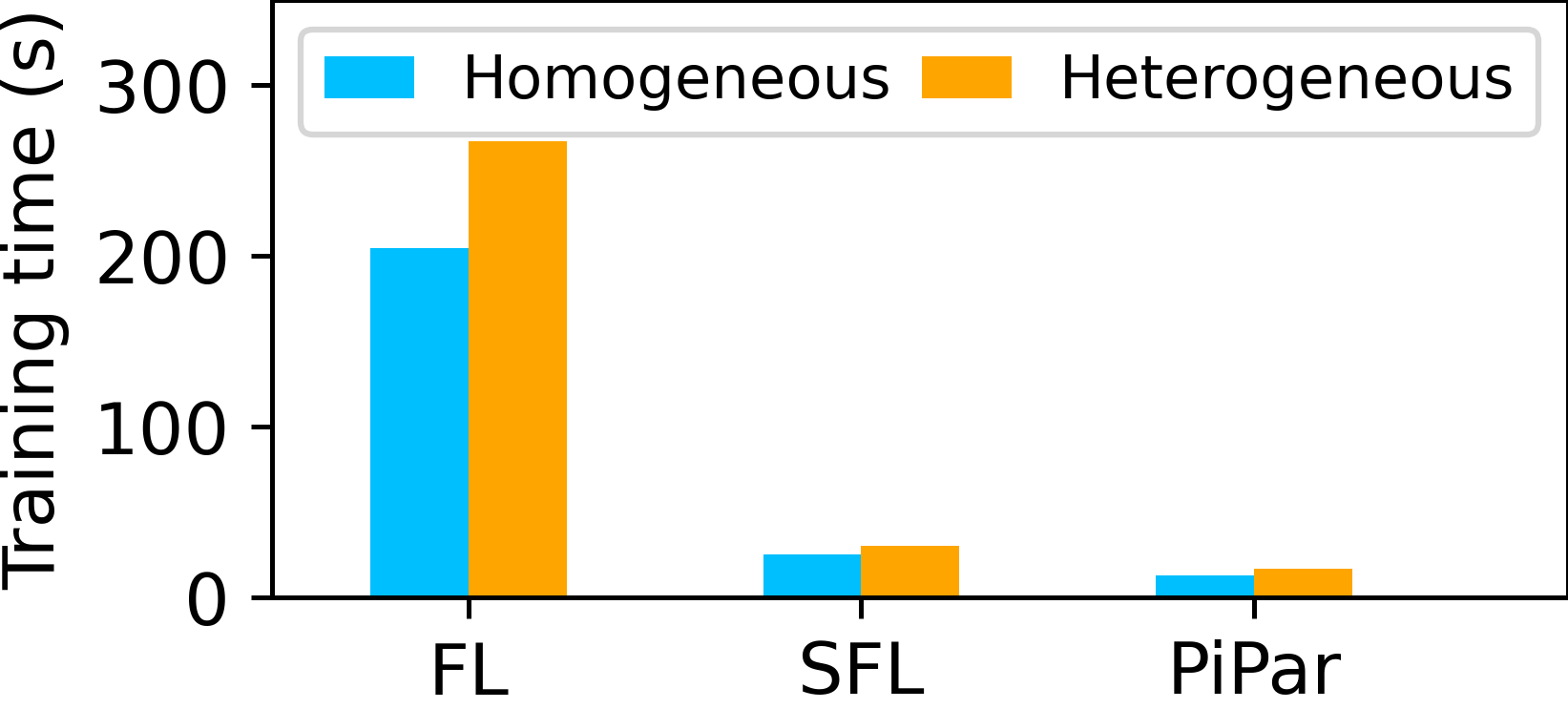}
	    \label{fig:resnet18-hetero-4g}
	    }
	\hfill
        \subfigure[ResNet-18 (4G+)]{
	    \includegraphics[width=0.3\textwidth]{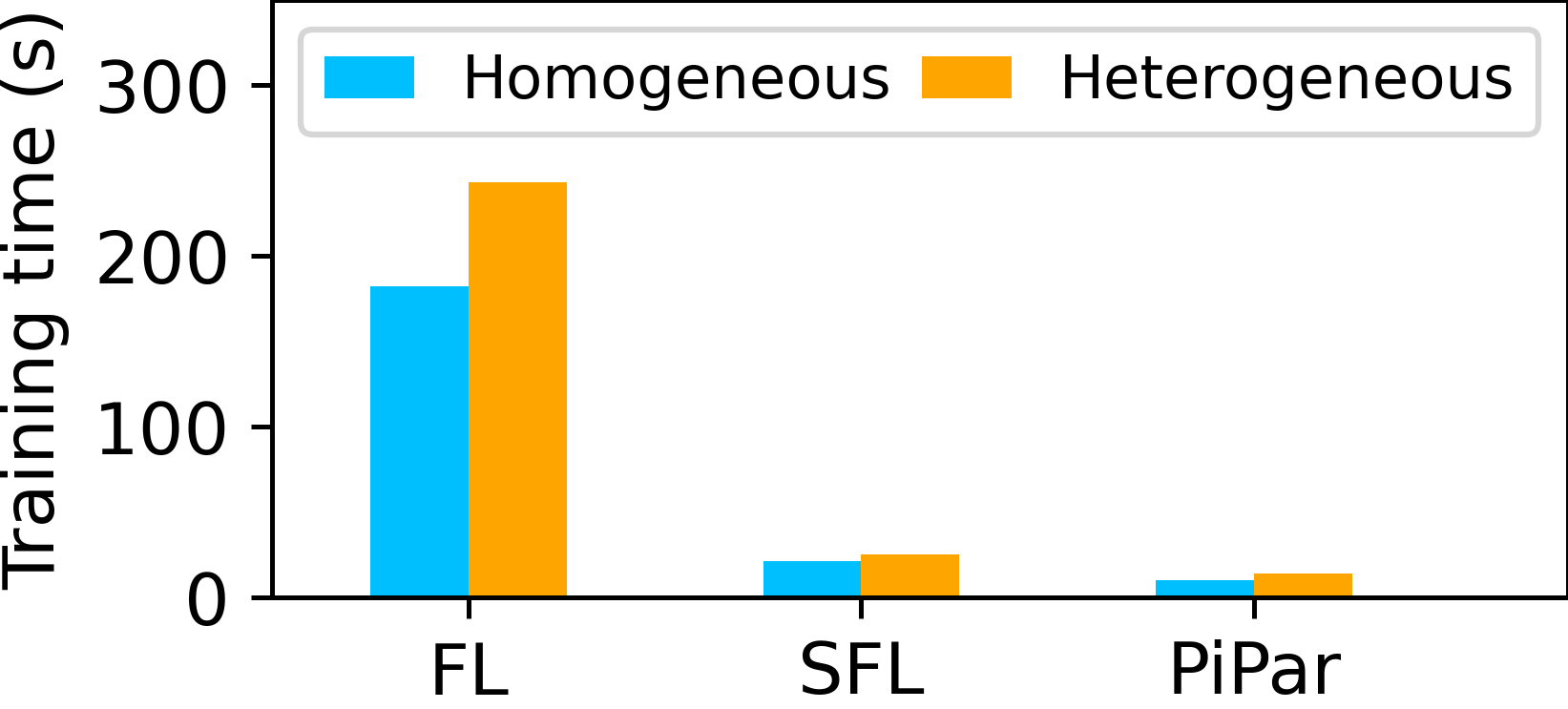}
	    \label{fig:resnet18-hetero-4gp}
	    }
	\hfill
        \subfigure[ResNet-18 (WiFi)]{
	    \includegraphics[width=0.3\textwidth]{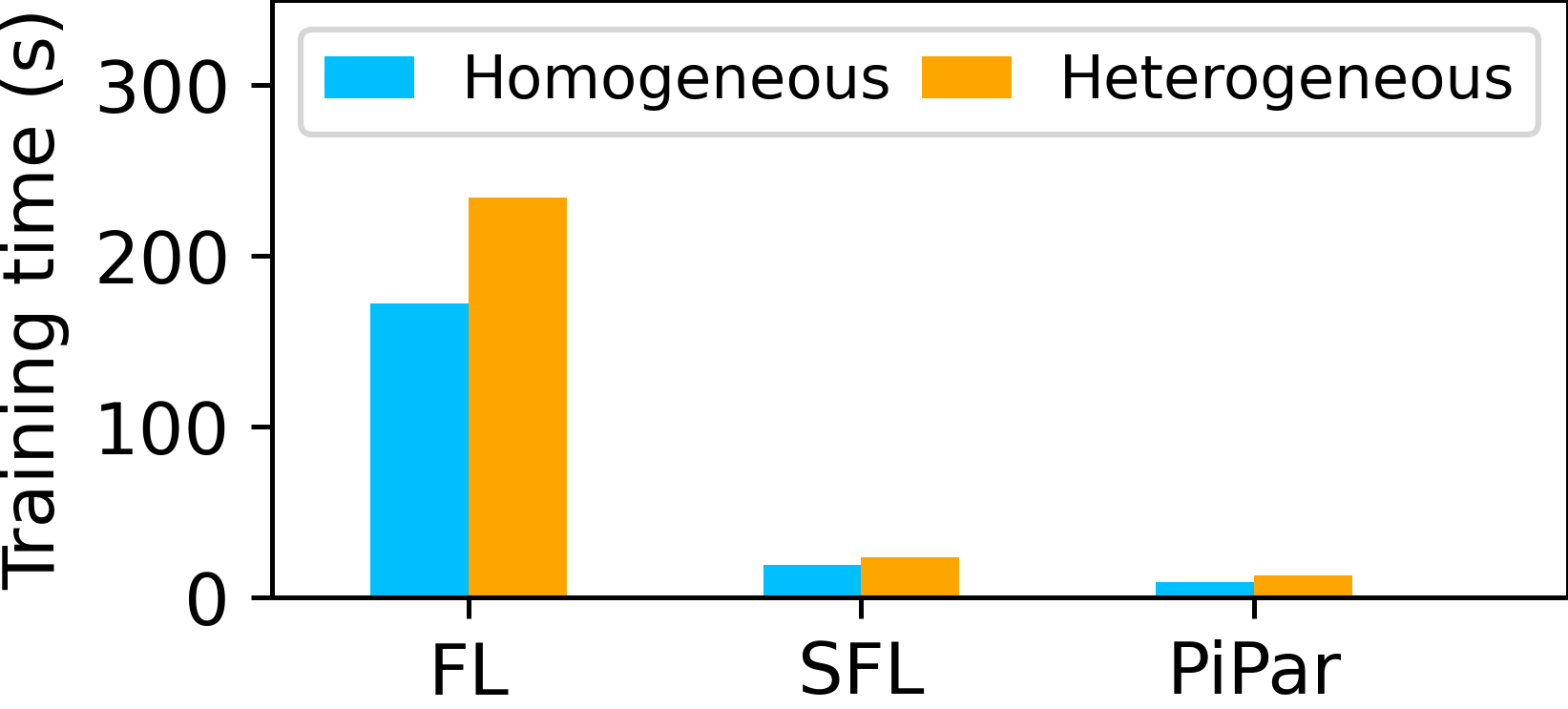}
	    \label{fig:resnet18-hetero-wifi}
	    }
	\hfill
        \subfigure[MobileNetV3-Small (4G)]{
	    \includegraphics[width=0.3\textwidth]{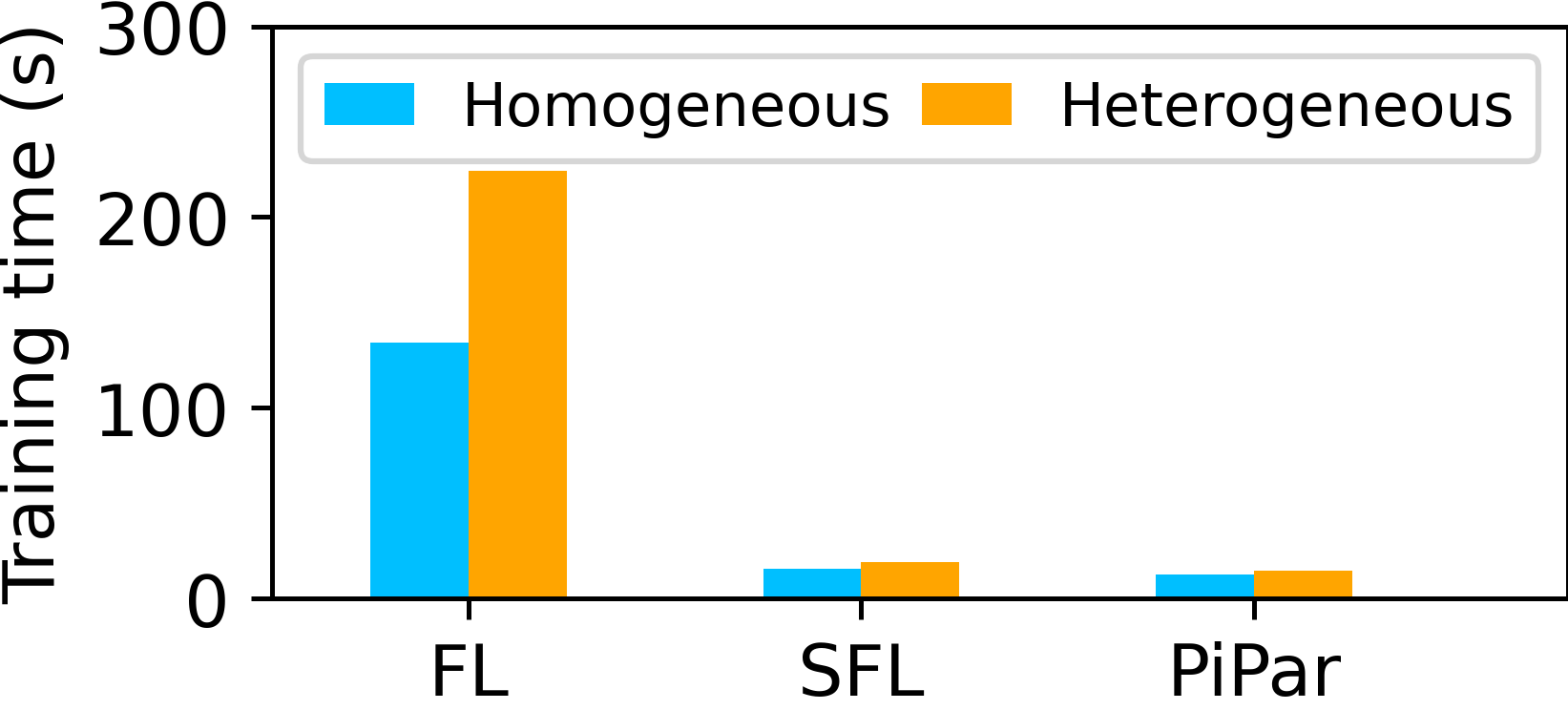}
	    \label{fig:mobilesmall-hetero-4g}
	    }
	\hfill
        \subfigure[MobileNetV3-Small (4G+)]{
	    \includegraphics[width=0.3\textwidth]{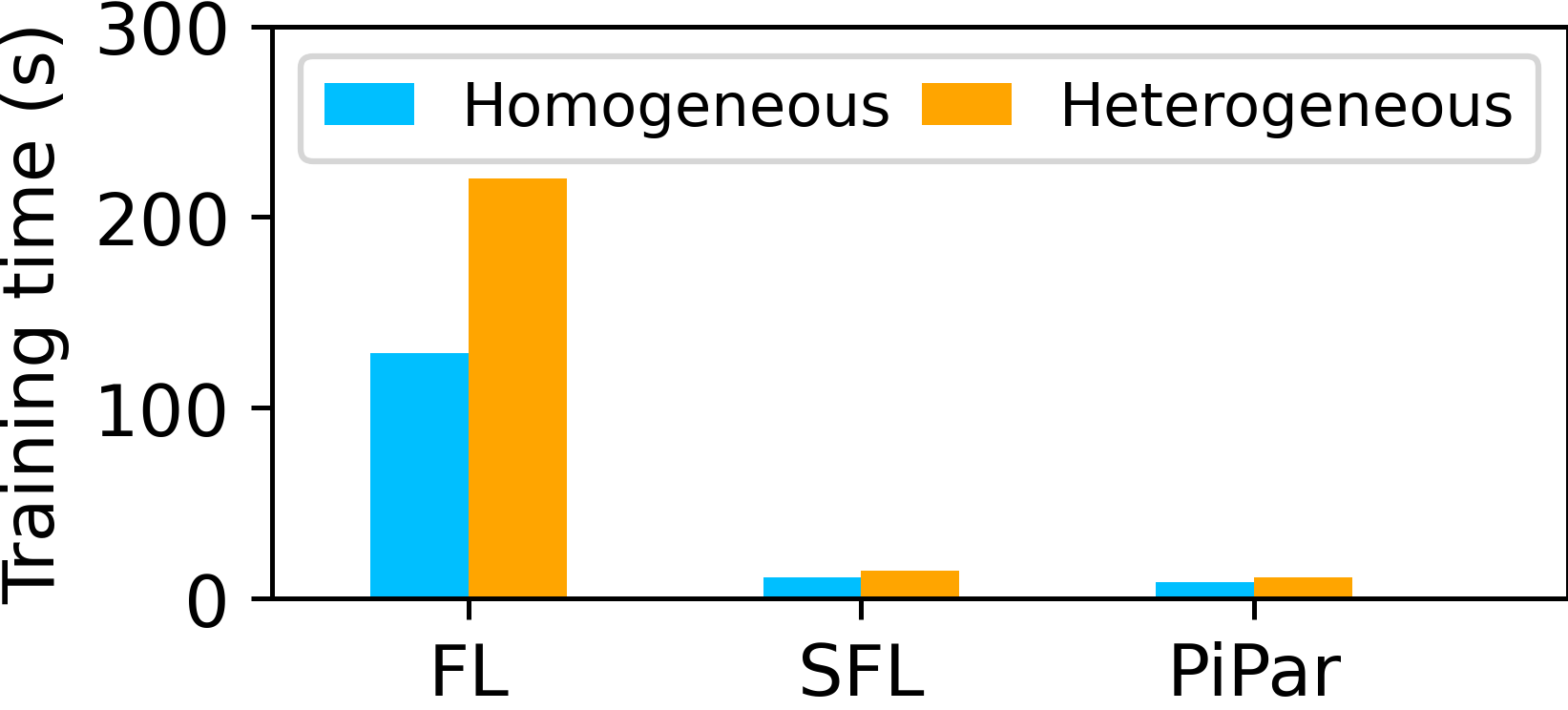}
	    \label{fig:mobilesmall-hetero-4gp}
	    }
	\hfill
        \subfigure[MobileNetV3-Small (WiFi)]{
	    \includegraphics[width=0.3\textwidth]{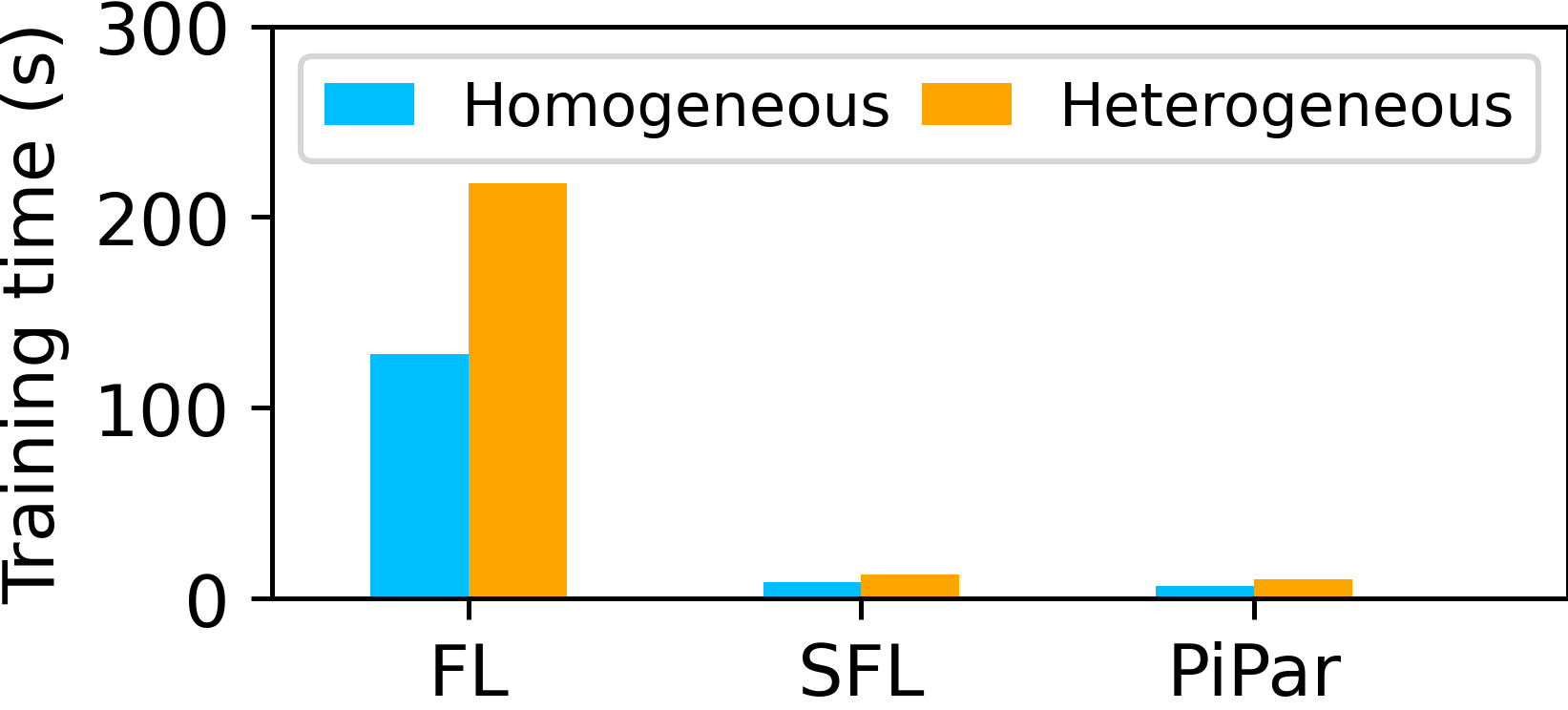}
	    \label{fig:mobilesmall-hetero-wifi}
	    }
        \hfill
	\caption{Training time per epoch for FL, SFL and \PiPar\ using small DNNs on CIFAR-10 under different network conditions on homogeneous and heterogeneous testbeds.}
	\label{fig:hetero}
\end{figure*}

Figure~\ref{fig:hetero} shows the training time per epoch for FL, SFL and \PiPar. Compared to the homogeneous testbed, the training time on the heterogeneous increases since there are slower devices. The faster devices have to wait for the stragglers before model aggregation. In all cases, \PiPar\ has lowest training time compared to FL and SFL on both homogeneous and heterogeneous testbeds. Specifically, on the heterogeneous testbed, \PiPar\ accelerates training of FL by up to 32$\times$ and SFL by up to 1.8$\times$. In addition, FL has a larger difference in performance between testbeds than SFL and \PiPar since the latter trains the last several layers of the DNN (as the DNN is partitioned) on the server, which is not affected by the heterogeneity of devices.

\subsubsection{Impact on Performance When Using Differential Privacy Methods}
\label{subsec:privacy}

Differential Privacy (DP)~\cite{dp, pixeldp} is used in CML methods to enhance privacy in CML by adding noise into data transferred between the devices and server. We consider the performance overhead introduced when using DP methods in FL, SFL and \PiPar. 

Two DP methods are considered. Firstly, classic DP~\cite{dp} is used to add noise to local models on devices before they are sent to the server to make them irreversible. Secondly, PixelDP\cite{pixeldp} adds an additional noise layer before the first layer of device models, which prevents activations from being restored to raw data via reverse engineering.

\begin{figure*}[tp]
	\centering
	\subfigure[VGG-5 (4G)]{
	    \includegraphics[width=0.3\textwidth]{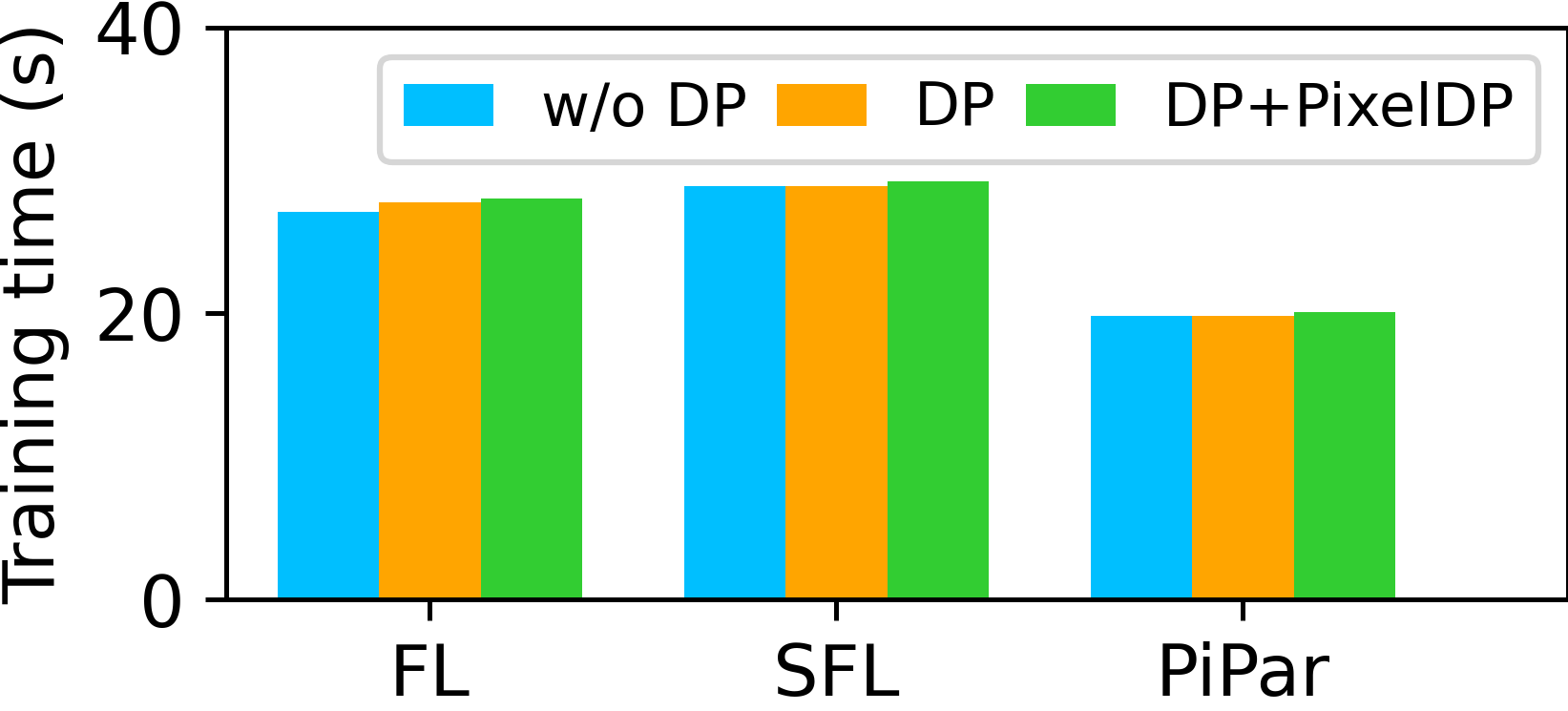}
	    \label{fig:vgg5-dp-4g}
	    }
	\hfill
        \subfigure[VGG-5 (4G+)]{
	    \includegraphics[width=0.3\textwidth]{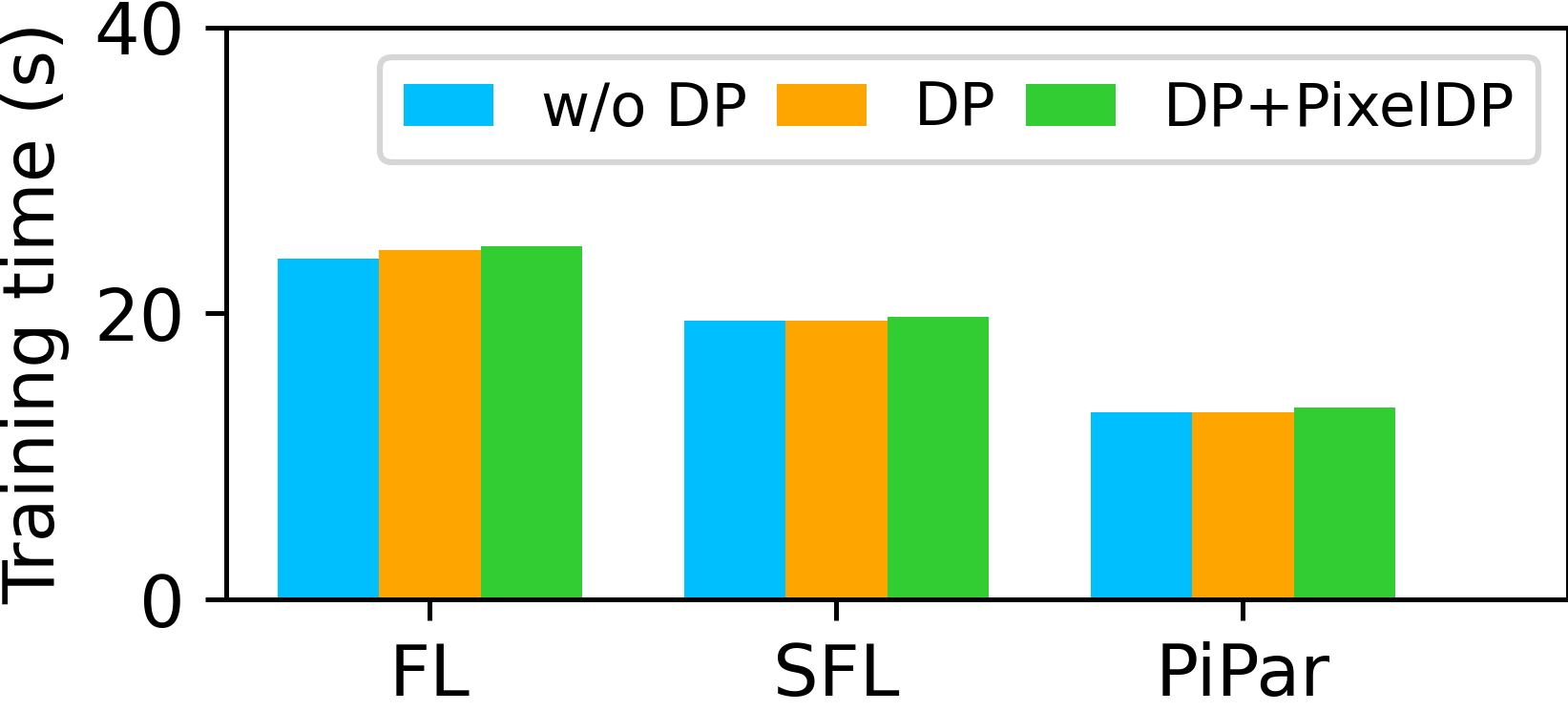}
	    \label{fig:vgg5-dp-4gp}
	    }
	\hfill
        \subfigure[VGG-5 (WiFi)]{
	    \includegraphics[width=0.3\textwidth]{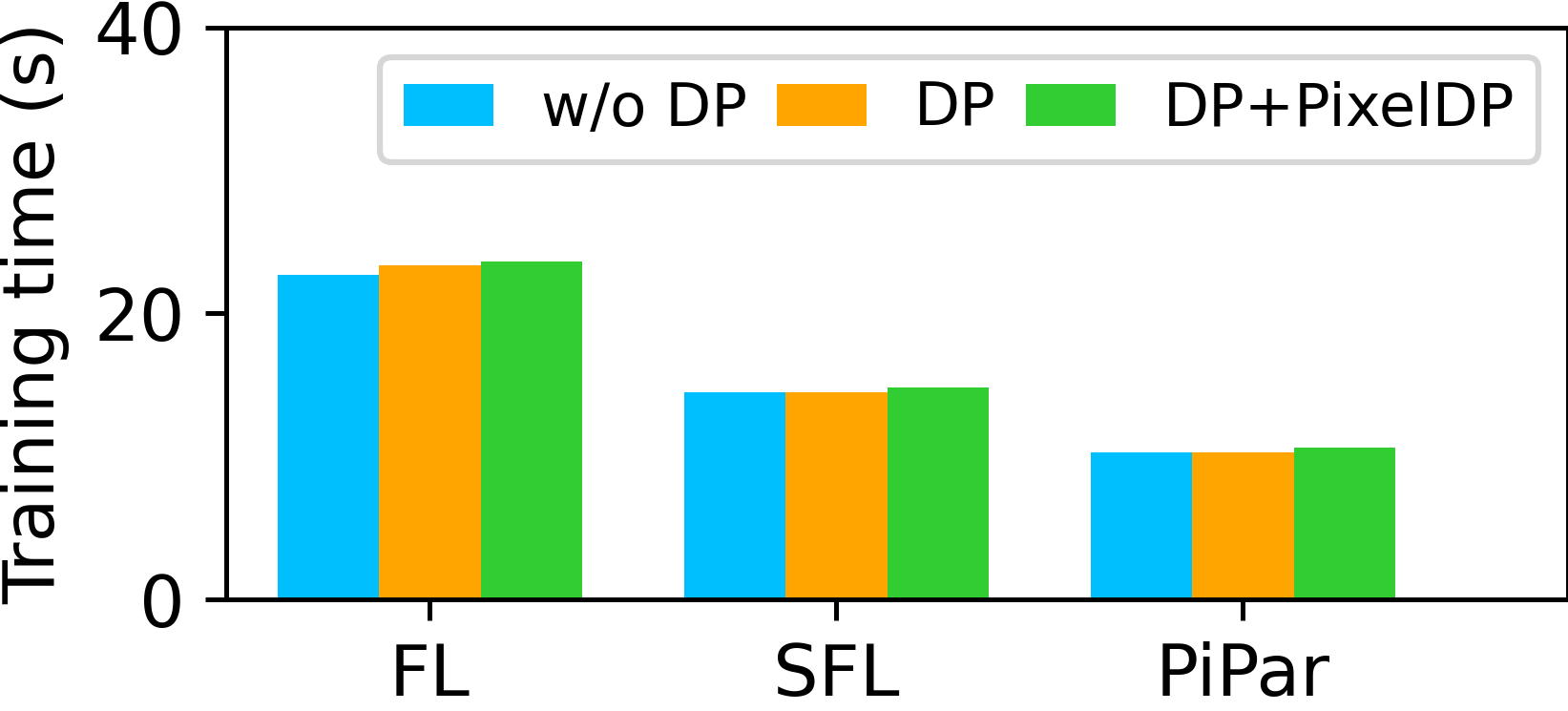}
	    \label{fig:vgg5-dp-wifi}
	    }
	\hfill
	\subfigure[ResNet-18 (4G)]{
	    \includegraphics[width=0.3\textwidth]{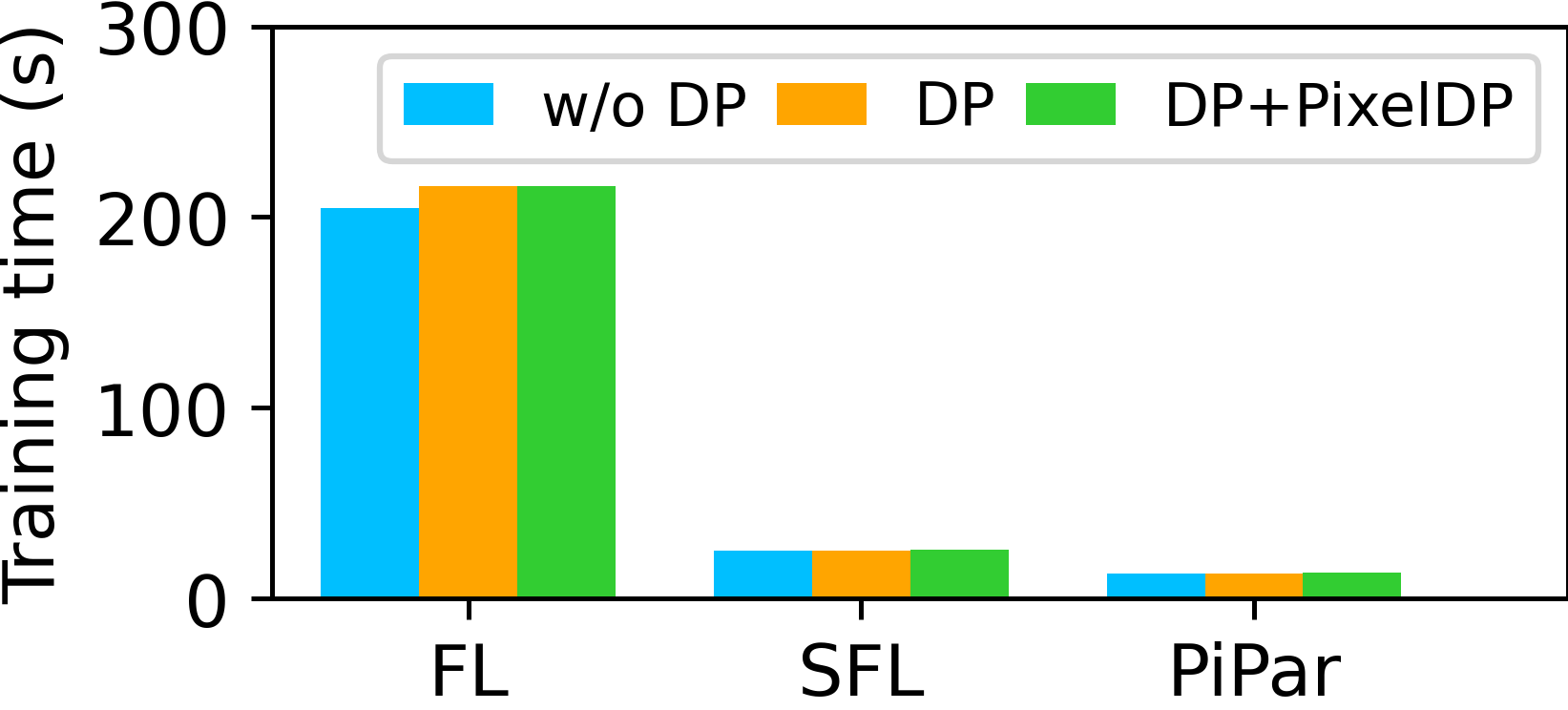}
	    \label{fig:resnet18-dp-4g}
	    }
	\hfill
        \subfigure[ResNet-18 (4G+)]{
	    \includegraphics[width=0.3\textwidth]{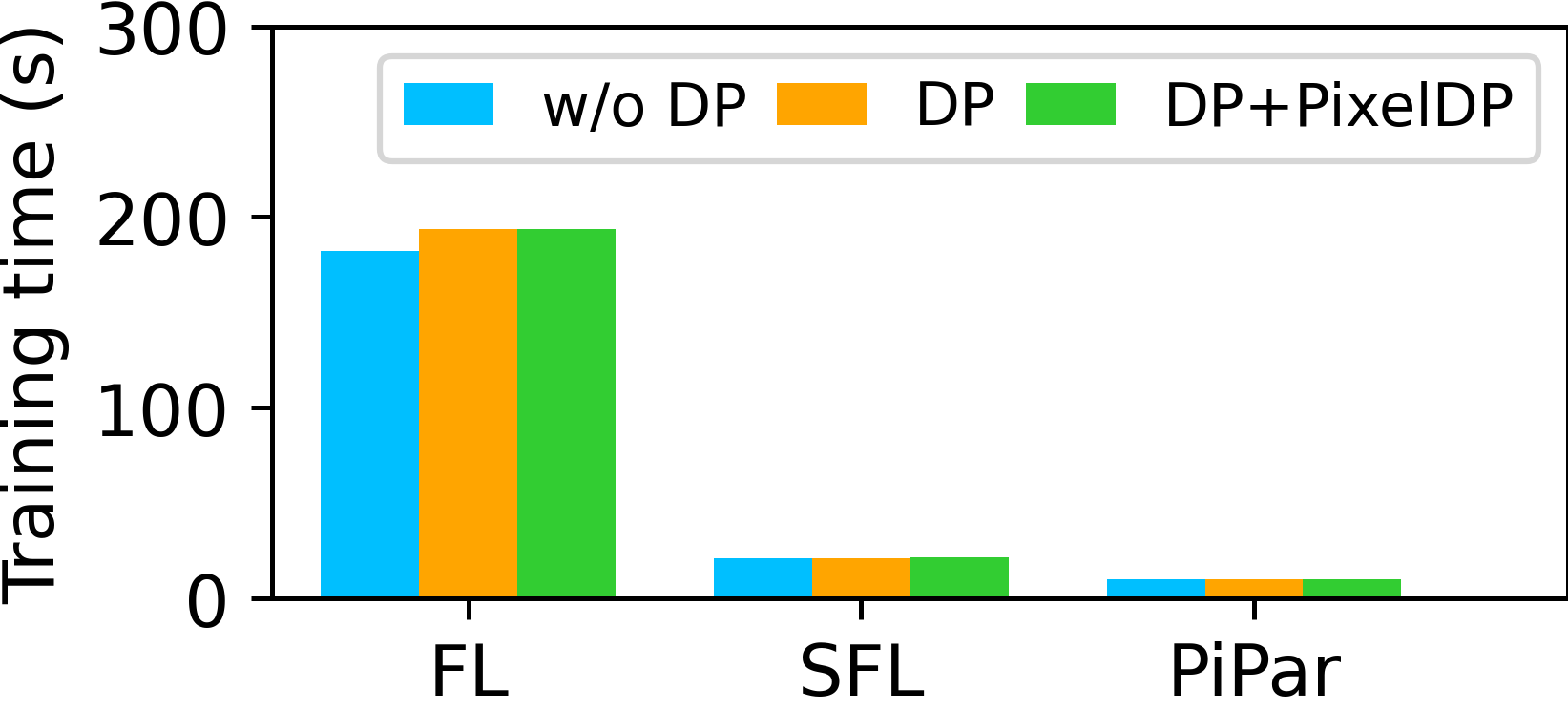}
	    \label{fig:resnet18-dp-4gp}
	    }
	\hfill
        \subfigure[ResNet-18 (WiFi)]{
	    \includegraphics[width=0.3\textwidth]{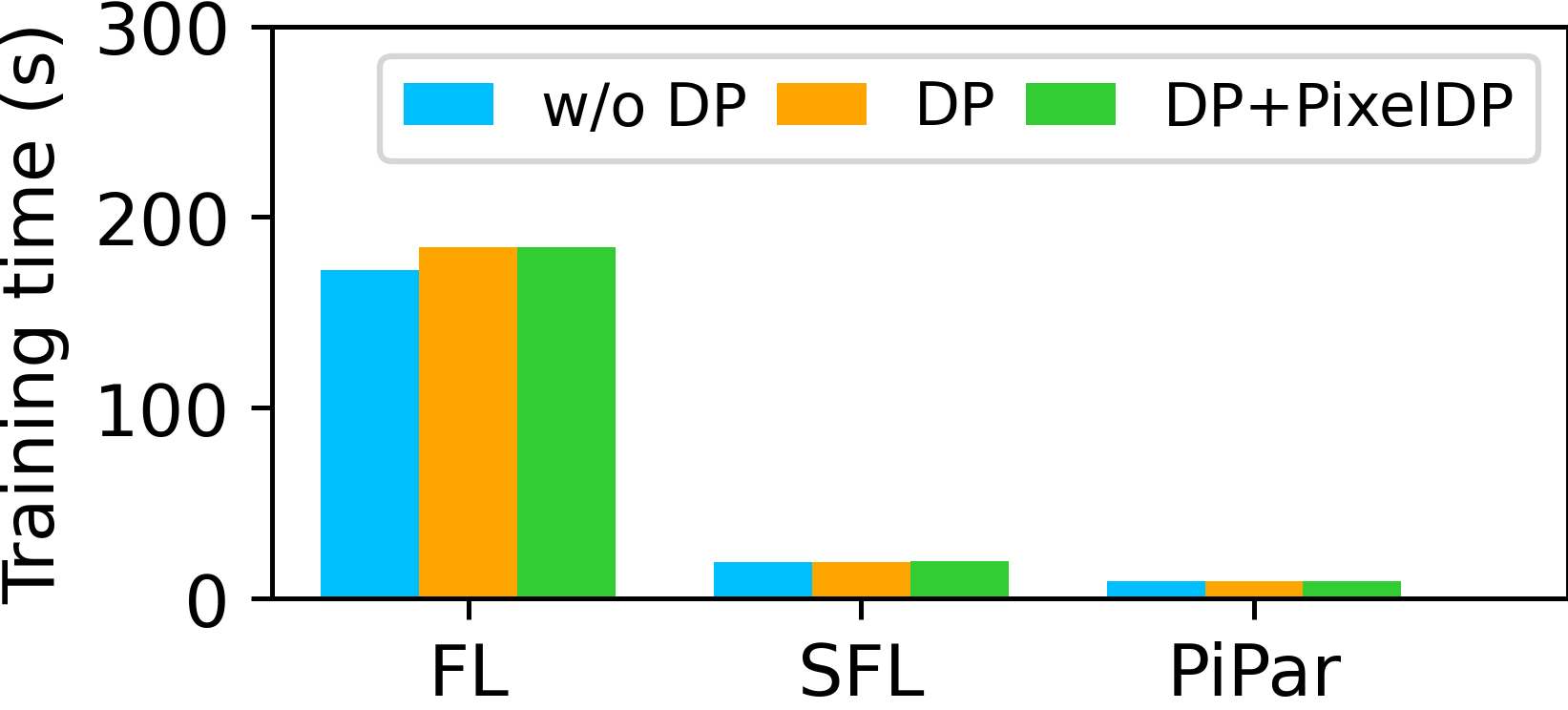}
	    \label{fig:resnet18-dp-wifi}
	    }
	\hfill
        \subfigure[MobileNetV3-Small (4G)]{
	    \includegraphics[width=0.3\textwidth]{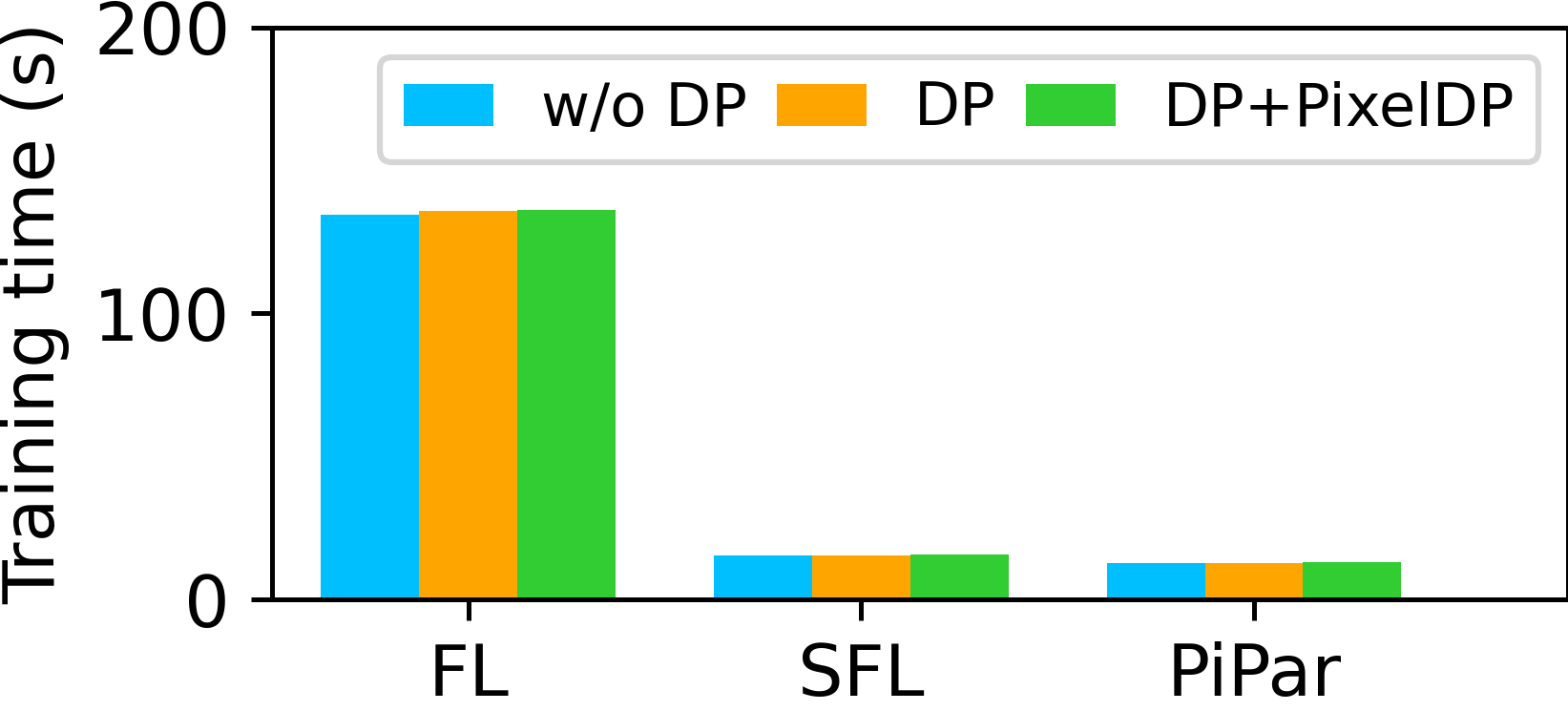}
	    \label{fig:mobilesmall-dp-4g}
	    }
	\hfill
        \subfigure[MobileNetV3-Small (4G+)]{
	    \includegraphics[width=0.3\textwidth]{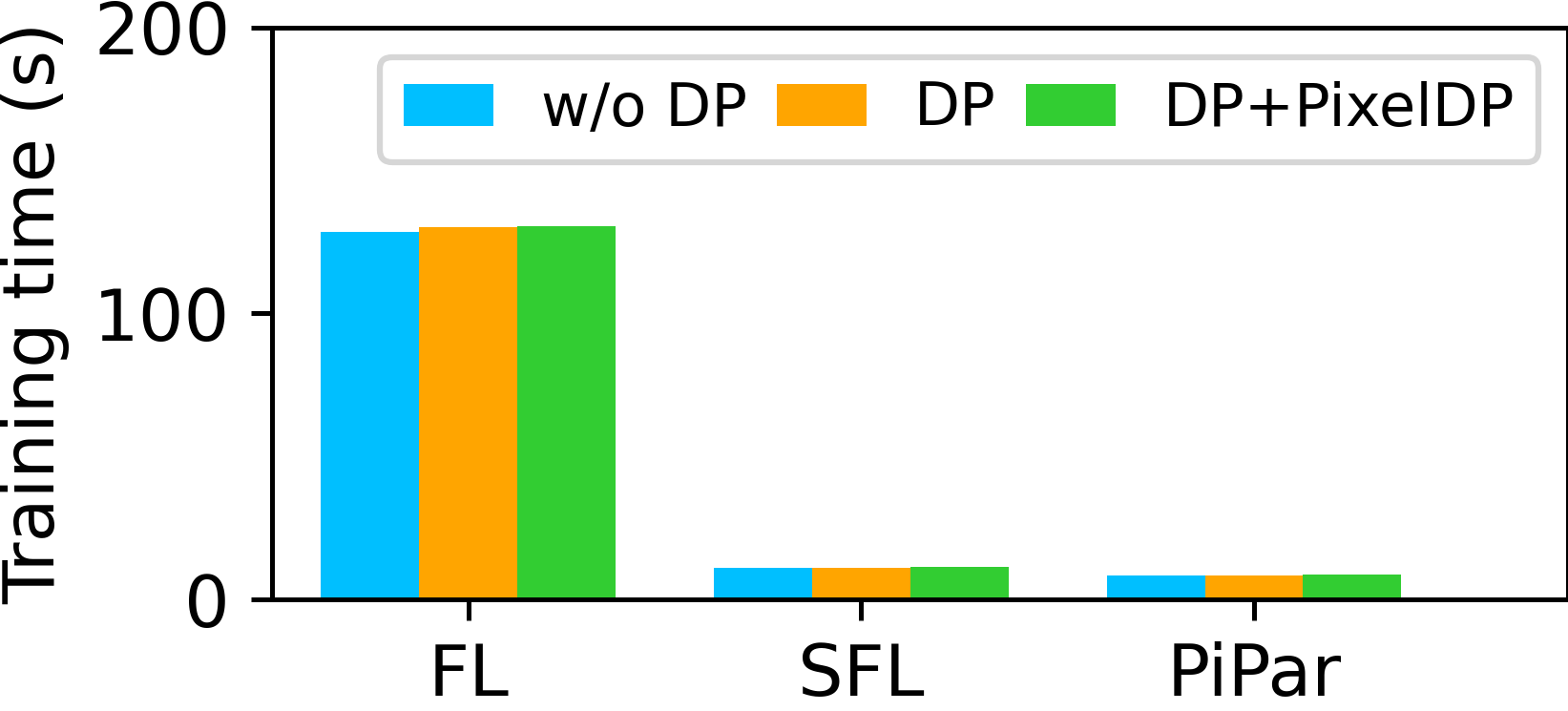}
	    \label{fig:mobilesmall-dp-4gp}
	    }
	\hfill
        \subfigure[MobileNetV3-Small (WiFi)]{
	    \includegraphics[width=0.3\textwidth]{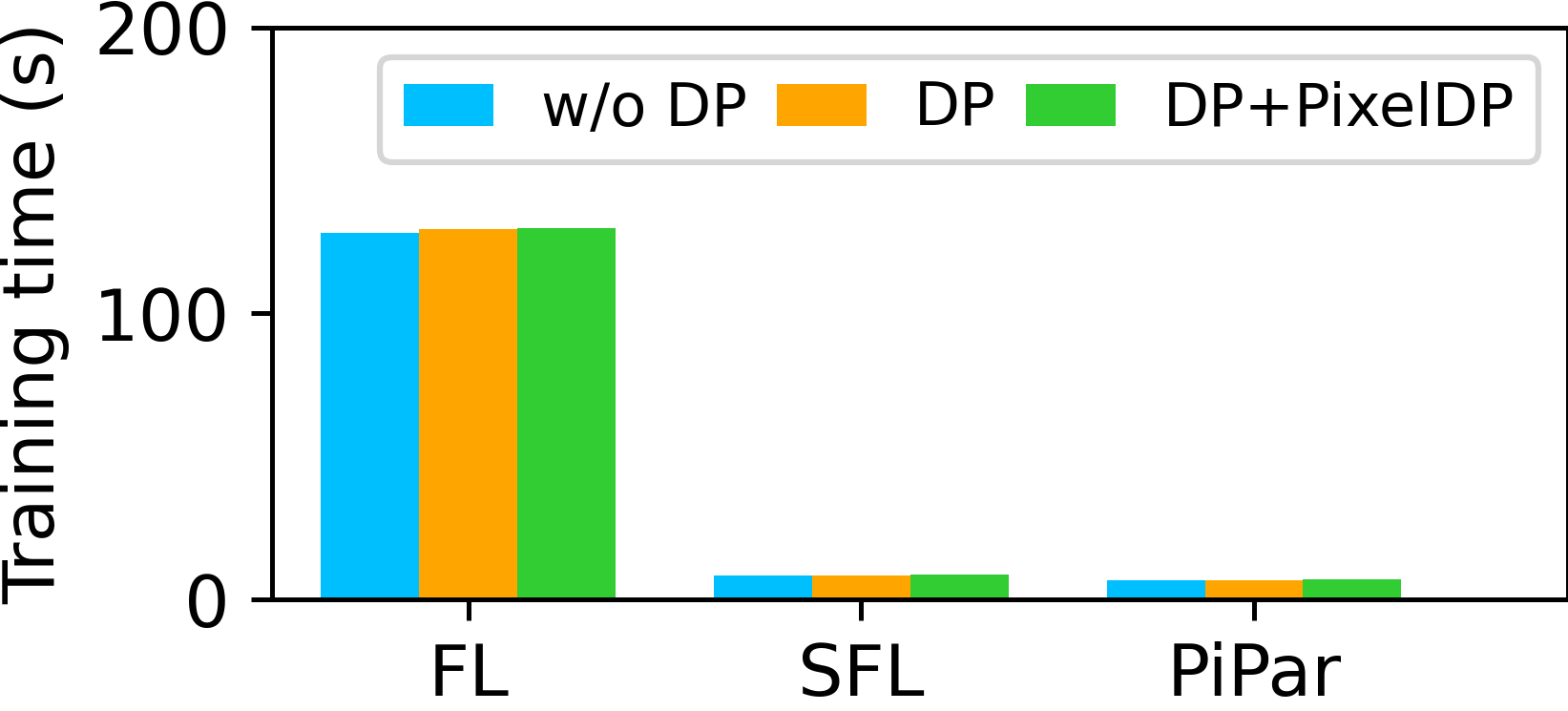}
	    \label{fig:mobilesmall-dp-wifi}
	    }
        \hfill
	\caption{Training time per epoch for FL, SFL and \PiPar\ using small DNNs on CIFAR-10 under different network conditions with and without differential privacy methods.}
	\label{fig:dp}
\end{figure*}

Figure~\ref{fig:dp} shows the training time per epoch for different CML methods with and without DP methods. Classic DP introduces an overhead of up to 11.7s to FL, 0.16s to SFL and 0.15s to \PiPar. Compared to SFL and \PiPar, FL has the largest overhead using DP because in FL the entire model is trained on the device and classic DP will add noise to each parameter in the model. The overhead of PixelDP on FL, SFL and \PiPar\ (up to 0.3s) is comparable since they use the same size of inputs and PixelDP adds noise to these inputs. The results highlight that the two DP methods applied to \PiPar\ do not introduce a larger overhead than FL and SFL.

\subsubsection{Impact of Changing Bandwidth on the Overhead for Automated Parameter Selection}
\label{subsec:bandwidth}

It was shown in Section~\ref{subsec:op-exp} that the automated parameter selection approach only needs to execute once before training in a stable network environment. Therefore, the overhead incurred is negligible. In this section, we measure the overhead of the approach in an unstable network where the bandwidth changes between 4G, 4G+ and WiFi conditions periodically in a controlled manner.

\begin{figure*}[tp]
	\centering
	\subfigure[VGG-5]{
	    \includegraphics[width=0.3\textwidth]{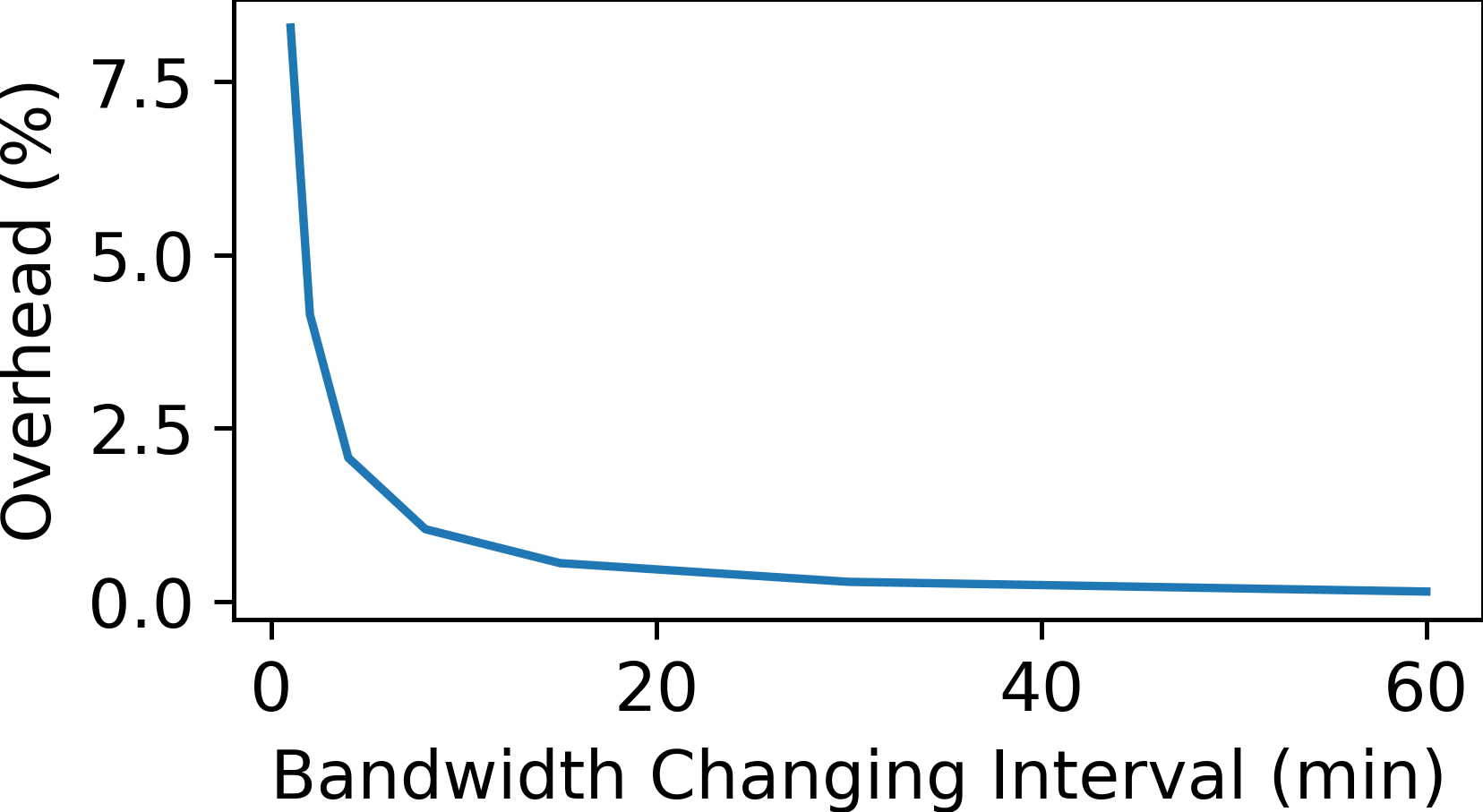}
	    \label{fig:vgg5-bc}
	    }
	\hfill
        \subfigure[ResNet-18]{
	    \includegraphics[width=0.3\textwidth]{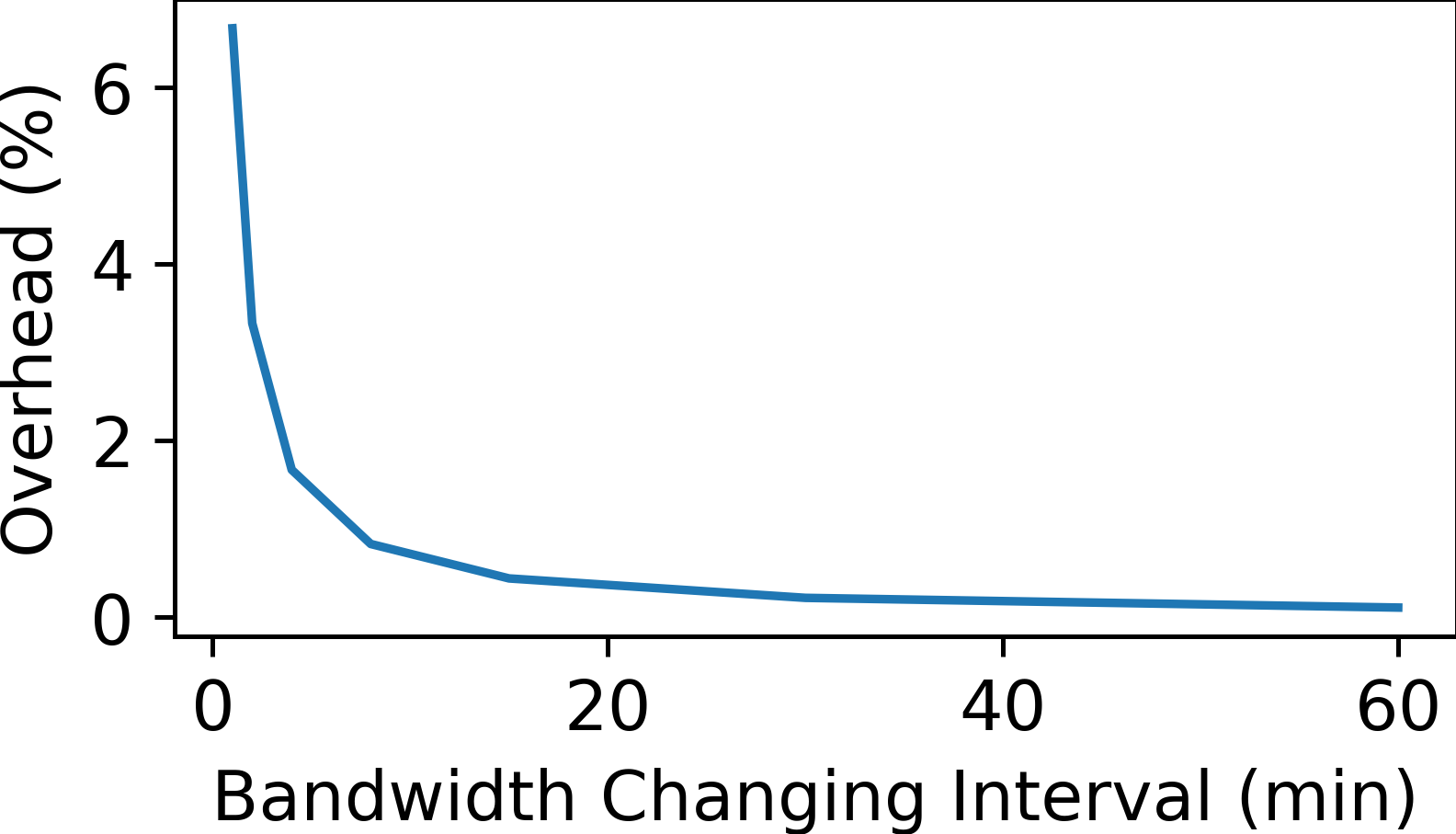}
	    \label{fig:resnet-18-bc}
	    }
	\hfill
        \subfigure[MobileNetV3-Small]{
	    \includegraphics[width=0.3\textwidth]{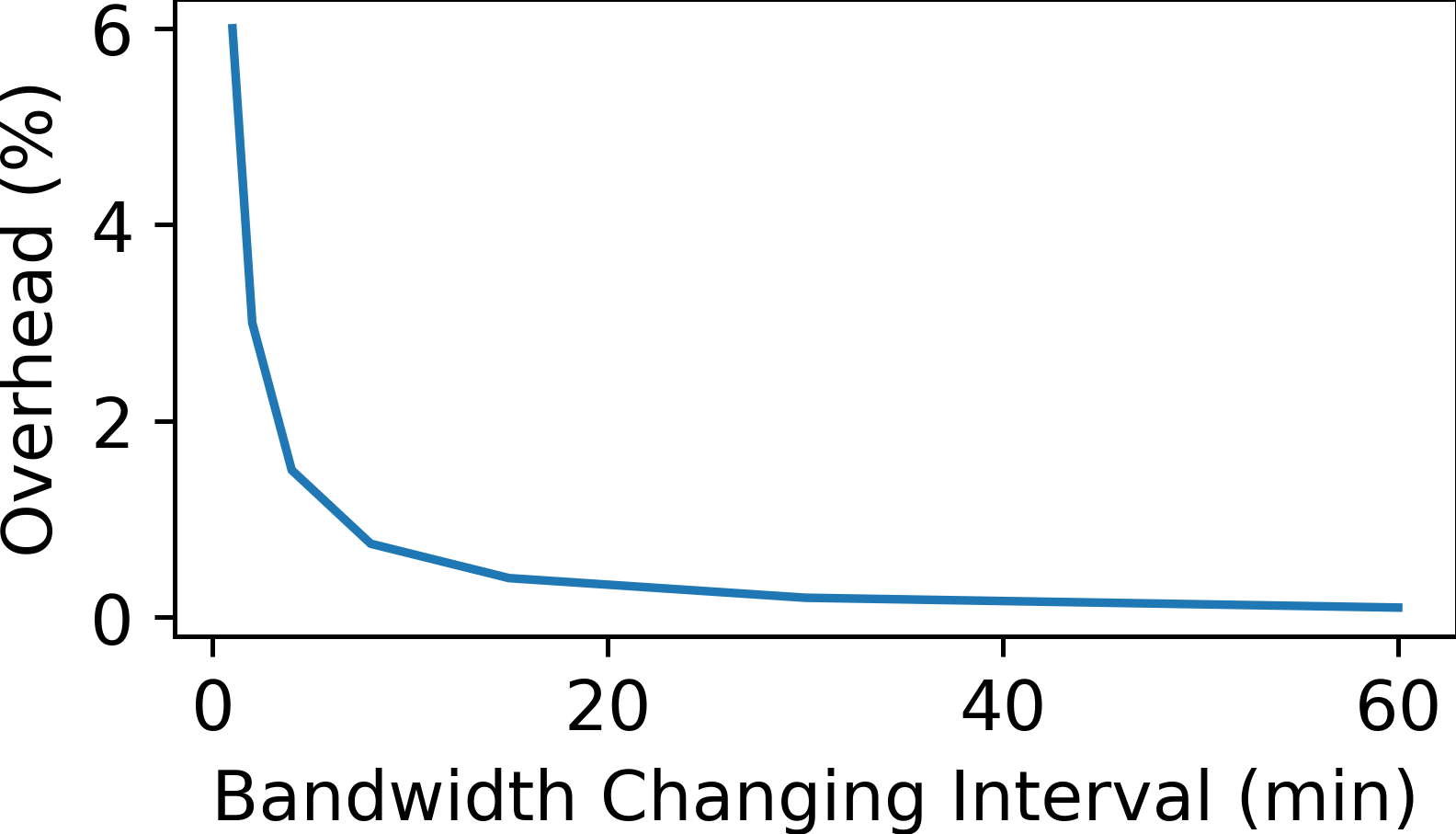}
	    \label{fig:mobilesmall-bc}
	    }
	\hfill
	\caption{The percentage overhead of the automated parameter selection approach with respect to training time for different intervals in which the bandwidth changes.}
	\label{fig:bc}
\end{figure*}

The overhead is measured for different intervals in which the bandwidth changes. The network is more unstable for smaller intervals. Figure~\ref{fig:bc} shows the percentage overhead for running the parameter selection approach with respect to training time for different intervals in which the bandwidth changes. The intervals considered range from 1 minute to 60 minutes. If the bandwidth changes every hour, the overhead of parameter selection is only 0.14\%, 0.11\% and 0.1\% of the training time for VGG-5, ResNet-18 and MobileNetV3-Small, respectively. Considering the worst case in the experiments, which is a change of bandwidth every minute, the approach overhead is up to 8.3\% of the training time. If the bandwidth change occurs on an average every 10 minutes or more then the overhead incurred is less than 1\%.

\section{Conclusion}
\label{sec:concl}
Deep learning models are collaboratively trained using paradigms, such as federated learning, split learning or split federated learning on a server and multiple devices. However, they are limited in that the computation and communication across the server and devices are inherently sequential. This results in low compute and network resource utilization and leads to idle time on the resources. We propose a novel framework, \PiPar, that addresses this problem for the first time by taking advantage of pipeline parallelism, thereby accelerating the entire training process. A novel training pipeline is developed to parallelize server-side and device-side computations as well as server-device communication. In the training pipeline, the DNN is split and deployed on the server and devices, and the training process on different mini-batches of data is re-ordered. A low overhead parameter selection approach is then proposed to maximize the resource utilization of the pipeline. Consequently, when compared to existing paradigms, our pipeline significantly reduces idle time on compute resources by up to 64.1$\times$ in training popular DNNs under different network conditions. An overall training speed up of up to 34.6$\times$ is observed. It is also experimentally demonstrated that \PiPar\ achieves performance benefits when incorporating differential privacy methods and operating in environments with heterogeneous devices and changing bandwidths.

\section*{Acknowledgment}
This work was sponsored by Rakuten Mobile, Inc., Japan.


\bibliographystyle{IEEEtran}
\bibliography{paper-arxiv-v1.bib}

\end{document}